\documentclass[11pt]{article}
\usepackage{amsfonts}
\usepackage{amssymb}
\usepackage{mathrsfs}
\usepackage{graphicx}
\usepackage{amsmath,amsthm,amssymb,amscd}
\usepackage[all]{xy}
\usepackage{subfig}
\usepackage[colorlinks,linktocpage=true,citecolor=blue]{hyperref}
\usepackage{multirow}
\usepackage{color}
\usepackage{float}
\usepackage{mathtools,slashed}
\usepackage{bbold}
\usepackage{bbm}
\usepackage{empheq}
\usepackage[normalem]{ulem}
\usepackage{cite}

\usepackage{amsxtra,graphics,epsfig,bm,tikz,xfrac,lscape}
\usetikzlibrary{decorations.pathmorphing}
\usetikzlibrary{decorations.markings}
\usetikzlibrary{arrows, decorations.markings, calc, fadings, decorations.pathreplacing, patterns, decorations.pathmorphing, positioning}

\usetikzlibrary{positioning,shapes}
\usetikzlibrary{chains}
\usetikzlibrary{arrows,fit,decorations.pathreplacing}
\tikzstyle{every picture}+=[remember picture]
\tikzstyle{na} = [baseline=-.5ex]

\DeclareMathOperator{\Vol}{Vol}
\usepackage{dsfont}
\def\RR{{\mathds{R}}}
\def\ZZ{{\mathds{Z}}}

\newcommand{\nn}{\nonumber}

\def\eqa{\begin{eqnarray}}
	\def\eqae{\end{eqnarray}}
\def\eq{\begin{equation}}
	\def\eqe{\end{equation}}
\def\be{\begin{equation}}
	\def\ee{\end{equation}}
\def\bea{\begin{eqnarray}}
	\def\eea{\end{eqnarray}}
\def\ba{\begin{array}}
	\def\ea{\end{array}}
\def\bd{\begin{displaymath}}
	\def\ed{\end{displaymath}}

\def\>{\rangle}
\def\<{\langle}

\def\i{\iota}
\def\j{\psi}

\numberwithin{equation}{section}

\newcommand{\fft}[2]{\frac{#1}{#2}}

\definecolor{darkblue}{rgb}{0,0,0.5}
\definecolor{darkred}{rgb}{0.5,0,0}
\definecolor{darkgreen}{rgb}{0,0.5,0}
\definecolor{orange}{rgb}{0.9,0.58,0}

\begin{document}
	
\begin{titlepage}
\hfill LCTP-22-03

\vskip 1 cm

\begin{center}
	
	\vskip .7cm
	
	{\Large \bf $c$\,-Functions in Flows Across Dimensions}\\
	
	\vskip .7cm
	
\end{center}

\vskip 1.7 cm

\begin{center}
	{\bf Alfredo Gonz\'alez Lezcano${}^a$, Junho Hong${}^b$, James T.  Liu${}^c$,}\\
	\vskip 0.5 cm
	{\bf Leopoldo A. Pando Zayas${}^{c,d}$  and Christoph F. Uhlemann${}^c$}\\
	\vskip .5 cm
	
\end{center}

\vskip .4cm 

\centerline{\it ${}^a$ Asia Pacific Center for Theoretical Physics}
\centerline{\it Postech, Pohang 37673, South Korea}

\bigskip

\centerline{\it ${}^b$ Institute for Theoretical Physics, KU Leuven}
\centerline{\it Celestijnenlaan 200D, B-3001 Leuven, Belgium}

\bigskip

\centerline{\it ${}^c$ Leinweber Center for Theoretical
	Physics,   Randall Laboratory of Physics}
\centerline{ \it The University of
	Michigan, Ann Arbor, MI 48109-1040, USA}

\bigskip

\centerline{\it ${}^d$ The Abdus Salam International Centre for Theoretical Physics}
\centerline{\it  Strada Costiera 11,  34014 Trieste, Italy}

\bigskip\bigskip

\vskip 1.5 cm
\begin{abstract}
We explore the notion of $c$-functions in renormalization group flows between theories in different spacetime dimensions. We discuss functions connecting central charges of the UV and IR fixed point theories on the one hand, and functions which are monotonic along the flow on the other. First, using the geometric properties of the holographic dual RG flows across dimensions and the constraints from the null energy condition, we construct a monotonic holographic $c$-function and thereby establish a holographic $c$-theorem across dimensions. Second, we use entanglement entropies for two different types of entangling regions in a field theory along the RG flow across dimensions to construct candidate $c$-functions which satisfy one of the two criteria but not both. In due process we also discuss an interesting connection between corner contributions to the entanglement entropy and the topology of the compact internal space. As concrete examples for both approaches, we holographically study twisted compactifications of 4d $\mathcal N=4$ SYM and compactifications of 6d $\mathcal N=(2,0)$ theories. 
\end{abstract}

\vskip  0.5 cm
\noindent
{\tt alfredo.gonzalez@apctp.org, junhophysics@gmail.com, jimliu@umich.edu,\\ lpandoz@umich.edu,  uhlemann@umich.edu}
\end{titlepage}

\tableofcontents

\newpage
\section{Introduction}\label{sec:intro}
Quantum field theories (QFTs) can typically be understood as renormalization group (RG) flows from a conformal field theory (CFT) in the UV to another CFT in the IR. The reduction in the effective number of degrees of freedom along the flow, due to coarse graining, is made precise by ``counting functions", which are monotonic along RG flows. Ideally, such functions interpolate between distinguished quantities in the CFTs, i.e.\ a-type central charges in even dimensions and sphere free energies in odd dimensions. In the following we will refer to both as central charges. The paradigmatic c-theorem establishing an RG monotone in 2d was shown by Zamolodchikov \cite{Zamolodchikov:1986gt}, the a-theorem in 4d by Komargodski/Schwimmer \cite{Cardy:1988cwa,Komargodski:2011vj}, and the 3d F-theorem by Casini/Huerta \cite{Jafferis:2010un,Klebanov:2011gs,Jafferis:2011zi,Casini:2012ei}.
Partial results for 6d can be found in \cite{Elvang:2012st,Cordova:2015fha,Heckman:2015axa} and for 5d in \cite{Jafferis:2012iv,Chang:2017cdx,Fluder:2020pym}.
The language of entanglement entropy (EE) has provided particularly useful perspectives on RG flows and allowed for a uniform proof of monotonicity results in $d=2,3,4$ \cite{Casini:2006es,Casini:2017roe,Casini:2017vbe}.

In this work we explore the notion of counting functions in RG flows across dimensions, meaning the compactification of a $D$-dimensional CFT, which is the UV fixed point, on a $D-d$ dimensional compact space, such that the IR fixed point is a $d$-dimensional CFT. The goal of this manuscript is to study $c$-functions along such RG flows across dimensions. This endeavor faces some immediate obstacles. Firstly, the aforementioned different nature of the ``central charges'' in even and odd dimensions, i.e.\ a-type conformal anomaly {\it versus} \ sphere free energy, poses a technical challenge for interpolating functions in flows from even to odd dimensions or vice versa. Secondly, and perhaps more importantly, central charges in different dimensions count different notions of degrees of freedom: a free scalar in 4d and a free scalar in 2d each gives a finite contribution to the central charge in 4d and in 2d, respectively. Yet, if the 4d scalar is compactified to 2d it gives rise to an infinite number of 2d scalar fields as Kaluza-Klein (KK) modes.
For functions interpolating between central charges in different dimensions this simple fact challenges intuitive arguments for monotonicity based on coarse graining: with different types of degrees of freedom counted in the UV and IR, there is no obvious reason for the change along the flow to be of definite sign.
One might hope that, despite this obstacle to intuitive arguments, some general notion of monotonicity for functions interpolating between the standard notion of central charges in different dimensions can be established. This is not the case, as we discuss more explicitly in the main part.
We therefore decouple the two criteria on our wish list for a $c$-function and consider two types of ``partial $c$-functions", which partially fulfill the criteria, namely
\begin{itemize}
	\item[(i)] $c_{\rm int}$ interpolating between central charges of the UV CFT$_D$ and the IR CFT$_{d}$,
    \item[(ii)] $c_{\rm mono}$ which are monotonic along RG flows across dimensions.
\end{itemize}
The first type of function extends the notion of central charge away from the fixed points. The second type of function captures the loss of degrees of freedom due to coarse-graining. In flows between CFTs in the same dimension, both criteria are satisfied simultaneously by the $c$-functions mentioned  in the opening paragraph. For flows across dimensions we settle for functions which (in general) satisfy one of the criteria but not both. 
The functions $c_{\rm mono}$ we will define change monotonically along the flows, but diverge at the UV fixed point. This is reminiscent of RG flows in 3d, where the free energy of free Maxwell theory diverges in the UV \cite{Giombi:2015haa}.

The tool we use for the exploration of partial $c$-functions in concrete RG flows across dimensions is AdS/CFT. Indeed, various holographic c-theorems were established first in holography as consequences of Einstein equations and the Null Energy Condition (NEC) \cite{Girardello:1998pd,Freedman:1999gp,Myers:2010xs,Myers:2010tj}. Holography has also provided a stage to entropically approach monotonicity theorems, via the Ryu-Takayanagi (RT) prescription for the holographic computation of entanglement entropy (EE) associated with sphere \cite{Casini:2011kv} and strip \cite{Myers:2012ed} entangling regions.

In this manuscript we construct partial $c$-functions in holographic RG flows across dimensions based on the NEC and the EE as it has been done in flows within the same dimension. There is a plethora of supergravity solutions describing the duals of RG flows interpolating between CFTs in different dimensions \cite{Maldacena:2000mw,Acharya:2000mu,Gauntlett:2000ng,Gauntlett:2001qs,Gauntlett:2001jj,Benini:2013cda,Benini:2015bwz,Bobev:2017uzs}, and we will use such solutions to explore partial $c$-functions in holographic RG flows across dimensions.

First, we construct a partial $c$-function in flows across dimensions by analyzing direct implications of the NEC as it was done before the advent of the entropic approach in flows within the same dimension. The result will be a new holographic $c$-theorem for flows across dimensions with a partial $c$-function in the sense of (ii) above with a divergent UV behavior. To the best of our knowledge, this approach was not considered before in flows across dimensions.

Next, to construct partial $c$-functions from the EE, we explore two types of spherical entangling regions: (a) ``all-spherical'' regions which are (generalizations of) spherical in all directions, and (b) ``wrapping'' regions which are spherical only in the non-compact directions, while wrapping the compact space on which the field theory is compactified. We then can build candidate $c$-functions from two ingredients, namely a choice of entangling region and a choice of differential operator which acts on the EE and extracts the candidate $c$-function:
\begin{itemize}
    \item From the ``all-spherical'' entangling region, one can, in principle, construct $c$-functions interpolating between the (a-type) central charges of the UV CFT$_D$ and the IR CFT$_d$ in the sense of (i) above by choosing the differential operators appropriately; we have examples showing that such function will not generically be monotonic. Nevertheless, such function can be shown to encode, in an interesting way, information about the topology of the compact space by means of corner contribution. 
    
    \item From the ``wrapping'' entangling region, we construct partial $c$-functions in the sense of (ii) above with a particular choice of differential operators acting on the EE. They are monotonic along the flow and connect to the lower-dimensional central charge in the IR  but they diverge in the UV. These $c$-functions count the natural notion of degrees of freedom from the lower-dimensional perspective. They make precise the intuition that the entire flow may be described from the lower-dimensional perspective as a flow from a $d$-dimensional UV theory with infinitely many fields (KK modes) to a more standard $d$-dimensional IR theory (which retains only the lowest KK modes). With a different choice of differential operator, we also construct interpolating $c$-functions in the sense of (i) above, which interpolate between linear combinations of central charges in the UV theory and the central charge in the IR theory.
\end{itemize}
We note that candidate holographic $c$-functions in RG flows across dimensions were previously studied in \cite{Macpherson:2014eza,Bea:2015fja,Legramandi:2021aqv}. These functions were inspired by the work of \cite{Klebanov:2007ws,Myers:2012ed} on the strip EE and defined as combinations of warp factors in the holographic metric which are local in the holographic radial coordinate.
Here we construct  partial $c$-functions explicitly from the EE for two different types of spherical entangling regions in RG flows across dimensions, by acting with appropriate differential operators on the EE's. These functions are non-local in the holographic radial coordinate but have the advantage of a clear field theory interpretation.

The rest of the manuscript is organized as follows. In section \ref{sec:NEC} we discuss the implications of the NEC for holographic RG flows across dimensions, which yields a partial holographic $c$-function in the sense of (ii) above. In section \ref{sec:EE} we first describe qualitatively the entangling regions involved in tracking EE along RG flows between CFTs in different dimensions. We then discuss partial $c$-functions extracted from the EE in the sense of either (i) or (ii) based on the qualitative description of entangling regions, and also construct those partial $c$-functions holographically. In section \ref{sec:EX} we provide partial holographic $c$-functions motivated either by the NEC or by the EE, in concrete examples of holographic RG flows between even-dimensional CFTs (or odd-dimensional AdS spaces). A summary and conclusions are given in section \ref{Sec:SumCon}. We relegate a number of technical details to Appendices.

\section{Holographic $c$-functions across dimensions from the NEC}\label{sec:NEC}
In the context of AdS/CFT, certain conformal field theories, CFT$_D$, are  dual  to supergravity solutions containing an AdS$_{D+1}$ factor. A central entry in the holographic dictionary roughly identifies the radial direction in AdS$_{D+1}$ with the energy scale in the field theory. 
It was understood early on that when the gravity solution is a domain wall that interpolates between two AdS$_{D+1}$ regions, its field theory interpretation is that of an RG flow \cite{Girardello:1998pd,Freedman:1999gp}. Based on the holographic identification of central charges \cite{Henningson:1998gx}, the authors of \cite{Freedman:1999gp} were able to demonstrate a holographic c-theorem by identifying a quantity that does not increase along the RG flow. This approach  also applies  to gravity theories with higher-curvature corrections, that is to gravity theory duals to a situation when the $a$ central charge and the $c$ central charge in four-dimensional field theories are not equal \cite{Myers:2010xs}. Similar considerations were extended to any dimension $d$ in \cite{Myers:2010tj}. Also refer to \cite{Chu:2019uoh} for a candidate $c$-function in holographic CFTs without Lorentz symmetry and the necessary and the sufficient conditions for its monotonicity.

Consequently, a supergravity solution interpolating between two AdS regions of different dimension is viewed as an RG flow between two CFTs of different dimensions. The metric takes the general form
\begin{equation}
	ds^2_D=e^{2f(z)}\left(\eta_{\mu\nu}dx^\mu dx^\nu+ dz^2\right) + e^{2g(z)} g_{ij}(y)dy^i dy^j,\label{flow:across}
\end{equation}
where $z$ is the holographic coordinate and the functions $f$ and $g$ have a particular behavior describing the interpolation. The coordinates $y^i$ described the compact manifold of dimension $p$ on which the higher dimensional CFT$_D$ is wrapped. The coordinates $\mu,\nu=0,.., (d-1)$, with $d=D-p$, ultimately describe the space where the IR CFT$_{d}$ lives. In the UV, the worldvolume of the CFT$_D$ is  collectively described by $(x^\mu, y^i)$.

To be rigorous, the metric (\ref{flow:across}) is not the most general geometry that describes holographic RG flows across dimensions. For example, there are known holographic RG flows where the $p$-dimensional `internal' geometry depends on the holographic radial coordinate $z$ in a non-separable way \cite{Anderson:2011cz,Bobev:2020jlb}. In this section we focus on the separable case (\ref{flow:across}) for simplicity; extending our analysis of holographic $c$-functions in separable flows (\ref{flow:across}) to more general non-separable flows is left for future research.

In this section we formulate holographic $c$-functions in holographic RG flows across dimensions (\ref{flow:across}) using the Null Energy Condition (NEC). In subsection \ref{sec:NEC:same} we first review the monotonicity of a holographic $c$-function in flows within the same dimension based on the NEC. In subsection \ref{sec:NEC:across} we move on to flows across dimensions and explore if one can formulate a monotonic holographic $c$-function using the NEC as in flows within the same dimension. Throughout this section, all the flows are assumed to be solutions of the Einstein equations.

\subsection{Flows within the same dimension}\label{sec:NEC:same}
For holographic domain wall flows within the same dimension, the NEC immediately yields a monotonic $c$-function along the flow \cite{Freedman:1999gp}.  Before considering flows across dimension, it is instructive to examine the proof of this holographic $c$-theorem.  For a planar AdS$_{D+1}$ domain wall flow, consider the metric
\begin{equation}
	ds^2=e^{2f(z)}(\eta_{\mu\nu}dx^\mu dx^\nu+dz^2).
\end{equation}
A simple calculation gives the following components of  the Riemann and Ricci  tensors
\begin{align}
	R_{\mu\nu\rho\sigma}&=-e^{2f}(f')^2(\eta_{\mu\rho}\eta_{\nu\sigma}-\eta_{\mu\sigma}\eta_{\nu\rho}),&R_{\mu\nu}&=-[f''+(D-1)(f')^2]\eta_{\mu\nu}\nn\\
	R_{\mu z\nu z}&=-e^{2f}f''\eta_{\mu\nu},&R_{zz}&=-Df''.
\end{align}
Note that the vacuum AdS solution is given by $f(z)=-\log(z/L)$, where $L$ is the AdS radius.  In this coordinate system, the UV is located at $z=0$ and the IR at $z=\infty$ so flow to the IR corresponds to increasing $z$.

We now consider the NEC, which is the statement that $T_{MN}\xi^M\xi^N\ge0$ where $\xi^M$ is a future directed null vector.  The Einstein equation then converts the condition on the stress-energy into a condition on the geometry, namely, $R_{MN}\xi^M\xi^N\ge0$.  Because of the planar symmetry, the only non-trivial condition comes from choosing the null vectors  $\xi=\partial/\partial t\pm\partial/\partial z$:
\begin{equation}
	R_{MN}\xi^M\xi^N=R_{tt}+R_{zz}=(D-1)[(f')^2-f'']\ge0.
\end{equation}
This can be expressed as
\begin{equation}
	(D-1)(e^{-f})''\ge0,
	\label{eq:apnec}
\end{equation}
which indicates that the function $e^{-f}$ cannot be concave down.  Noting that vacuum AdS corresponds to $e^{-f}=z/L$, we can define an effective AdS radius
\begin{equation}
	L_{\mathrm{eff}}(z)=\fft1{(e^{-f})'},
	\label{eq:Leff}
\end{equation}
along with a $c$-function
\begin{equation}
	c(z)=\fft{L_{\mathrm{eff}}(z)^{D-1}}{G_N},
	\label{eq:cfuncz}
\end{equation}
where $G_N$ is the $(D+1)$-dimensional Newton's constant.  Application of the NEC, (\ref{eq:apnec}), then demonstrates
\begin{equation}
	c'(z)=-(D-1)\fft{L_{\mathrm{eff}}(z)^D}{G_N}\left(\fft1{L_{\mathrm{eff}}(z)}\right)'=-(D-1)\fft{L_{\mathrm{eff}}(z)^D}{G_N}(e^{-f})''\le0,
\end{equation}
so that $c(z)$ is monotonic non-increasing along flows to the IR.

Instead of working with the $c$-function, (\ref{eq:cfuncz}), it is perhaps more direct to consider the flow of the effective AdS radius, (\ref{eq:Leff}).  In this case, we find
\begin{equation}
	L_{\mathrm{eff}}(z)'=-L_{\mathrm{eff}}^2\left(\fft1{L_{\mathrm{eff}}(z)}\right)'=-L_{\mathrm{eff}}^2(e^{-f})''\le0,
	\label{eq:Leffdec}
\end{equation}
which is an equivalent statement that the effective AdS radius is non-increasing along flows to the IR.  One way to visualize such flows is to consider $e^{-f}$ as a function of $z$.  This function starts at zero when $z=0$ (the UV) and increases in a manner that is never concave down as $z$ grows to the IR.  The local slope of this curve then gives the inverse of the effective AdS radius at that point along the flow.  Since the slope is monotonic increasing by the NEC, the effective radius is monotonic decreasing, which is the essence of the statement (\ref{eq:Leffdec}).

\subsection{A holographic $c$-theorem across dimensions}\label{sec:NEC:across}
We now attempt to generalize this analysis to flows across dimensions.  Along these lines, we now consider a holographic domain-wall flow from AdS$_{D+1}$ to AdS$_{d+1}$ with the metric
\begin{equation}
	ds^2=e^{2f(z)}(g_{\mu\nu}(x)dx^\mu dx^\nu+dz^2)+e^{2g(z)}g_{ij}(y)dy^idy^j,
	\label{eq:mnmet}
\end{equation}
where $\mu,\nu=0,1,\ldots,d-1$ and $i,j=1,2,\ldots,D-d$.  Note that we allow for curved slices of AdS$_{d+1}$ as well as a curved internal manifold, generalizing (\ref{flow:across}).

This more general metric yields the Ricci tensor
\begin{align}
	\hat R^\mu_\nu&=e^{-2f}R^\mu_\nu-e^{-2f}[f''+f'((d-1)f'+(D-d)g')]\delta^\mu_\nu,\nn\\
	\hat R^i_j&=e^{-2g}R^i_j-e^{-2f}[g''+g'((d-1)f'+(D-d)g')]\delta^i_j,\nn\\
	\hat R^z_z&=-e^{-2f}[df''+(D-d)(g''+g'(g'-f'))],
\end{align}
where $\hat R^M_N$ is computed in the full $(D+1)$-dimensional metric, while the unhatted Riccis are computed with the corresponding $g_{\mu\nu}$ and $g_{ij}$ metrics.  We now consider the NEC,   $R_{MN}\xi^M\xi^N\ge0$ where this time we can choose the null vector to lie along either $t$-$x$, $t$-$z$ or $t$-$y$.  The result is
\begin{align}
	R^x_x-\hat R^t_t&=e^{-2f}(R^x_x-R^t_t)&\ge0,\nn\\
	\hat R^z_z-\hat R^t_t&=-e^{-2f}[R^t_t+(d-1)(f''-(f')^2)+(D-d)(g''+g'(g'-2f'))]&\ge0,\nn\\
	\hat R^y_y-\hat R^t_t&=e^{-2g}R^y_y+e^{-2f}[-R^t_t+f''-g''+(f'-g')((d-1)f'+(D-d)g')]&\ge0.
\end{align}
The first condition is just the NEC in the spacetime of $g_{\mu\nu}$, while the remaining two provide non-trivial constraints on the flow.

We now specialize to flat slicings of AdS$_{d+1}$ and take the `internal' metric, $g_{ij}$, to have constant curvature
\begin{equation}
	R_{\mu\nu}=0,\qquad R_{ij}=\kappa\fft{D-d-1}{\ell^2}g_{ij},
\end{equation}
where $\kappa=1$, $0$ or $-1$ for positive, flat or negative curvature, respectively. Analysis for more general internal metric with a non-constant curvature will be left for future research. Then we have two conditions
\begin{subequations}
	\begin{align}
		\mbox{NEC1 :}&\quad-(d-1)(f''-(f')^2)-(D-d)(g''+g'(g'-2f'))&\ge0,\label{eq:NEC1}\\
		\mbox{NEC2 :}&\quad(f'-g')'+(f'-g')((d-1)f'+(D-d)g')+\kappa\fft{D-d-1}{\ell^2}e^{2f-2g}&\ge0.\label{eq:NEC2}
	\end{align}\label{eq:NECs}%
\end{subequations}
Before investigating these conditions, we first recall the parameters of the flow.

Assuming a flow from AdS$_{D+1}$ in the UV to AdS$_{d+1}$ in the IR, we have the boundary conditions at the endpoints of the flow
\begin{align}
	\mbox{UV ($z=0$) :}\quad&(e^{-f})'=(e^{-g})'=\fft1{L_{\mathrm{UV}}},\nn\\
	\mbox{IR ($z=\infty$) :}\quad&(e^{-f})'=\fft1{L_{\mathrm{IR}}},\qquad (e^{-g})'=0.
	\label{eq:UVIRasymp}
\end{align}
In addition, we can define the \textit{unnormalized} holographic central charges
\begin{equation}
	a_{\mathrm{UV}}=(L_{\mathrm{UV}})^{D-1},\qquad a_{\mathrm{IR}}=\ell^{D-d}e^{(D-d)g_{\mathrm{IR}}}(L_{\mathrm{IR}})^{d-1},
	\label{eq:aUVaIR}
\end{equation}
where $g_{\mathrm{IR}}=\lim_{z\to\infty}g(z)$.  Here the factor $(\ell e^{g_{\mathrm{IR}}})^{D-d}$ arises from the volume of the internal space.  Since the lower-dimensional factor $e^{2f}$ governs the AdS geometry in the UV and the IR, we can continue to use the definition (\ref{eq:Leff}) for the effective AdS radius along the flow.  This allows us to write
\begin{equation}
	a_{\mathrm{UV}}=L_{\mathrm{eff}}(z)^{D-1}\Big|_{z\to0},\qquad a_{\mathrm{IR}}=\ell^{D-d}e^{(D-d)g(z)}L_{\mathrm{eff}}(z)^{d-1}\Big|_{z\to\infty}.
\end{equation}

Since we are interested in flows to the IR, the form of $a_{\mathrm{IR}}$ suggests that we define a holographic $c$-function of the form
\begin{equation}
	\tilde c(z)\equiv\ell^{D-d}e^{(D-d)g(z)}L_{\mathrm{eff}}(z)^{d-1}=\fft{\ell^{D-d}e^{(D-d)g(z)}}{((e^{-f(z)})')^{d-1}}.
\end{equation}
However, this is not the only function that approaches $a_{\mathrm{IR}}$ in the IR.  In particular, since $g(z)$ approaches a fixed constant in the IR, we could alternatively define the local holographic (LH) $c$-function
\begin{empheq}[box=\fbox]{equation}
	c_\text{LH}(z)=\fft{\ell^{D-d}}{((e^{-\tilde f(z)})')^{d-1}}.
	\label{eq:cfunction?}
\end{empheq}
where $\tilde f(z)$ is an `effective' IR warp factor
\begin{equation}
	\tilde f(z)\equiv f(z)+\fft{D-d}{d-1}g(z).
	\label{eq:tildef}
\end{equation}
We will examine the properties of this partial $c$-function below.  However, here we note that this flows to $a_{\mathrm{IR}}$ in the IR, but blows up in the UV because $e^{-\tilde f(z)}\sim(z/L_{\mathrm{UV}})^{\fft{D-1}{d-1}}$, resulting in $c_\text{LH}(z)\sim1/z^{D-d}$ as $z\to0$. Thus this does \textit{not} interpolate between UV and IR central charges, except when $D=d$, where the flow preserves dimension.  From the IR or lower dimensional point of view, this blowing up is not a surprise, as the internal space decompactifies in the UV.  The higher dimensional theory in the UV, when viewed by a lower dimensional observer, then has an infinite number of lower-dimensional degrees of freedom.  Alternatively, we can consider using $L_{\mathrm{eff}}(z)$ in (\ref{eq:Leff}) as proxy for the central charges.  This can be thought of as taking a UV perspective, as $L_{\mathrm{eff}}(z)$ is directly related to $a_{\mathrm{UV}}$.  However, in the IR, this does not yield $a_{\mathrm{IR}}$ as it does not capture the volume of the internal space.

While neither $c_\text{LH}(z)$ in (\ref{eq:cfunction?}) nor $L_{\mathrm{eff}}(z)$ in (\ref{eq:Leff}) flow from $a_{\mathrm{UV}}$ to $a_{\mathrm{IR}}$, we can nevertheless ask whether they are monotonic along the flow.  For $c_\text{LH}(z)$, the question is whether it flows monotonically from infinity in the UV to $a_{\mathrm{IR}}$ in the IR.  For $L_{\mathrm{eff}}(z)$ one might hope to find it to decrease along the flow, just as it would in flows preserving dimension.  

Before addressing these questions, we point out a few general features of the NEC inequalities, (\ref{eq:NECs}). The NEC2 (\ref{eq:NEC2}) can be rewritten as
\begin{equation}
	\mbox{NEC2 :}\quad(e^{(d-1)f+(D-d)g}(f'-g'))'\ge-\kappa\fft{D-d-1}{\ell^2}e^{(d+1)f+(D-d-2)g}.
\end{equation}
The right-hand side vanishes for $\kappa=0$, while it is strictly positive for $\kappa=-1$.  Thus for these two cases we can conclude that the function
\begin{equation}
	\mathcal C=e^{(d-1)f+(D-d)g}(f'-g'),
\end{equation}
is monotonically non-decreasing towards the IR in the sense $\mathcal C'\ge0$ for $\kappa=0$ and $\mathcal C'>0$ for $\kappa=-1$.  Furthermore, assuming the IR asymptotics, (\ref{eq:UVIRasymp}), we see that
\begin{equation}
	\mathcal C(z)\sim-\fft{e^{(D-d)g_{\mathrm{IR}}}}{L_{\mathrm{IR}}}\left(\fft{L_{\mathrm{IR}}}{z-z_0}\right)^d<0\qquad(z\to\infty),
\end{equation}
where $z_0$ is a constant `phase shift'. Since this is negative and a non-decreasing function towards the IR, we see that $\mathcal C$ must also be negative in the UV.  This shows that $f'<g'$ in the UV and that this is a strict inequality, at least for the $\kappa=0$ and $\kappa=-1$ cases.  Heuristically, NEC2 (\ref{eq:NEC2}) provides a constraint on how fast the slopes $f'$ and $g'$ can diverge from each other along the flow, but otherwise does not relate directly to the central charges $a_{\mathrm{UV}}$ and $a_{\mathrm{IR}}$.

We now turn to NEC1 (\ref{eq:NEC1}), which reduces to the simple expression (\ref{eq:apnec}) when $D=d$, and which hence may be expected to be more closely related to the central charges.  Note that NEC1 can be rewritten in terms of the effective warp factor $\tilde f$ as
\begin{equation}
	(e^{-\tilde f})''\ge\fft{(D-1)(D-d)}{(d-1)^2}e^{-\tilde f}(g')^2.
\end{equation}
Since $e^{-\tilde f}$ is positive and $(g')^2$ is non-negative, we find
\begin{equation}
	(e^{-\tilde f})''\ge0,
	\label{eq:tf''}
\end{equation}
which is a generalization of (\ref{eq:apnec}).  This is all that is needed to demonstrate that the local holographic $c$-function defined in (\ref{eq:cfunction?}) is monotonic along flows to the IR in that
\begin{empheq}[box=\fbox]{equation}
	c_\text{LH}(z)'\le0.
\end{empheq}
This is the closest we come to a holographic $c$-theorem across dimensions. Hence the LH $c$-function (\ref{eq:cfunction?}) corresponds to a partial $c$-function in the sense of (ii) introduced in section \ref{sec:intro}, but here we name it $c_\text{LH}$ instead of $c_\text{mono}$ to distinguish (\ref{eq:cfunction?}) from another partial $c$-function in the sense of (ii) which will be constructed by exploring the EE in the following sections \ref{sec:EE} and \ref{sec:EX}. Observe that (\ref{eq:cfunction?}) is defined as a local function in the bulk geometry: the partial $c$-functions from the entropic approach, on the other hand, will be constructed based on RT surfaces in the bulk geometry and thereby given as non-local functions. 

The partial $c$-function (\ref{eq:cfunction?}), however, does not interpolate between $a_{\mathrm{UV}}$ and $a_{\mathrm{IR}}$ since it diverges in the UV. This can be understood as the inherent difficulty of defining the number of degrees of freedom in a theory which grows additional dimensions in the UV. A simple way to visualize this is to consider KK reducing a scalar field, which naturally corresponds to infinitely many fields in the lower-dimensional theory. Notice, however, that in the lower-dimensional theory only the zero mode is massless and that all other fields are massive. This situation might provide a window for a well-defined answer to the question of how many degrees of freedom contribute below certain scale.

\section{(Holographic) $c$-functions across dimensions from EE}\label{sec:EE}
In this section we discuss interpolating $c$-functions in the sense of (i), which connect central charges of a higher-dimensional UV theory to those of a lower-dimensional IR theory, and counting $c$-functions in the sense of (ii), which are monotonic along the RG flow across dimensions. We will extract these partial $c$-functions from the EE associated with two different entangling regions. In subsection \ref{sec:EE:FT} we first explore them along the RG flow across dimensions in the field theory side. In subsection \ref{sec:EE:holo} we investigate them in holographic RG flows across dimensions.

\subsection{$c$-functions across dimensions from EE}\label{sec:EE:FT}
Let us first briefly review the construction of a $c$-function along the RG flow between CFTs within the same dimension in terms of the EE for a spherical entangling region. The starting point is the expansion of the EE for a spherical entangling region with respect to a dimensionless ratio $\fft{R}{\epsilon}$ where $R$ and $\epsilon$ denote the radius of the entangling region and the UV short distance cutoff, respectively. To be specific, for a $D$-dimensional CFT on flat space $\RR^{1,D-1}$, the EE for a $(D-1)$-dimensional spatial ball of radius $R$, $B^{D-1}$, takes the general form \cite{Ryu:2006bv,Ryu:2006ef}
\begin{equation}
\label{Eq:EE-General}
    S_\text{EE}(R;B^{D-1},\epsilon)=\mu_{D-2}\fft{R^{D-2}}{\epsilon^{D-2}}+\mu_{D-4}\fft{R^{D-4}}{\epsilon^{D-4}} + \ldots +
    \begin{cases}
	(-1)^{\frac{D}{2}-1} \, 4A \, \log (R/\epsilon) &(D~\text{even})\\
	(-1)^{\frac{D-1}{2} }\,\,F & (D~\text{odd})
    \end{cases}~.
\end{equation}
The central charge is then given by the universal part, namely $A$ in even dimensions and $F$ in odd dimensions.

For a 2-dimensional QFT along the RG flow ($D=2$), the $c$-function was constructed in terms of a logarithmic derivative of the EE as \cite{Casini:2006es} 
\begin{equation}
    c(R)=R\partial_R S_\text{EE}(R;B^1,\epsilon).\label{c:CT}
\end{equation}
This 2-dimensional $c$-function (\ref{c:CT}) interpolates between the UV and IR central charges at the conformal fixed points and decreases monotonically along the RG flow.

For $D>2$, the construction of a $c$-function from the entropic point of view becomes more subtle. The authors of \cite{Liu:2012eea}, for example, considered a function
\begin{equation}
    c_\text{LM}(R)=\begin{cases}
        \fft{1}{(d-2)!!}R\partial_R(R\partial_R-2)\cdots(R\partial_R-(d-2))S_\text{EE}(R;B^{D-1},\epsilon) & ($D$\text{ even})\\
        \fft{1}{(d-2)!!}(R\partial_R-1)(R\partial_R-3)\cdots(R\partial_R-(d-2))S_\text{EE}(R;B^{D-1},\epsilon) & ($D$\text{ odd})
    \end{cases},\label{c:LM}
\end{equation}
which interpolates between the UV and IR central charges by construction but does not necessarily have a monotonic behavior along RG flows for $D\geq4$ in general\footnote{One can construct a monotonic interpolating $c$-function for $D=4$ as in \cite{Komargodski:2011vj} but it is not motivated by the EE so we do not discuss it in this manuscript.}. The authors of \cite{Casini:2017vbe}, on the other hand, used strong subadditivity together with the expansion (\ref{Eq:EE-General}) to show that a suitably subtracted EE, $\Delta S_\text{EE}(R)$, satisfies the inequality
\begin{align}
	\label{Eq:MainSubAd}
	R \partial^2_{R}\, \Delta S_\text{EE}(R)-(D-3) \partial_R\,\Delta  S_\text{EE}(R)&\leq 0~,
\end{align}
and thereby established a monotonically decreasing function
\begin{align}
	c_\text{CTT}(R)=R \partial_R\Delta S(R) - (D-2) \Delta S(R)~\label{c:CTT}
\end{align}
along RG flows within the same dimension $D\leq 4$. This function (\ref{c:CTT}), however, does not interpolate between the  UV and IR central charges precisely even though it encodes that information.

\medskip

To explore partial $c$-functions along the RG flow between CFTs in different dimensions, we consider a $D$-dimensional UV CFT on $\RR^{1,d-1}\times M_{D-d}$ where a $d$-dimensional IR CFT is naturally obtained by compactifying the UV CFT on $M_{D-d}$. The above discussion on $c$-functions along RG flows within the same dimension then suggest to take a spatial spherical entangling region of $\RR^{1,d-1}\times M_{D-d}$ and explore the corresponding EE to construct an appropriate partial $c$-function along RG flows across dimensions. When the radius $R$ of a spherical entangling region is small compared to the scales introduced by $M_{D-d}$, namely a curvature scale and additional scales associated with non-trivial cycles (e.g. the lengths of the $S^1$'s in a torus when $M_{D-d}=T^{D-d}$), the geometry looks effectively flat and therefore the corresponding EE will have the same expansion as (\ref{Eq:EE-General}). As we increase the radius of an entangling region to probe the IR CFT, however, the EE will present different behaviors due to geometric differences between $\RR^{1,d-1}\times M_{D-d}$ and $\RR^{1,D-1}$. We describe this qualitatively new feature of EE and its implication  for partial $c$-functions in detail. 

\subsubsection{CFT on $\RR^{1,1}\times S^1$}
For simplicity, we start with a CFT on $\RR^{1,1}\times S^1$ ($D=3$, $d=2$, and $M_{D-d}=S^1$) as a simple model to illustrate the salient points, and generalize the discussion afterwards. 
For such a 3d CFT, the finite part of the EE of a spherical region would yield the free energy as $c$-function,
but this will serve as a model to illustrate the features of the entangling surfaces of interest also for flows, e.g., from 4d to 2d.
A constant-time slice in this 3d geometry is a cylinder, $\RR\times S^1$, which is flat, and we set the circumference of the $S^1$ to one.
The UV spherical region discussed above then corresponds to 
\begin{align}\label{eq:regionA-disc}
	A&=\lbrace (x,y)\in \RR\times S^1 \,|\, x^2+y^2\leq R^2\rbrace~,
\end{align}
with $R \ll 1$. To probe the IR physics of the QFT we increase $R$, leaving the definition of the region unchanged.
In this process the finite length of the $S^1$ becomes important eventually.
\begin{figure}
\centering
	\begin{tikzpicture}
		\foreach \i in {0,...,3} \draw[dashed] (\i,0.4) -- (\i,3.6);
		
		\draw (0.5,2) circle (5pt);
		\draw (1.5,2) circle (5pt);
		\draw (2.5,2) circle (5pt);
	\end{tikzpicture}
\hskip 20mm
	\begin{tikzpicture}
		\draw (0.5,2) circle (15pt);
		\draw (1.5,2) circle (15pt);
		\draw (2.5,2) circle (15pt);
		
		\draw[white,fill=white]  (-0.25,1.82) rectangle (3.2,2.18);
		
		\foreach \i in {0,...,3} \draw[dashed] (\i,0.4) -- (\i,3.6);
	\end{tikzpicture}
\hskip 20mm
	\begin{tikzpicture}
		\draw (0.5,2) circle (30pt);
		\draw (1.5,2) circle (30pt);
		\draw (2.5,2) circle (30pt);
	
		\draw[white,fill=white]  (-0.7,1.08) rectangle (3.7,2.92);
		
		\foreach \i in {0,...,3} \draw[dashed] (\i,0.4) -- (\i,3.6);
	
	\end{tikzpicture}

\caption{The cylinder $\RR\times S^1$ as quotient $\RR\times \RR/\ZZ$. The quotient identifies the dashed vertical lines, each region between dashed lines is a fundamental domain. The vertical direction is non-compact. Left: for small $R$ the region $A$ in (\ref{eq:regionA-disc}) takes the same form as in $\RR^2$ (in each fundamental domain). Center: when $R$ exceeds the length of the $S^1$, the discs from adjacent fundamental domains intersect, forming a region which wraps the $S^1$ and has {\it corners}. Right: for large $R$ the corners smooth out and the region $A$ in (\ref{eq:regionA-disc}) approaches a product form, in which it wraps an interval in $\RR$ and the entire $S^1$ at each point of the interval.\label{fig:cylinder-corners}
}
\end{figure}
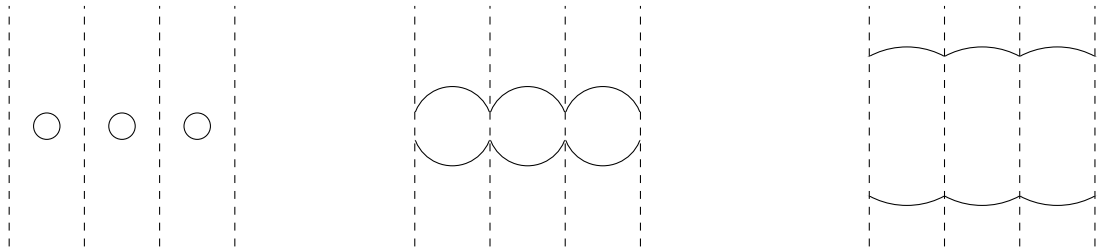
This is illustrated in Fig.~\ref{fig:cylinder-corners}: when $2R$ reaches the length of the $S^1$ the topology of the entangling region changes. While the region $A$ is topologically a disc for small $R$, it turns into an annulus with a non-contractible cycle for $2R>1$.

As shown in Fig.~\ref{fig:cylinder-corners}, when $2R$ reaches the size of the $S^1$ the entangling region in addition develops corners. These corners lead to new universal contributions to the EE, which are logarithmically divergent \cite{Bueno:2015rda,Bueno:2015xda}. Their coefficient is set by the opening angle of the corner. We note that the onset of these corners is sudden: they are infinitely sharp for $2R= 1$. As $R$ is increased further, the opening angle increases. In the limit $R\rightarrow\infty$ the opening angle approaches $\frac{\pi}{2}$, the corner smoothes out and the corner contribution to the EE vanishes.

Since we started with a 3d CFT, no logarithmically divergent term is present in the UV and the universal contribution to the EE in flat space would be given by the finite part. For the compactified theory we have an additional universal piece, which is a logarithmically divergent term whose coefficient is discontinuous across $2R=1$: it picks up a sudden corner contribution at $2R=1$, which then decreases as $R$ is further increased and ultimately vanishes for $R\rightarrow\infty$. The topology of the compact space thus imprints itself on the running of the log coefficient in the EE of a spherical region from the UV to the IR.

In the limit $R\rightarrow\infty$ the form of the region $A$ simplifies again: it approaches a product form where $A$ consists of an interval in $\RR$ and includes the entire $S^1$ at each point of the interval. 
That is, for large $R$ the region $A$ approaches a different type of region, $A^\prime$, defined by
\begin{align}\label{eq:regionAprime}
	A^\prime&=\lbrace (x,y)\in\RR\times S^1 \,\vert\, x^2\leq R^2\rbrace~.
\end{align}
This is the natural spherical region from the perspective of the 2d IR CFT.

\begin{figure}
\centering
	\begin{tikzpicture}
		\foreach \i in {0,...,3} \draw[dashed] (\i,0.4) -- (\i,3.6);
		
		\draw (0,2.16) -- (3,2.16);
		\draw (0,1.84) -- (3,1.84);
	\end{tikzpicture}
	\hskip 20mm
	\begin{tikzpicture}		
		\foreach \i in {0,...,3} \draw[dashed] (\i,0.4) -- (\i,3.6);		
		\draw (0,2.55) -- (3,2.55);
		\draw (0,1.45) -- (3,1.45);
	\end{tikzpicture}
	\hskip 20mm
	\begin{tikzpicture}
		\foreach \i in {0,...,3} \draw[dashed] (\i,0.4) -- (\i,3.6);		
		\draw (0,3.0) -- (3,3.0);
		\draw (0,1.0) -- (3,1.0);		
	\end{tikzpicture}
	
	\caption{$\RR\times S^1=\RR\times \RR/\ZZ$ as in Fig.~\ref{fig:cylinder-corners}. This figure shows the ``wrapping'' region (\ref{eq:regionAprime}) for different values of $R$, to be contrasted with the ``all-spherical'' region (\ref{eq:regionA-disc}) in Fig.~\ref{fig:cylinder-corners}.\label{fig:wrapping region}
	}
\end{figure}
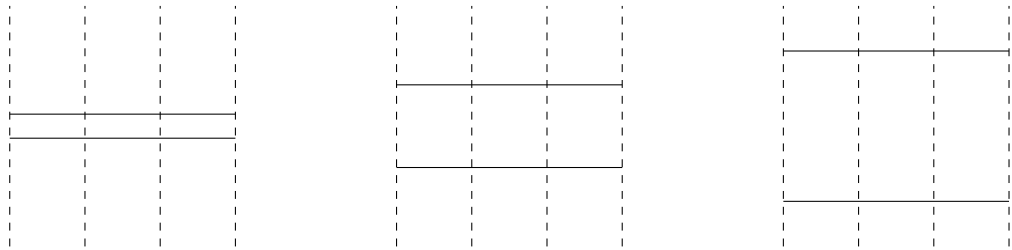

One may take an alternative perspective inspired by the IR point of view, and focus entirely on the region $A^\prime$ in (\ref{eq:regionAprime}).
This region emerged as a limit of the spherical region (\ref{eq:regionA-disc}) for large $R$, but we can take (\ref{eq:regionAprime}) as a definition and consider it for all $R$, as illustrated in Fig.~\ref{fig:wrapping region}.
This region is spherical in the non-compact directions and includes at each point the entire $S^1$. The boundary of this region is always smooth, and it does not capture the topology of the compact space through corner contributions. Different values of $R$ still capture physics at different scales in the CFT. However, since the region always contains the entire $S^1$ the EE always contains IR physics on the scale of the $S^1$ size.  
In the UV, this region does not become fully spherical, so the EE does not cleanly capture the $a$-type central charge in even dimensions or the sphere free energy in odd dimensions for the higher dimensional theory. On the other hand, this ``wrapping'' region (\ref{eq:regionAprime}) can still capture a combination of $a$- and $c$-type central charges, as we will discuss below.

\subsubsection{CFT on $\RR^{1,d-1}\times M_{D-d}$}
We now generalize the discussion to a CFT on $\RR^{1,d-1}\times M_{D-d}$ with compact spaces $M_{D-d}$ of generic dimension and possibly with curvature. A natural curved-space generalization of the ``all-spherical'' region (\ref{eq:regionA-disc})  is 
\begin{align}\label{eq:A-q}
	A_q&=\lbrace p\in \RR^{d-1}\times M_{D-d} \,|\, d(p,q)\leq R\rbrace
\end{align}
where $q$ is a reference point and $d(p,q)$ denotes the geodesic distance between $p$ and $q$.
The qualitative features are similar to the discussion of the previous toy model: in the UV, when $R$ is small compared to the curvature scale and the length of all cycles in $M_{D-d}$, the geometry is  indistinguishable from flat space, and the EE associated with $A_q$ approaches the EE of a spherical region in flat space. In the deep IR, when $R$ is large compared to all other scales, we expect the region $A_q$ to approach a product form, in which it is spherical in the non-compact directions  and includes $M_{D-d}$ in its entirety at each point of the disc. 

\begin{figure}
\centering
	\begin{tikzpicture}[scale=1.3]
		\draw (0,0) circle (1);
	     \def\phi{10};
		  \foreach \latitude in {-50, -20,...,80} {
			\pgfmathsetmacro\verticaloffset{cos(\phi)*sin(\latitude)};
			\pgfmathsetmacro\radius{cos(\latitude)};
			\pgfmathsetmacro\blcolor{100*(\latitude+90)/180};
			
			\tikzset{xyplane/.estyle={cm={1, 0, 0, cos(90 + \phi), (0, \verticaloffset)}}}
			\draw [xyplane,blue!\blcolor!red] (\radius,0) arc (0:180:\radius);
			\draw [xyplane, dashed,blue!\blcolor!red] (\radius,0) arc (360:180:\radius);
		}
	\end{tikzpicture}
\hskip 10mm
\begin{tikzpicture}[scale=0.83]
			\foreach \i in {-0.5,0.5,1.5} \foreach \j in {-0.5,0.5,1.5} \draw[red] (\i,\j) circle (0.55);
			
			\foreach \i in {-0.5,0.5,1.5} \draw[white,fill=white] (\i-0.23,-1.15) rectangle (\i+0.23,2.15);
			\foreach \i in {-0.5,0.5,1.5} \draw[white,fill=white] (-1.15,\i-0.23) rectangle (2.15,\i+0.23);
			
			\foreach \i in {-0.5,0.5,1.5} \foreach \j in {-0.5,0.5,1.5} \draw[blue] (\i,\j) circle (0.2);
			
			\foreach \i in {0,...,1} \draw[dashed] (\i,-1) -- (\i,2);
			\foreach \i in {0,...,1} \draw[dashed] (-1,\i) -- (2,\i);
\end{tikzpicture}
\hskip 10mm
\begin{tikzpicture}
	\node at (0,0) {\includegraphics[width=0.16\linewidth]{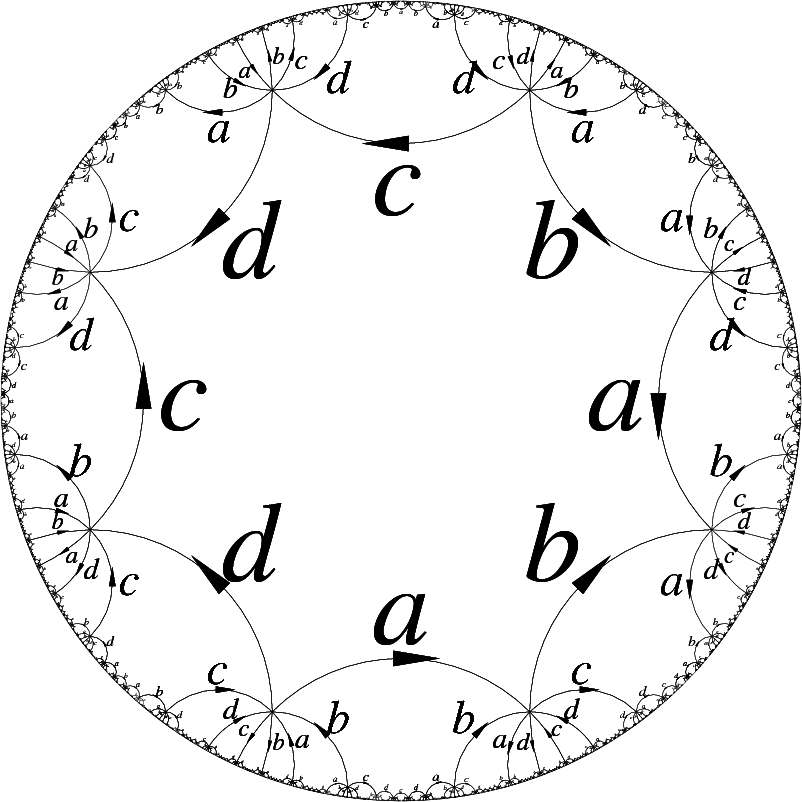}};
	\draw[blue] (0,0) circle (0.2);
	\draw[red] (0,0) circle (0.75);
\end{tikzpicture}
	\caption{Left: Regions $A_q$ defined in (\ref{eq:A-q}) for $M_{D-d}=S^2$, centered at the north pole, for different values of $R$. At each point of $A_q$ on the sphere, $A_q$ includes a disc in $\RR^{d-2}$ whose radius depends on the distance from the north pole. Center: Torus as $\RR^2/\ZZ^2$, with a region $A_q$, centered at the center of the fundamental domain, shown for a small $R$ (blue) and large $R$ (red). Right: Higher-genus Riemann surface as quotient of the Poincar\'e disc with a small-$R$ region (blue) and a large-$R$ region (red). Upon increasing $R$ further the topology of $A_q$ changes.\label{fig:Sigma-corners}}
\end{figure}
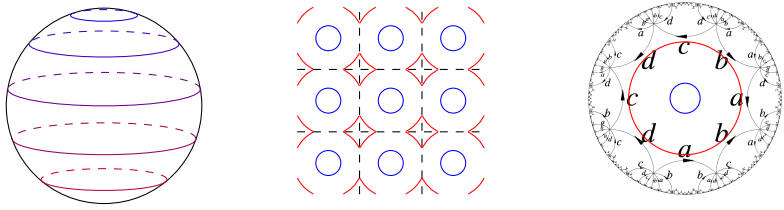

For general $2R$ smaller than the length of the shortest cycle, the region $A_q$ may be deformed compared to flat space if $M_{D-d}$ is curved, but the boundary of $A_q$ stays smooth. Once $2R$ exceeds the length of the shortest cycle, the region may develop corners. 
In general, a conical singularity in odd dimensions (where the finite part is universal in flat space) leads to $\log$ terms, while in even dimensions (where the $\log$ term is universal in flat space) conical singularities lead to $\log^2$ terms \cite{Bueno:2015lza}. This suggests that different coefficients in the EE can be used to detect the topology of the spacetime and the central charges.
For an illustration see Fig.~\ref{fig:Sigma-corners}.

In the IR limit, where $R$ is larger than all other scales, we expect the surface $A_q$ to approach, as in the toy model before, a product form which can be written as
\begin{align}\label{eq:A-q-prime}
	A_q^\prime&=\left\lbrace (\vec{x},y)\in \RR^{d-1}\times M_{D-d} \,\big\vert\, |\vec{x}-\vec{x_q}|^2 \leq R^2\right\rbrace~.
\end{align}
This region is spherical in the non-compact directions, $\vec{x}$,  and contains, at each point of the spherical region, the compact space $M_{D-d}$ in its entirety.
As before, we may turn things around and consider the region $A_q^\prime$ for arbitrary $R$.
This region never becomes ``all-spherical'' in the UV, and the associated EE always contains some IR physics on the scale of $M_{D-d}$.
In 4d a combination of the UV central charges can still be extracted from the EE for small $R$, though it is now a combination of $a$ and $c$, weighted by the integrated curvature of the compact space.
When the compact space is flat (e.g.,  a torus), no central charge can be extracted. Such regions $A^\prime_q$ are, nevertheless, smooth and computationally tractable, and lead to a notion of monotonicity. We will discuss them below and in section \ref{sec:EE:holo} for AdS$_5 \to$ AdS$_3$ and AdS$_7 \to$ AdS$_3$ flows.

\subsubsection{Central charges across dimensions and monotonicity}
We now discuss the physics of partial $c$-functions extracted from the EE for the two different types of entangling regions (\ref{eq:A-q}) and (\ref{eq:A-q-prime}) studied above. To be specific, we investigate if the resulting partial $c$-functions are monotonic along the RG flow across dimensions or interpolating the UV and IR central charges. We will focus on flows from 4d to 2d for concreteness where $M_{D-d}$ corresponds to a Riemann surface of genus $\mathfrak g$ as $M_{D-d}=\Sigma_{\mathfrak g}$. 

\medskip

The EE for the ``all-spherical'' region $A_q$, (\ref{eq:A-q}), in the UV approaches the EE of a spherical region in flat 4d space. As such, we can extract the $a$-type central charge from the coefficient of the log term in (\ref{Eq:EE-General}).
In the deep IR, the ``all-spherical" region $A_q$ wraps the compact space and becomes spherical in the non-compact directions, so the EE should also detect the 2d central charge.
This should allow to define an interpolating $c$-function. The arguments establishing monotonicity of the combination in (\ref{c:CTT}), however, rely on 4d Poincar\'e symmetry, which is broken in compactifications.

One may ask if there can in principle be a monotonically decreasing function which interpolates between the 4d $a$-type central charge in the UV and the 2d central charge in the IR. This is not the case. Let us provide a counterexample by considering 4d $\mathcal N=4$ SYM compactified on a Riemann surface \cite{Benini:2013cda}. The UV $a$-type central charge is set by the dimension of the gauge group, $d_G$.
The compactification is characterized, in addition to the Riemann surface $\Sigma_{\mathfrak g}$ with curvature $R=2\kappa$ where $\kappa\in\lbrace 0,\pm 1\rbrace$, by 3 flux parameters $p^I\,(I=1,2,3)$ constrained by $p^1+p^2+p^3=-\kappa$. The IR central charge is given by \cite[(3.12)]{Benini:2013cda}
\begin{align}\label{eq:cIR-N4SYM}
	c^{}_{2d}&= \frac{-12\eta_{\mathfrak g} d_G\,p^1p^2p^3}{(p^1)^2+(p^2)^2+(p^3)^2-2(p^1p^2+p^2p^3+p^3p^1)}~.
\end{align}
Here $\eta_{\mathfrak g}=2|\mathfrak{g}-1|$ for genus $\mathfrak{g}\neq 1$ and $\eta_{\mathfrak g}=1$ for a torus. The values for $p^I$ which lead to a non-trivial IR CFT can be found in \cite{Benini:2013cda}. For all $\kappa$ one can make $p^1$, $p^2$ large, which makes $p^3$ large as well.
The numerator in (\ref{eq:cIR-N4SYM}) is cubic in the $p^I$, the denominator quadratic. As a result, the IR central charge can be made arbitrarily large by choosing the fluxes accordingly. The UV $a$-type central charge, on the other hand, is fixed by $d_G$. 
The IR central charge can thus,  in particular, be larger than the UV central charge, so that a non-increasing function cannot connect the 4d central charge in the UV to the 2d central charge in the IR.
This is not surprising: the 4d and 2d central charges count different notions of degrees of freedom -- each 4d field (counted by the UV central charge) can be seen as an infinite family of 2d fields (counted by the IR central charge), and how many of the latter survive in the IR depends on the fluxes.

\medskip

We now turn to the ``wrapping'' region $A_q^\prime$, (\ref{eq:A-q-prime}). This region does not become fully spherical in the UV and, consequently,  does not cleanly detect the $a$-type central charge. Instead, it detects the linear combination (see the end of subsection~\ref{sec:EE:5d:EE} for details)
\begin{align}\label{eq:EE-Aq-UV}
	S_{\rm EE}(A_q^\prime)&=\# \frac{\Vol[\Sigma_{\mathfrak g}]}{\epsilon^2} - \left[\frac{3a-c}{3\pi}\int_{\Sigma_{\mathfrak g}}R_{\Sigma_{\mathfrak g}}\right]\log\frac{R}{\epsilon}+\mathcal O(\epsilon^0).
\end{align}
We note that in 4d the $c$-type central charge is not in general monotonic, so the combination $3a-c$ is not a natural $c$-function from that perspective. Moreover, for a torus compactification the UV log term vanishes, and the EE associated with the region $A_q^\prime$ does not detect any 4d UV central charge.
In the IR limit, on the other hand, the regions $A_q^\prime$ and $A_q$ approach the same shape, so both EE associated with $A_q^\prime$ and $A_q$ detect the 2d IR central charge. Then, motivated by a 2-dimensional $c$-function (\ref{c:CT}), we propose a monotonic partial $c$-function across dimensions (4d$\to$2d) as
\begin{align}\label{eq:SEE-Sq-dR}
	c_{\rm mono}(R)\equiv R \,\partial_R\, S_{\rm EE}(A'_q)
\end{align}
in terms of the EE for the ``2d spherical'' or ``wrapping'' region $A_q^\prime$. The leading divergence in (\ref{eq:EE-Aq-UV}) is independent of $R$ and is eliminated by this derivative, and we obtain a finite quantity along the flow.
As we will see below, this derivative blows up in the UV limit, $R\rightarrow 0$, due to the {\it finite} terms indicated by the ellipsis in (\ref{eq:EE-Aq-UV}).
This blow-up makes sense physically \cite{Bea:2015fja}: the function $c_{\rm mono}(R)$ is, in spirit, a 2d central charge. We can view the 4d UV CFT (and the entire flow) as a 2d QFT comprising (in the UV infinite) towers of KK modes on $\Sigma_{\mathfrak g}$.
It is a rather unusual 2d QFT, but has 2d Poincar\'e invariance.
In the UV limit, where the fields truly become 4d fields representing infinite KK towers of 2d fields, a 2d notion of central charge should diverge\footnote{We will also consider an interpolating $c$-function extracted from the EE for the region $A_q^\prime$ in subsections~\ref{sec:EE:5d:EE} and \ref{sec:EE:7d:UVIR}, which is not necessarily monotonic.}.

\medskip

To sum up, we have described two choices for EE's which capture aspects of the UV and IR theories and of the RG flow.
The EEs for the ``all-spherical'' regions $A$ in (\ref{eq:regionA-disc}) and $A_q$ in (\ref{eq:A-q}), illustrated in Figs.~\ref{fig:cylinder-corners} and \ref{fig:Sigma-corners}, respectively, have the following features:
\begin{itemize}
	\item  In the UV and IR they encode the natural notions of central charge (through the finite part in odd dimensions and the log term in even dimensions)
	\item  Along the flow they detect the topology of the compact space through corner contributions (leading to $\log$ terms in odd dimensions and to $\log^2$ terms in even dimensions)
\end{itemize}
One may define a function interpolating between the UV and IR central charges from this EE, but we argued that one can not expect such a function to be generally monotonic.

The key features of the EE associated with the ``wrapping'' regions $A^\prime$ in (\ref{eq:regionAprime}) and $A_q^\prime$ in (\ref{eq:A-q-prime}), in contrast, are
\begin{itemize}
    \item They provide a natural extension of the notion of central charge in the lower-dimensional IR fixed point theory to the RG flow. This notion is monotonic but diverges in the UV.
	\item  In the UV they detect (in even dimensions) a combination of the $a$ and $c$ central charges, weighted by the integrated curvature of the compact space.
\end{itemize}
Along the flow the entangling region $A_q^\prime$ is smooth without corners.  The quantity $S_{\rm EE}(A_q^\prime)$ is the natural EE to consider if one views the higher-dimensional CFT as an infinite set of lower-dimensional fields.

Comparing to the two notions of partial $c$-functions defined in the introduction, $S_{\rm EE}(A_q)$ is thus naturally suited for defining an interpolating function $c_{\rm int}$. From $S_{\rm EE}(A_q^\prime)$ one can naturally define a monotonic function $c_{\rm mono}$, which is divergent in the UV. Since $S_{\rm EE}(A_q^\prime)$ still detects a linear combination of $a$ and $c$ UV central charges in even dimensions, one can also define a notion of interpolating $c$-function from it. We will discuss this problem holographically in the following subsection~\ref{sec:EE:holo}.

\subsection{Holographic $c$-functions across dimensions from EE} \label{sec:EE:holo}
In the previous subsection \ref{sec:EE:FT}, we have considered two types of entangling regions (\ref{eq:A-q}) and (\ref{eq:A-q-prime}) in a $D$-dimensional CFT on $\RR^{1,d-1}\times M_{D-d}$ and discussed properties of partial $c$-functions extracted from the associated EE's. In this subsection we discuss the same problem from a holographic perspective by exploring the EE given as the RT surfaces for boundary entangling regions (\ref{eq:A-q}) and (\ref{eq:A-q-prime}) in subsections \ref{sec:EE:holo:dissecting} and \ref{sec:EE:holo:wrap} respectively. The RT surfaces will be constructed in bulk geometries that holographically represent the RG flows between boundary CFTs in different dimensions, which is described by a metric of the form (\ref{eq:mnmet}): here we specialize the metric (\ref{eq:mnmet}) further to 
\begin{equation}
	ds^2=e^{2f(z)}(-dt^2+dz^2+dr^2+r^2d\Omega^2_{d-2})+e^{2g(z)}ds_{M_{D-d}}^2, \label{metric}
\end{equation}
with the following asymptotic AdS behaviors
\begin{equation}
	\begin{alignedat}{4}
		z&\to 0~&:\qquad&f(z)\to\log(L_\text{UV}/z),\qquad& g(z)&\to \log(L_\text{UV}/z),\\
		z&\to \infty~&:\qquad&f(z)\to\log(L_\text{IR}/z),\qquad& g(z)&\to g_{\rm IR}.\label{hor/asymp}
	\end{alignedat}
\end{equation}
%

\subsubsection{Entangling region dissecting $M_{D-d}$}\label{sec:EE:holo:dissecting}
As discussed in subsection~\ref{sec:EE:FT}, the entangling region (\ref{eq:A-q}) undergoes topology changes as the size is increased. This imprints itself also on the holographic realization, as we discuss now.

For reference, we start with flows within the same dimension, described by  (\ref{metric}) with $D=d$. The EE for a $(D-1)$-dimensional spatial ball of radius $R$, $B^{D-1}$, can be computed holographically by the area of an RT minimal surface homologous to the boundary entangling region $B^{D-1}$ as \cite{Ryu:2006bv,Ryu:2006ef}
\begin{equation}
	S_\text{EE}(R;B^{D-1},\epsilon)=\frac{{\rm Vol}(S^{D-2})}{4G^{(D+1)}_N}\min_{\{r(z=\epsilon)=R\}}\bigg[\int_\epsilon^{z_0}dz\, r^{D-2}\,e^{(D-1)f(z)}\sqrt{1+ r'(z)^2}\bigg],\label{EE:area:AdS}
\end{equation}
where $G^{(D+1)}_N$ is the $(D+1)$-dimensional Newton's constant and $S^{D-2}$ comes from the boundary of the entangling surface, $S^{D-2}=\partial B^{D-1}$. Here $\epsilon$ is a UV cutoff realized in the bulk by restricting the radial coordinate to $z\geq\epsilon$ and $z=z_0$ denotes the deepest point in the bulk that the minimal surface probes along the holographic radial coordinate.
From this holographic expression for the EE one can then define $c$-functions by forming the combinations in (\ref{c:CTT}).

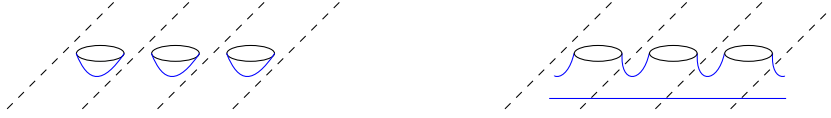
\begin{figure}
	\centering
	\begin{tikzpicture}[x={(1cm,0cm)}, y={(0.41cm,0.41cm)}, z={(0cm,1cm)}]
		\foreach \i in {0,...,3} \draw[dashed] (\i,-1.8,0) -- (\i,1.8,0);
		\foreach \i in {0,...,2} \draw (0.5+\i,0,0) ellipse (9pt and 3pt);
		
		\foreach \i in {0,...,2} \draw[blue] (0.18+\i,0) .. controls (0.75+\i,-1) and (0.95+\i,-1) .. (0.82+\i,0);		
	\end{tikzpicture}
	\hskip 20mm
	\begin{tikzpicture}[x={(1cm,0cm)}, y={(0.41cm,0.41cm)}, z={(0cm,1cm)}]
		
		\foreach \i in {-1,...,2} \draw[blue] (0.82+\i,0) .. controls (1.2+\i,-1) and (1.5+\i,-1) .. (1.18+\i,0);		
		\draw[white,fill=white]  (-0.3,0,-0.5) rectangle (-0.09,0,0.1);
		\draw[white,fill=white]  (-0.3+3.29,0,-0.5) rectangle (-0.09+3.3,0,0.1);
		
		\foreach \i in {0,...,3} \draw[dashed] (\i,-1.8,0) -- (\i,1.8,0);
		\foreach \i in {0,...,2} \draw (0.5+\i,0,0) ellipse (9pt and 3pt);
		
		\draw[blue] (-0.15,0,-0.6) -- (3,0,-0.6);
	\end{tikzpicture}
	
	\caption{Possible phase transition between ``flat space" and ``wrapping" surfaces. The vertical direction is the holographic radial coordinate, the plane represents the conformal boundary.\label{fig:RT-trans}}
\end{figure}

For flows across dimensions there are new effects even when the compact space is flat.
As discussed in section~\ref{sec:EE}, the topology of the entangling region (\ref{eq:A-q}) changes as its size is increased, as illustrated for a CFT on $\RR^{1,1}\times S^1$ in Fig.~\ref{fig:cylinder-corners}. A holographic dual for such a CFT would be described by a geometry of the form (\ref{metric}),
\begin{align}\label{eq:ds2-flow-S1}
	ds^2&=e^{2f(z)}(ds^2_{\RR^{1,1}}+dz^2)+e^{2g(z)}ds^2_{S^1}~.
\end{align}
In the UV it approaches $AdS_4$, in the IR it reduces to $AdS_3$ with an $S^1$ internal space.
For the holographic computation of the EEs we seek minimal surfaces which are anchored on the $\RR^{1,1}\times S^1$ conformal boundary.
In the UV the RT surface for a small spherical region $A$ becomes identical to a surface for a spherical region in $AdS_4$, i.e.\ a special case of (\ref{EE:area:AdS}). 
As the size of $A$ is increased, the RT surface in the flow solution (\ref{eq:ds2-flow-S1}) gets deformed compared to a surface in $AdS_4$, since the warp factors of the $S^1$ part and the $\RR^2$ part are generally different.
When the size of the region $A$ approaches the size of the $S^1$, one anticipates phase transitions between topologically distinct RT surfaces. These are illustrated in Fig.~\ref{fig:RT-trans}.
Upon further increasing the size of $A$ the entangling region itself changes topology, as illustrated in Fig.~\ref{fig:cylinder-corners}.  Region $A$ also develops corners which impact the form of the RT surfaces as well.
In the deep IR limit, for very large region $A$, the RT surface approaches a form where it wraps the $S^1$ entirely and splits the $\RR^{1,1}$ part in such a way that it computes the EE of a spherical region in the lower-dimensional CFT (in 2d an interval).
This is the natural surface for extracting the central charge of the lower-dimensional CFT: One views (\ref{eq:ds2-flow-S1}) as an $AdS_3$ solution, with an internal space which decompactifies in the UV. The natural ``all-spherical'' RT surface in $AdS_3$  then wraps the $S^1$.\footnote{%
	In 10d string theory setups the ``usual" spherical RT surface wraps the internal space. Here the $S^1$ becomes the internal space in the IR and should thus be wrapped entirely in the IR solution.}

As a further example we consider flows from AdS$_5$ to AdS$_3$, realized by wrapping a 4d theory on a Riemann surface of genus $\mathfrak g$ and the length scale $\ell$. In this case, the metric (\ref{metric}) is specialized to
\bea
ds^2=e^{2f(z)}(-dt^2 +dz^2 +dr^2 )+\ell^2e^{2g(z)} (d\theta^2 +S(\theta)^2 d\phi^2),\label{metric:5d:theta}
\eea
where $S(\theta)=\{\sin\theta, \theta, \sinh \theta\}$ for a Riemann surface of genus zero, one or higher. 
The region $\mathcal A_q$ in (\ref{eq:A-q}) with the reference point $q$ at the origin in $\RR^2$ and on $\Sigma_{\mathfrak g}$ at $\theta=0$ is then given by
\begin{equation}
	\mathcal A=\left\{p\in\RR^2\times\Sigma_{\mathfrak g}\,|\,d(p,q)\leq R,~q\text{ is the point at }r=\theta=0\right\},\label{A:partial}
\end{equation}
where the geodesic distance $d$ is defined with respect to the field theory metric.
For small $R$ the associated EE is given by the area of a minimal surface in the bulk (\ref{metric:5d:theta}) which can be parametrized by $z=z(r,\theta)$,  as
\begin{equation}
	S_\text{EE}(R;\mathcal A)=\fft{1}{4G_N^{(5)}}\min\left[2\pi\int drd\theta\,S(\theta)\ell e^{f(z)+g(z)}\sqrt{\ell^2e^{2g(z)}(1+(\partial_r z)^2)+e^{2f(z)}(\partial_\theta z)^2}\right],\label{EE:partial}
\end{equation}
When $R$ becomes comparable to the size of $\Sigma_{\mathfrak g}$, the entangling region and minimal surface will undergo topology changes, of the type illustrated in Fig.~\ref{fig:Sigma-corners}.
For $\Sigma_{\mathfrak g}=S^2$ the parametrization of the surface in terms of $z(r,\theta)$ should remain sufficient, but for $\mathfrak{g}\geq 1$ the surfaces will depend on $\phi$ as well.
The extremality condition either way leads to a PDE.

In the strict UV limit ($R\to0$), the problem reduces to a spherical surface in $AdS_5$. For small $\mathcal A$ the embedding $z(r,\theta)$ stays near the conformal boundary, where $f(z)\approx g(z)$. 
Moreover, the region $\mathcal A$ only extends to small values of $\theta$, so that $S(\theta)\approx \theta$.
Then the embedding $z(r,\theta)$ only depends on the radial coordinate in the combined space of $\RR^2$ and $\Sigma_{\mathfrak g}$, $\tilde r\equiv\sqrt{r^2+\ell^2\theta^2}$, and we can take $z(r,\theta)\mapsto z(\tilde r)$. In (\ref{EE:partial}) we then have 
\begin{equation}
	\ell^2(\partial_rz)^2+(\partial_\theta z)^2 \mapsto (\ell^2(\partial_r\tilde r)^2+(\partial_\theta\tilde r)^2)(z')^2 = \ell^2(z')^2,
\end{equation}
where the prime denotes a derivative with respect to $\tilde r$. Upon taking into account the Jacobian 
\begin{equation}
	2\pi\int drd\theta\,\ell^2\theta \mapsto 4\pi\int d\tilde r\,\tilde r^2~,
\end{equation}
this turns equation (\ref{EE:partial}) into equation (\ref{EE:area:AdS}) with $D=4$.
We naturally recover the higher-dimensional spherical surface.

In the example section \ref{sec:EX} we will not consider these surfaces explicitly, and focus on surfaces wrapping the compact space studied in the following subsection \ref{sec:EE:holo:wrap} instead, for which the extremality condition leads to an ODE.

\subsubsection{Entangling region wrapping $M_{D-d}$}\label{sec:EE:holo:wrap}
We now turn to the case where an entangling region wraps $M_{D-d}$ completely as in (\ref{eq:A-q-prime}): to be specific, we set the entangling region as $B^{d-1}\times M_{D-d}$ with a $(d-1)$-dimensional spatial ball $B^{d-1}$ of radius $R$. The associated EE is given as a minimal surface area in the bulk (\ref{metric}) as
\begin{equation}
\begin{split}
    &S_\text{EE}(R;B^{d-1}{\times}M_{D-d},\epsilon)\\
    &=\fft{\text{vol}[S^{d-2}]\text{vol}[M_{D-d}]}{4G^{(D+1)}_N}\min_{r(z=\epsilon)=R}\bigg[\int_\epsilon^{z_0}dz\,r^{d-2}\,e^{(d-1)f(z)+(D-d)g(z)}\sqrt{1+r'(z)^2}\bigg].
\end{split}
\end{equation}
The combination $(d-1)f(z)+(D-d)g(z)$ incorporates both the AdS$_{d+1}$ warp factor and the volume factor of the wrapped $M_{D-d}$.  This is, in fact, proportional to the effective warp factor defined in (\ref{eq:tildef}), which allows us to write
\begin{equation}
    S_\text{EE}(R;B^{d-1}{\times}M_{D-d},\epsilon)
    =\fft{\text{vol}[S^{d-2}]\text{vol}[M_{D-d}]}{4G^{(D+1)}_N}\min_{r(z=\epsilon)=R}\bigg[\int_\epsilon^{z_0}dz\,r^{d-2}\,e^{(d-1)\tilde f(z)}\sqrt{1+r'(z)^2}\bigg].\label{EE:area}
\end{equation}

For the special case $d=2$, the minimal surface area (\ref{EE:area}) takes a simple form, since $\partial B^1$ just consists of two points. To begin with, (\ref{EE:area}) reduces for $d=2$ to
\begin{equation}
    S_\text{EE}(R;B^1{\times}M_{D-2},\epsilon)=\fft{2\text{vol}[M_{D-2}]}{4G^{(D+1)}_N}\min_{r(z=\epsilon)=R}\bigg[\int_\epsilon^{z_0}dz\,e^{\tilde f(z)}\sqrt{1+r'(z)^2}\bigg],\label{EE:area:5d}
\end{equation}
where now
\begin{equation}
    \tilde f(z)=f(z)+(D-2)g(z)\qquad(d=2).
\end{equation}
Note the factor $2$ resulting from the fact that the boundary of a one-dimensional ball (=real interval), $\partial B^1$, consists of two points.
Applying the variational principle to the integral in (\ref{EE:area:5d}) with respect to $r(z)$, one can determine a function $r(z)$ that parametrizes a minimal surface as
\begin{equation}
    r''(z)+(1+r'(z)^2)\tilde f'(z)r'(z)=0\quad\to\quad r'(z)=\pm\fft{Ce^{-\tilde f(z)}}{\sqrt{1-(Ce^{-\tilde f(z)})^2}}\label{eq:r}
\end{equation}
with an integration constant $C$. Based on isotropy and smoothness at the tip of a minimal surface we introduce the turning point, $z_0$. The surface satisfies $r'(z)\to-\infty$ as $z\to z_0$, which chooses the appropriate sign in (\ref{eq:r}) to be negative and also fixes the constant $C$ as
\begin{equation}
    C=e^{\tilde f_0}\qquad\text{where}\qquad \tilde f_0\equiv \tilde f(z_0).
\end{equation}
Now we introduce a function
\begin{equation}
\mathcal F(z)\equiv e^{\tilde f_0-\tilde f(z)},\label{eq:F}
\end{equation}
in terms of which the minimal surface area (\ref{EE:area:5d}) reads
\begin{equation}
    S_\text{EE}(R;B^1{\times}M_{D-2},\epsilon)=\fft{\text{vol}[M_{D-2}]}{4G^{(D+1)}_N}\,2e^{\tilde f_0}\int_\epsilon^{z_0}\fft{dz}{\mathcal F(z)\sqrt{1-\mathcal F(z)^2}}.\label{EE:area:5d:2}
\end{equation}
The EE (\ref{EE:area:5d:2}) is written as a function of the turning point, $z_0$, but can readily be expressed in terms of the radius of the entangling surface, $R$. This is achieved by following the relation between $z_0$ and $R$ derived from (\ref{eq:r}):
\begin{equation}
    r(z)=-\int_{z_0}^zdz\,\fft{\mathcal F(z)}{\sqrt{1-\mathcal F(z)^2}}\quad\to\quad R=\int_0^{z_0}dz\,\fft{\mathcal F(z)}{\sqrt{1-\mathcal F(z)^2}}.\label{z:to:R}
\end{equation}
%

\section{Examples}\label{sec:EX}
In this section we study the twisted compactification of 4d $\mathcal N=4$ SYM on a Riemann surface $\Sigma_{\mathfrak{g}}$ and 6d $\mathcal N=(2,0)$ compactified on a hyperbolic 4-manifold $\mathbb{H}_4$ or on a torus $T^2$ as concrete examples. Our main tool is holography. The holographic duals of these compactifications are described by metrics of the form (\ref{metric}) with the asymptotic behaviors (\ref{hor/asymp}). In subsections~\ref{sec:EX:5d}, \ref{sec:EX:7d}, and \ref{sec:EX:7-5d} we discuss various notions of interpolating and monotonic $c$-functions discussed in section \ref{sec:EE} along these holographic RG flows dual to the above mentioned partially compactified CFTs respectively. 

\subsection{Flows from AdS$_5$ to AdS$_3$: partially compactified $\mathcal N=4$ SYM theory}\label{sec:EX:5d}
In this subsection we formulate holographic $c$-functions in flows from AdS$_5$ to AdS$_3$, namely 
\begin{equation}
    ds^2=e^{2f(z)}(-dt^2+dz^2+dr^2)+e^{2g(z)}ds^2_{M_2}\label{metric:5d}
\end{equation}
from (\ref{metric}) with $D=4$, $d=2$, and the asymptotic behavior described in (\ref{hor/asymp}). For concreteness, we further specialize to the case where the metric (\ref{metric:5d}) is a classical solution of the 5d $\mathcal N=2$ gauged STU model (see Appendix \ref{App:STU} for a brief summary). For the metric (\ref{metric:5d}) together with two physical scalar fields and three vector fields to satisfy all the equations of motion of the 5d $\mathcal N=2$ gauged STU model given in (\ref{N=2:sugra:Einstein}), (\ref{N=2:sugra:scalar}), (\ref{N=2:sugra:vector}), and the Bianchi identity (\ref{N=2:sugra:Bianchi}), we take the following Ansatz
\begin{equation}
\begin{split}
	ds^2&=e^{2f(z)}(-dt^2+dz^2+dr^2)+e^{2g(z)}\ell^2e^{2h_{\mathfrak g}(x,y)}(dx^2+dy^2),\\
	A^I&=-p^I\omega_{\mathfrak g}\quad\to\quad F^I=-p^Ie^{2h_{\mathfrak g}(x,y)}dx\wedge dy,\\
	X^I&=X^I(z),  
\end{split}\label{ansatz:BS}
\end{equation}
where the 2d manifold $M_2$ in (\ref{metric:5d}) is assumed to be a Riemann surface of genus $\mathfrak g$ and the length scale $\ell$. Here we have introduced a function $h_{\mathfrak g}$ and a one-form $\omega_{\mathfrak g}$ that characterize the Riemann surface $\Sigma_{\mathfrak g}$ as ($\kappa=1$, $\kappa=0$, and $\kappa=-1$ for $\mathfrak g=0$, $\mathfrak g=1$, and $\mathfrak g>1$ respectively)
\begin{equation}
e^{h_{\mathfrak g}(x,y)}=\begin{cases}
	\fft{2}{1+x^2+y^2} & (\mathfrak g=0)\\
	\sqrt{2\pi} & (\mathfrak g=1)\\
	\fft1y & (\mathfrak g>1)
\end{cases},\qquad
\omega_{\mathfrak g}=\begin{cases}
	\fft{2(-ydx+xdy)}{1+x^2+y^2} & (\mathfrak g=0)\\
	\pi(-ydx+xdy) & (\mathfrak g=1)\\
	\fft{dx}{y} & (\mathfrak g>1)
\end{cases}.
\end{equation}

To present the behaviors of holographic $c$-functions in flows (\ref{ansatz:BS}) more explicitly, we will consider magnetically charged BPS AdS$_5$ black string (numerical) solutions in 5d $\mathcal N=2$ gauged STU model. The BPS equations (\ref{N=2:sugra:BPS}) for a magnetic black string Ansatz (\ref{ansatz:BS}) are given as
\begin{equation}
\begin{split}
	0&=\fft{dg}{d\rho}-\fft13\left(X^1+X^2+X^3-e^{-2g}(\fft{p^1}{X^1}+\fft{p^2}{X^2}+\fft{p^3}{X^3})\right),\\
	0&=\fft{df}{d\rho}-\fft13\left(X^1+X^2+X^3+\fft12e^{-2g}(\fft{p^1}{X^1}+\fft{p^2}{X^2}+\fft{p^3}{X^3})\right),\\
	0&=\fft{d\phi^1}{d\rho}-\fft{\sqrt6}{3}\left(X^1+X^2-2X^3+\fft12e^{-2g}(\fft{p^1}{X^1}+\fft{p^2}{X^2}-\fft{2p^3}{X^3})\right),\\
	0&=\fft{d\phi^2}{d\rho}-\sqrt2\left(X^1-X^2+\fft12e^{-2g}(\fft{p^1}{X^1}-\fft{p^2}{X^2})\right),\\
	0&=p^1+p^2+p^3+\kappa,
\end{split}\label{N=2:sugra:BPS:reduced}
\end{equation}
where $\kappa=1,0,-1$ for $\mathfrak g=0$, $\mathfrak g=1$, $\mathfrak g>1$ respectively. Here we have introduced a holographic radial coordinate $\rho$ as
\begin{equation}
	d\rho=-e^fdz,\label{z:to:rho}
\end{equation}
and set $\boldsymbol{g}=\ell=1$ (and thereby $L_\text{UV}=1$ from (\ref{fgphi:UV:perturbative})) for simplicity. One can solve the BPS equations (\ref{N=2:sugra:BPS:reduced}) numerically for $g(\rho),f(\rho),\phi^1(\rho),\phi^2(\rho)$ and we provided several examples in Appendix \ref{App:STU}. Also refer to \cite{Benini:2013cda,Donos:2011pn} for similar numerical investigations. See \cite{Maldacena:2000mw,Klemm:2000nj,Cacciatori:2003kv,Bernamonti:2007bu,Bobev:2014jva,Azzola:2018sld} for analytic black string solutions with some special configurations of magnetic charges.

In \ref{sec:EE:5d:NEC}, we investigate the LH $c$-function (\ref{eq:cfunction?}) in flows (\ref{ansatz:BS}). Then in \ref{sec:EE:5d:EE}, we propose two different holographic $c$-functions in the same flows from the entropic point of view, namely EE $c$-functions, by exploring an effective logarithmic contribution to the holographic EE (\ref{EE:area:5d:2}) in flows (\ref{ansatz:BS}). In both subsections, we will present the behavior of LH/EE $c$-functions explicitly for numerical BPS AdS$_5$ black string solutions found by solving (\ref{N=2:sugra:BPS:reduced}).

\subsubsection{Local Holographic  $c$-function}\label{sec:EE:5d:NEC}
\begin{figure}[t]
\centering
    \includegraphics[width=0.5\textwidth]{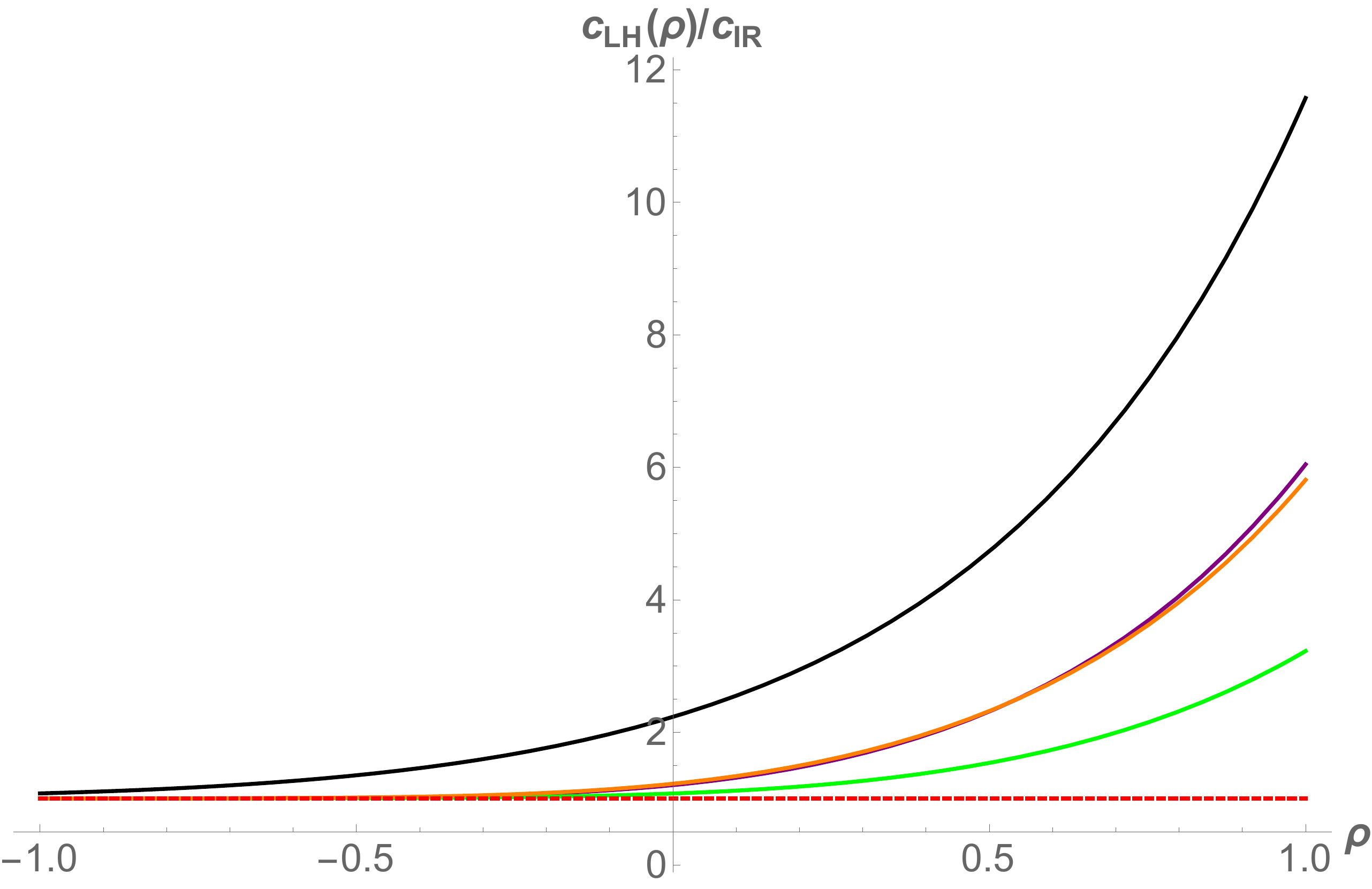}
    \caption{Holographic $c$-functions (\ref{c:5d:NEC:xi}) divided by the corresponding IR asymptotic values $c_\text{IR}=c_\text{LH}(-\infty)$ for flows from AdS$_5$ to AdS$_3$. See (\ref{BPS:near-horizon}) for the explicit values of $c_\text{IR}=L_\text{IR}e^{2\tilde G_0}$. Black/green/purple/orange curves represent $(p^1,p^2,\kappa)$=$(1/3,1/4,-1)$,$(2,3,-1)$,$(1,2,1)$,$(1,2,0)$ cases respectively. They all approach 1 (dashed red line) in the IR limit $\rho\to-\infty$.}\label{cc:across:NEC}
\end{figure}

The LH $c$-function (\ref{eq:cfunction?}) in flows from AdS$_5$ to AdS$_3$, (\ref{ansatz:BS}), reduces to
\begin{equation}
    c_\text{LH}(z)=\fft{\ell^2}{(e^{-f(z)-2g(z)})'}.\label{c:5d:NEC}
\end{equation}
To present the behavior of the LH $c$-function (\ref{c:5d:NEC}) in flows from AdS$_5$ to AdS$_3$ more explicitly, we consider it in numerical BPS AdS$_5$ black string solutions found by solving the system (\ref{N=2:sugra:BPS:reduced}).  In these BPS black string solutions, the LH $c$-function (\ref{c:5d:NEC}) reads (recall $\ell=1$ for these solutions)
\begin{equation}
    c_\text{LH}(z)\quad\to\quad c_\text{LH}(\rho)=\fft{e^{2g(\rho)}}{f'(\rho)+2g'(\rho)},\label{c:5d:NEC:xi}
\end{equation}
in terms of a new holographic radial coordinate $\rho$ introduced in (\ref{z:to:rho}). In Fig. \ref{cc:across:NEC} we show the behavior of the LH $c$-function (\ref{c:5d:NEC:xi}) for AdS$_5$ black strings with different horizons characterized by $\kappa\in\{1,0,-1\}$ and various configurations of magnetic charges $(p^1,p^2,p^3)$ under the constraint $p^1+p^2+p^3=-\kappa$. Note that the LH $c$-function (\ref{c:5d:NEC:xi}) has the main property that we expect, namely,  decreases monotonically as the flow approaches IR ($\rho\to-\infty$) starting from the infinite UV behavior ($\rho\to\infty$) in these examples. 
%

\subsubsection{EE $c$-functions}\label{sec:EE:5d:EE}
We now construct two different EE $c$-functions in flows from AdS$_5$ to AdS$_3$ by exploring an effective logarithmic contribution to the holographic EE (\ref{EE:area:5d:2}) in flows (\ref{ansatz:BS}).

To extract an effective logarithmic contribution to the EE (\ref{EE:area:5d:2}) in flows (\ref{ansatz:BS}), one should expand the EE (\ref{EE:area:5d:2}) with respect to the radius $R$ by solving the equations of motion (\ref{BS:Einstein}) and (\ref{BS:scalar}) for the metric functions $f(z)$ and $g(z)$ in the Ansatz (\ref{ansatz:BS}), substituting them into (\ref{EE:area:5d:2}), and then using the relation (\ref{z:to:R}). This is highly involved for a finite $R$ in general. To get intuition, we focus on the two limits where analytical results can be achieved: the small entangling region limit ($R/L_\text{UV}\ll1$) and the large entangling region limit ($R/L_\text{IR}\gg1$). Reference \cite{Bea:2015fja} provided a similar asymptotic analysis for theories with the `universal twist' \cite{Bobev:2017uzs}.

For a small entangling region ($R/L_\text{UV}\ll1$), the corresponding minimal surface in the bulk will not probe the deep IR region along the holographic radial coordinate. In other words, the turning point,  $z_0$, in the minimal surface area (\ref{EE:area:5d:2}) will be small compared to the AdS$_5$ radius $L_\text{UV}$ and therefore we can use both $R/L_\text{UV}\ll1$ and $z_0/L_\text{UV}\ll1$ in the small entangling region limit. This means that the small $z/L_\text{UV}$ expansion of the $\mathcal F$ function introduced in (\ref{eq:F}) is sufficient to accurately evaluate the EE using the formula (\ref{EE:area:5d:2}) in the small entangling region limit. As a result, following details summarized in \ref{App:EE:5d}, one can derive the UV expansion of EE (\ref{EE:area:5d:2}) as
\begin{equation}
\begin{split}
	S_\text{EE}(R;B^1{\times}\Sigma_{\mathfrak g},\epsilon)&=\fft{L_\text{UV}^3\text{vol}[\Sigma_{\mathfrak g}]}{2G^{(5)}_N}\left(\fft{\ell^2}{2\epsilon^2}-\fft{\pi^\fft32\Gamma(\fft23)^3}{2\Gamma(\fft16)^3}\fft{\ell^2}{R^2}-\fft{\kappa}{3}\log\fft{R}{\epsilon}+\mathcal O((\fft{R}{\ell})^0)\right).\label{area:UV:4}
\end{split}
\end{equation}
Some remarks on the UV expansion of EE (\ref{area:UV:4}) are in order:
\begin{itemize}

    \item The first term in the UV expansion (\ref{area:UV:4}) is consistent with the area law, namely, it is proportional to the area of the entangling surface $\partial B^1\times\Sigma_{\mathfrak g}$ divided by the UV cutoff square, namely $\fft{l^2}{\epsilon^2}$ where $\ell$ denotes the radius of a Riemann surface $\Sigma_{\mathfrak g}$. Note that $\partial B^1$ simply corresponds to two points and therefore the first term in the UV expansion (\ref{area:UV:4}) is proportional to $\fft{l^2}{\epsilon^2}$ but independent of $R$ the radius of $B^1$. 
    \item The second term in the UV expansion (\ref{area:UV:4}) is absent in (\ref{Eq:EE-General}) and this is hardly surprising since there is no additional length scale like the radius of a Riemann surface, $\ell$, when we consider the EE for a spherical entangling surface in flows within the same dimension. In other words, an extra parameter of length dimension, $\ell$, allows for the UV expansion of EE to have a term proportional to $\fft{\ell^2}{R^2}$ as in (\ref{area:UV:4}).
    \item The constant term of order $(\fft{R}{\ell})^0$ in (\ref{area:UV:4}) can be computed explicitly but it does not yield a universal contribution as scaling the UV cutoff $\epsilon$ can change such a contribution arbitrarily.

\end{itemize}

For a large entangling region ($R/L_\text{IR}\gg1$), the corresponding minimal surface in the bulk will probe the deep IR region along the holographic radial coordinate. In other words, $z_0$ in the minimal surface area (\ref{EE:area:5d:2}) will approach the horizon $z=\infty$. Hence the large entangling region limit $R/L_\text{IR}\gg1$ can also be written as $L_\text{IR}/z_0\ll1$. We will evaluate the EE (\ref{EE:area:5d:2}) in this limit. Following  details summarized in \ref{App:EE:5d}, one can derive the IR expansion of EE (\ref{EE:area:5d:2}) as
\begin{equation}
\begin{split}
	S_\text{EE}(R;B^1{\times}\Sigma_{\mathfrak g},\epsilon)=\fft{\ell^2\text{vol}[\Sigma_{\mathfrak g}]}{2G^{(5)}_N}L_\text{IR}e^{2\tilde G_0}\log\fft{R}{\Lambda}+\mathcal O((\fft{L_\text{IR}}{R})^0).\label{area:IR:3}
\end{split}
\end{equation}
%

\subsubsection*{Monotonic $c$-function}
If we view the flow from the two-dimensional point of view, the entropic treatment suggests to consider a holographic $c$-function (\ref{eq:SEE-Sq-dR}) introduced in \cite{Casini:2006es}, which we repeat here as
\begin{empheq}[box=\fbox]{equation}
	c_\text{mono}(R)\equiv R\, \partial_R \, S_\text{EE}(R;B^1{\times}\Sigma_{\mathfrak g},\epsilon).\label{cc:across:mono}
\end{empheq}
This partial $c$-function (\ref{cc:across:mono}) yields the coefficient of a universal logarithmic contribution in the IR expansion of EE, (\ref{area:IR:3}). The same holographic $c$-function was constructed in flows within the same dimension using the EE for a strip region on the boundary \cite{Myers:2012ed}.

The holographic $c$-function (\ref{cc:across:mono}), however, does \emph{not} correspond to the coefficient of an effective logarithmic contribution to the EE (\ref{EE:area:5d:2}), which we have considered as a natural candidate for holographic $c$-functions. This is because a simple logarithmic derivative of an EE does not yield the coefficient of a universal logarithmic contribution in the UV expansion of EE, (\ref{area:UV:4}). Hence the  holographic $c$-function (\ref{cc:across:mono}) matches a 2d central charge in the IR ($c_\text{mono}(R{\to}\infty)=c_\text{IR}/3$ to be precise) but does not reproduce a 4d central charge in the UV. In fact, the  holographic $c$-function in equation (\ref{cc:across:mono}) blows up in the UV due to the 2nd term proportional to $\fft{\ell^2}{R^2}$ in the UV expansion (\ref{area:UV:4}). As we have already discussed in section \ref{sec:NEC}, such a divergent UV behavior is natural in that a 4d UV CFT has an infinite number of degrees of freedom when viewed by a 2d observer. 

\begin{figure}[t]
	\centering
	\includegraphics[width=0.46\textwidth]{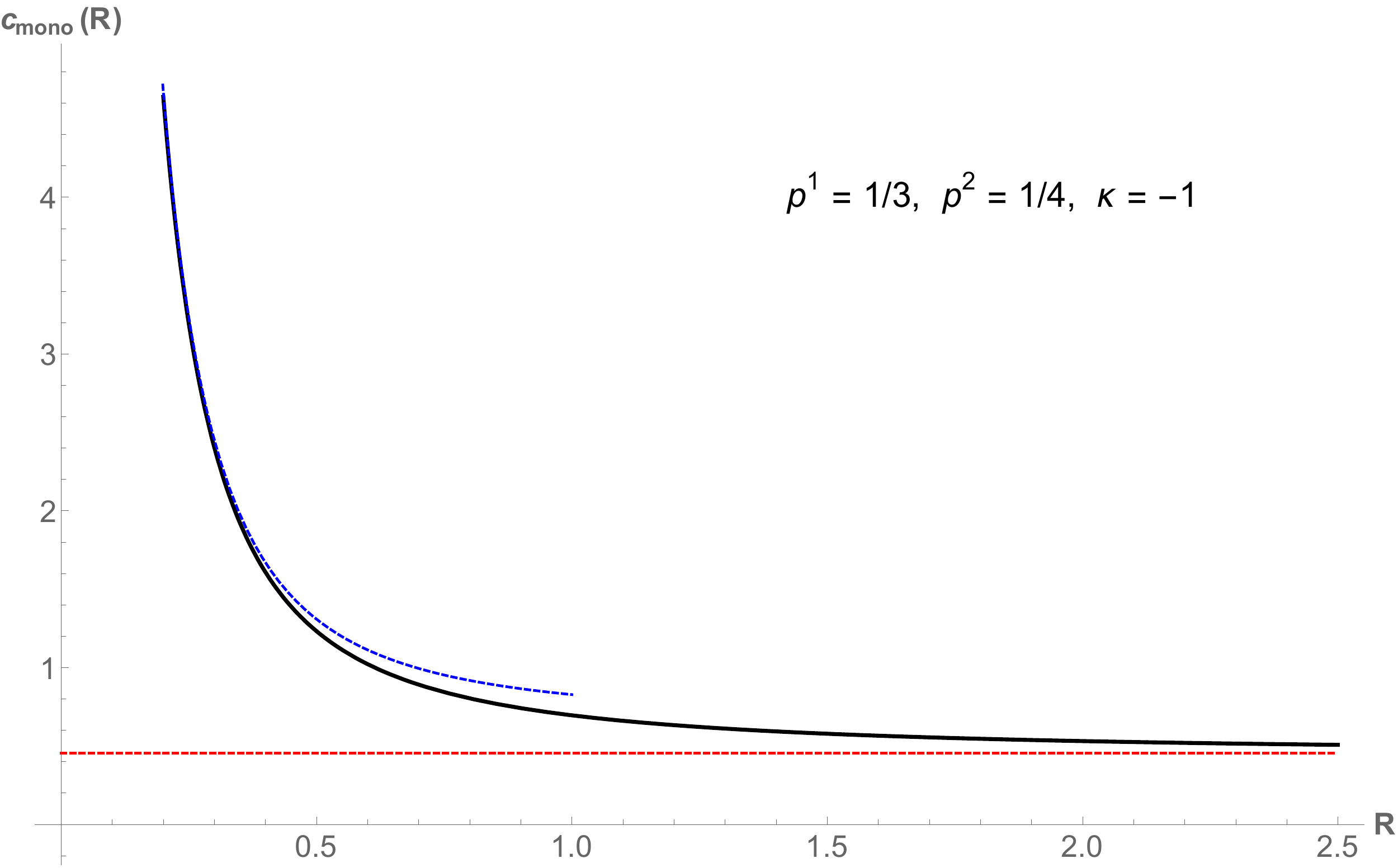}~~
    \includegraphics[width=0.46\textwidth]{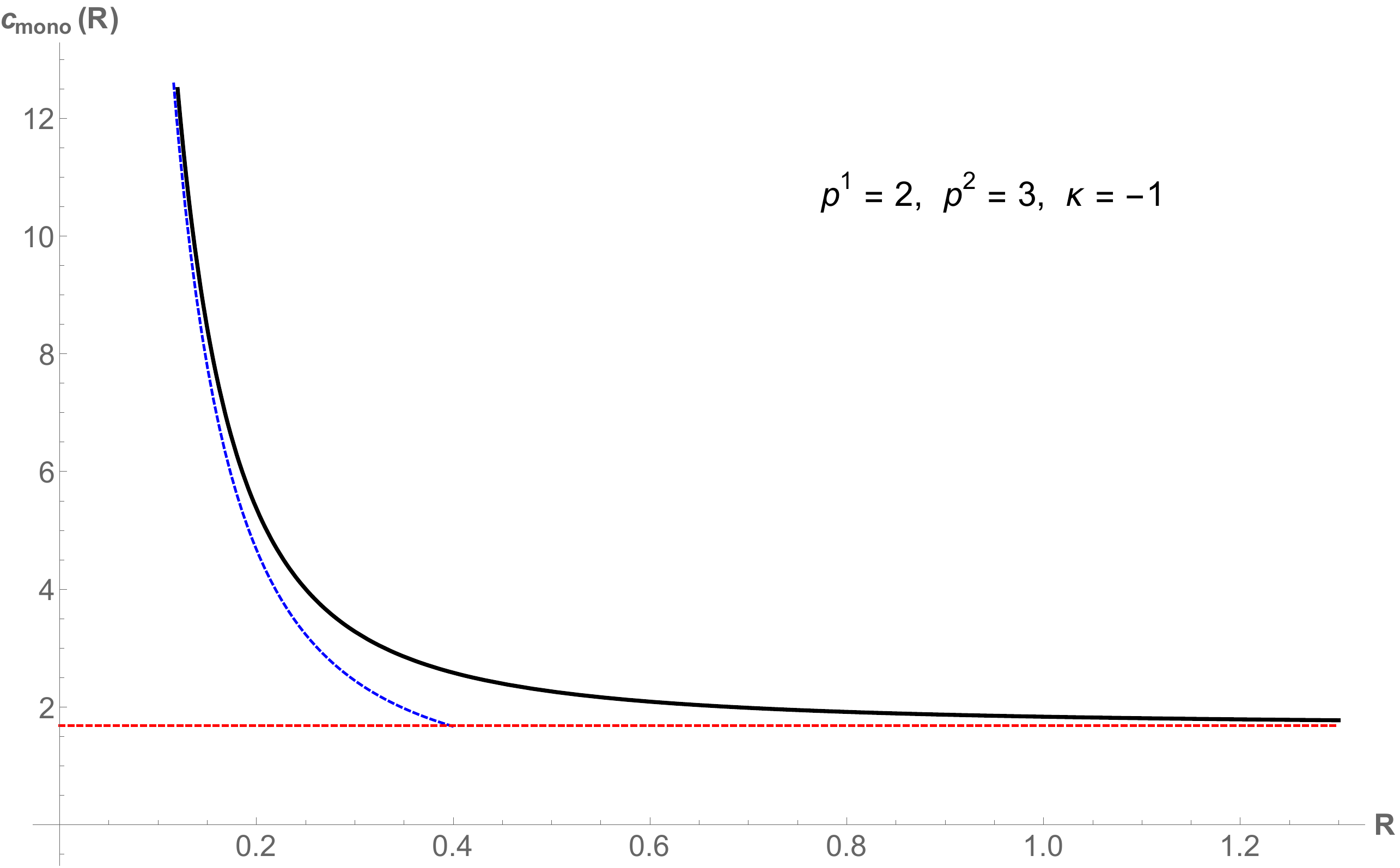}\\
    \includegraphics[width=0.46\textwidth]{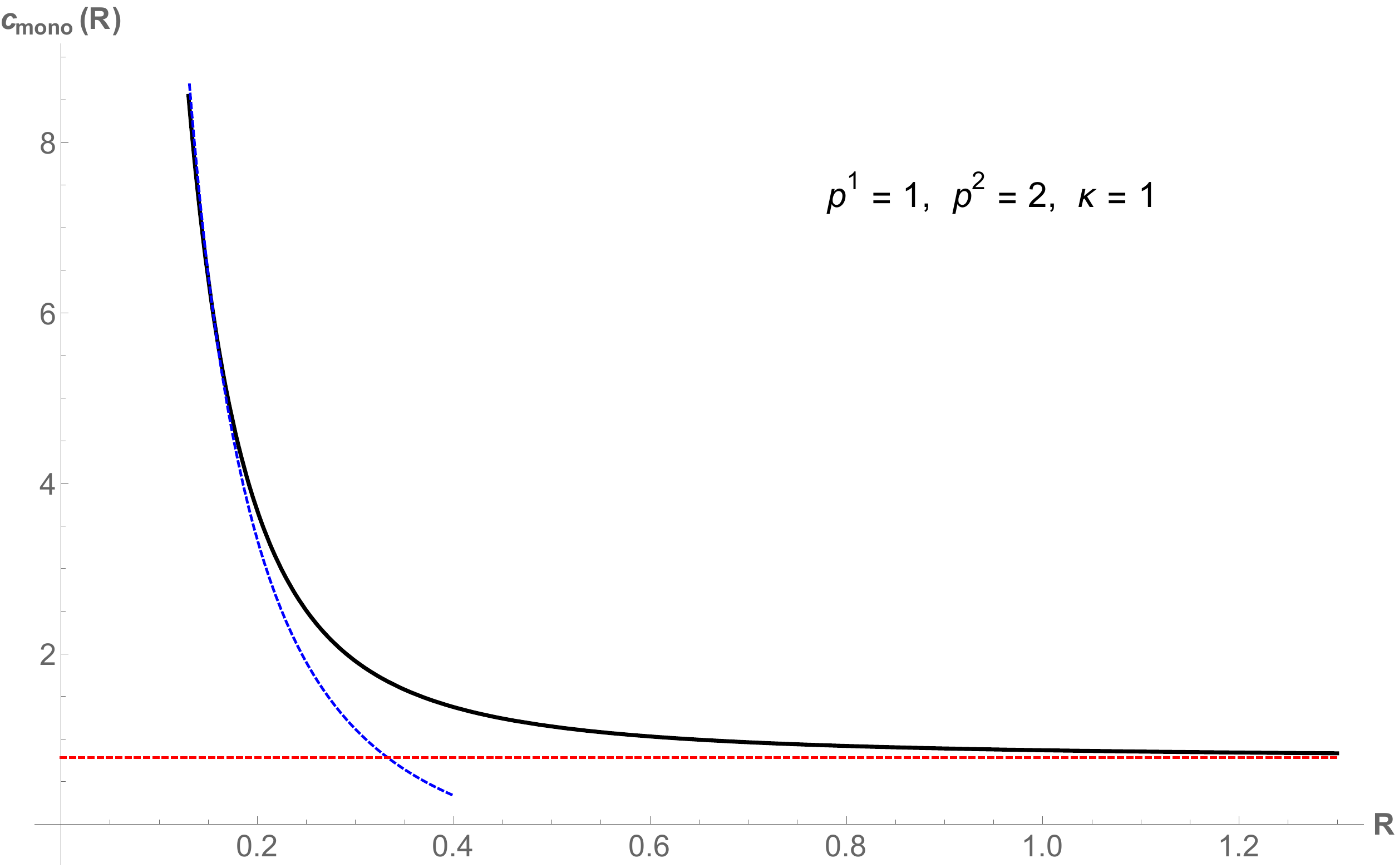}~~
    \includegraphics[width=0.46\textwidth]{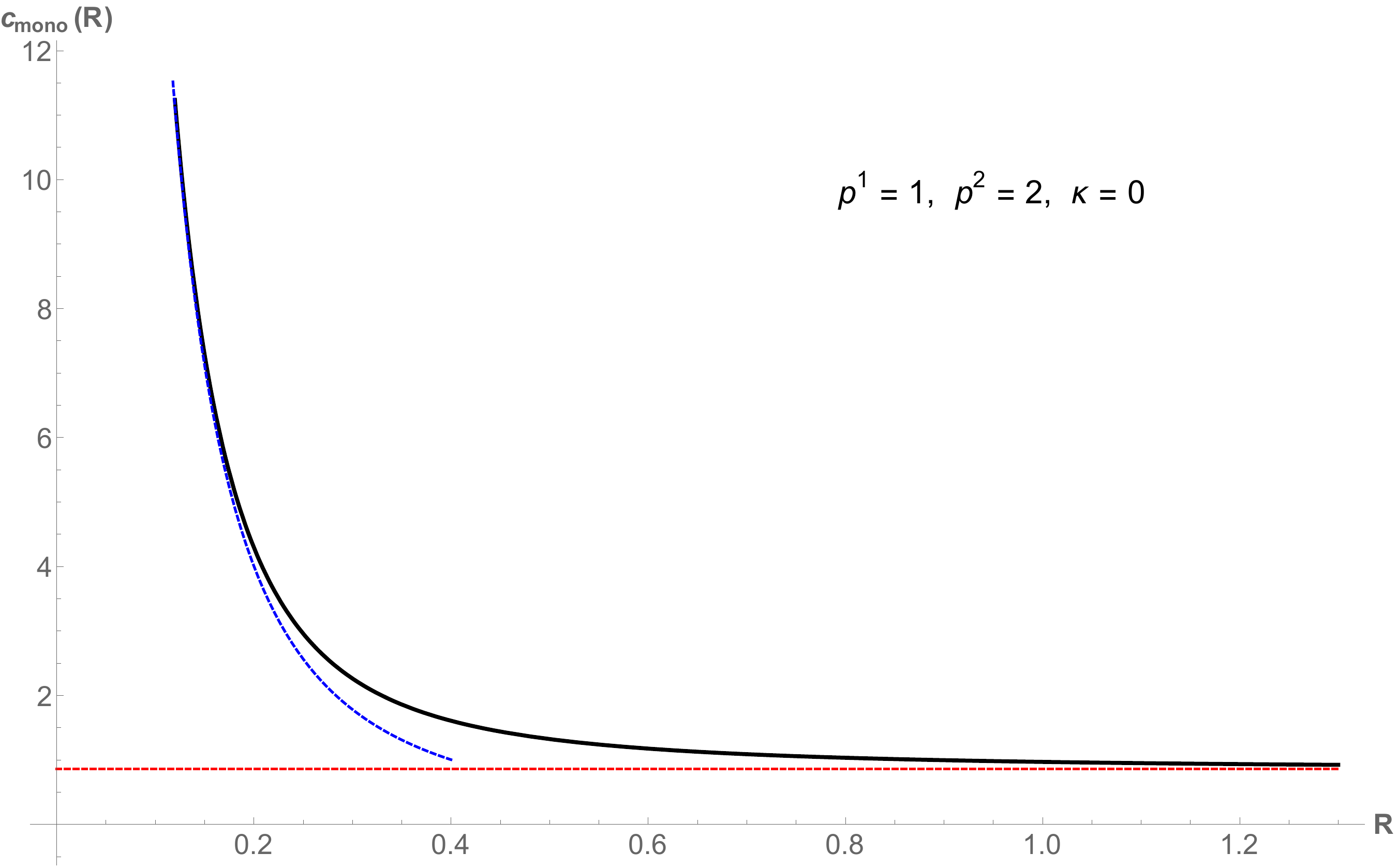}
	\caption{Monotonic holographic $c$-functions (\ref{cc:across:mono}) for flows from AdS$_5$ to AdS$_3$ (set $\fft{\text{vol}[\Sigma_{\mathfrak g}]}{4G^{(5)}_N}=1$ for presentation). Red dashed lines represent the asymptotic IR values (\ref{c:mono:IR}) and blue dashes curves represent a universal $\fft{1}{R^2}$-divergent UV behavior (\ref{c:mono:UV}).}\label{cc:across:5to3:mono}
\end{figure}

To present the behavior of the monotonic holographic $c$-function (\ref{cc:across:mono}) in flows from AdS$_5$ to AdS$_3$ more explicitly, we consider it in numerical BPS AdS$_5$ black string solutions found by solving (\ref{N=2:sugra:BPS:reduced}). Using the numerical solutions, we evaluated the EE (\ref{EE:area:5d:2}) with $D=4$ for different values of the radius $R$ of the entangling region. Based on these data, we derived the  monotonic holographic $c$-function (\ref{cc:across:mono}) for a given magnetically charged BPS AdS$_5$ black string solution. Fig. \ref{cc:across:5to3:mono} shows the behavior of the resulting holographic $c$-functions (\ref{cc:across:mono}) for various magnetically charged AdS$_5$ black strings. The monotonic holographic $c$-function (\ref{cc:across:mono}) starts from the IR value
\begin{equation}
	c_\text{mono}(R\to\infty)=\fft{\text{vol}[\Sigma_{\mathfrak g}]}{4G_N^{(5)}}\fft{4p^1p^2p^3}{-(p^1)^2-(p^2)^2-(p^3)^2+2(p^1p^2+p^2p^3+p^3p^1)},\label{c:mono:IR}
\end{equation}
and diverges in the UV regime following a universal curve
\begin{equation}
    c_\text{mono}(R\to0)=\fft{\text{vol}[\Sigma_{\mathfrak g}]}{2G_N^{(5)}}\fft{\pi^\fft32\Gamma(\fft23)^3}{\Gamma(\fft16)^3}\fft{1}{R^2}+\mathcal O(1).\label{c:mono:UV}
\end{equation}
The asymptotic behaviors (\ref{c:mono:IR}) and (\ref{c:mono:UV}) are derived by substituting IR and UV expansions of the EE, namely (\ref{area:IR:3}) and (\ref{area:UV:4}), into the definition (\ref{cc:across:mono}) and using $\boldsymbol{g}=\ell=L_\text{UV}=1$ for numerical BPS black string solutions we have constructed. In (\ref{c:mono:IR}) we have also used a near-horizon BPS solution \cite{Benini:2013cda}
\begin{equation}
\begin{split}
	L_\text{IR}&=\fft{2\left(p^1p^2p^3(-p^1+p^2+p^3)(p^1-p^2+p^3)(p^1+p^2-p^3)\right)^\fft13}{-(p^1)^2-(p^2)^2-(p^3)^2+2(p^1p^2+p^2p^3+p^3p^1)},\\
	e^{2\tilde G_0}&=\left(\fft{(p^1)^2(p^2)^2(p^3)^2}{(-p^1+p^2+p^3)(p^1-p^2+p^3)(p^1+p^2-p^3)}\right)^\fft13,
\end{split}\label{BPS:near-horizon}
\end{equation}
where $\tilde G_0$ has been introduced in (\ref{fgphi:IR}) as a term in the IR expansion of the background metric.
%

\subsubsection*{Interpolating $c$-function}
This time we construct an alternative holographic $c$-function for a flow from AdS$_5$ to AdS$_3$. This function interpolates between combinations of central charges across dimensions. Our proposal reads
\begin{empheq}[box=\fbox]{equation}
	c_\text{int}(R)\equiv\fft{1}{2} R\partial_R{\left(R\partial_R+2\right)}S_\text{EE}(R;B^1{\times}\Sigma_{\mathfrak g},\epsilon)~.\label{cc:across:int}
\end{empheq}
The holographic central charge (\ref{cc:across:int}) corresponds to the coefficient of an effective logarithmic contribution to the EE (\ref{EE:area:5d:2}). In UV and IR regimes, (\ref{cc:across:int}) indeed yields the coefficient of a universal logarithmic contribution to the EE, which are closely related to the 4d central charges of a boundary CFT and the 2d central charge of a horizon CFT, respectively. One can prove this property explicitly by substituting the UV expansion (\ref{area:UV:4}) and the IR expansion (\ref{area:IR:3}) into the proposal (\ref{cc:across:int}) as
\begin{equation}
\begin{alignedat}{2}
    c_\text{int}(R\to0)&=-\fft{\kappa\text{vol}[\Sigma_{\mathfrak g}]}{3}\fft{L_\text{UV}^3}{2G^{(5)}_N}&~~&=-\fft{4\kappa\text{vol}[\Sigma_{\mathfrak g}]}{3\pi}a_\text{UV},\\
    c_\text{int}(R\to\infty)&=\fft{\ell^2L_\text{IR}e^{2\tilde G_0}}{2G^{(5)}_N}\text{vol}[\Sigma_{\mathfrak g}]&~~&=\fft{1}{3}c_\text{IR},
\end{alignedat}\label{cc:across:5to3:int:endpoints}
\end{equation}
where we have used $a_\text{UV}=\fft{\pi L_\text{UV}^3}{8G_N^{(5)}}$ for a 4d $a$ central charge \cite{Bobev:2017uzs}: note that the first operator $R\partial_R$ extracts the constant coefficient of $\log R$ term and the second one $R\partial_R+2$ eliminates the $\fft{1}{R^2}$-leading term in the UV expansion (\ref{area:UV:4}). The overall difference by a multiplicative factor between central charges and a holographic $c$-function in (\ref{cc:across:5to3:int:endpoints}) is consistent with the literature: the difference in UV will be derived below and the difference in IR is consistent with the observation in \cite{Ryu:2006bv,Ryu:2006ef}.

\begin{figure}[t]
\centering
\includegraphics[width=0.46\textwidth]{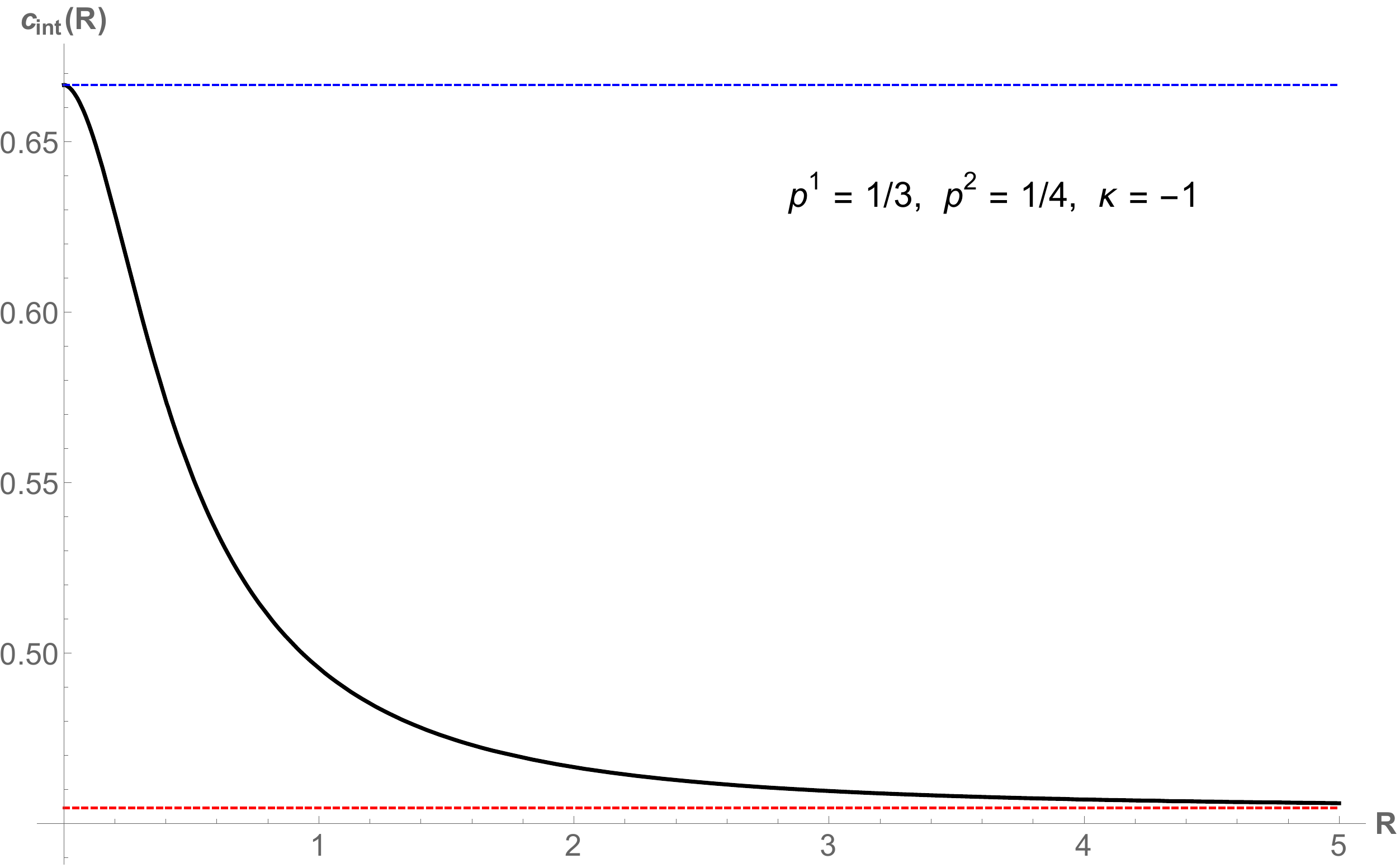}~~
\includegraphics[width=0.46\textwidth]{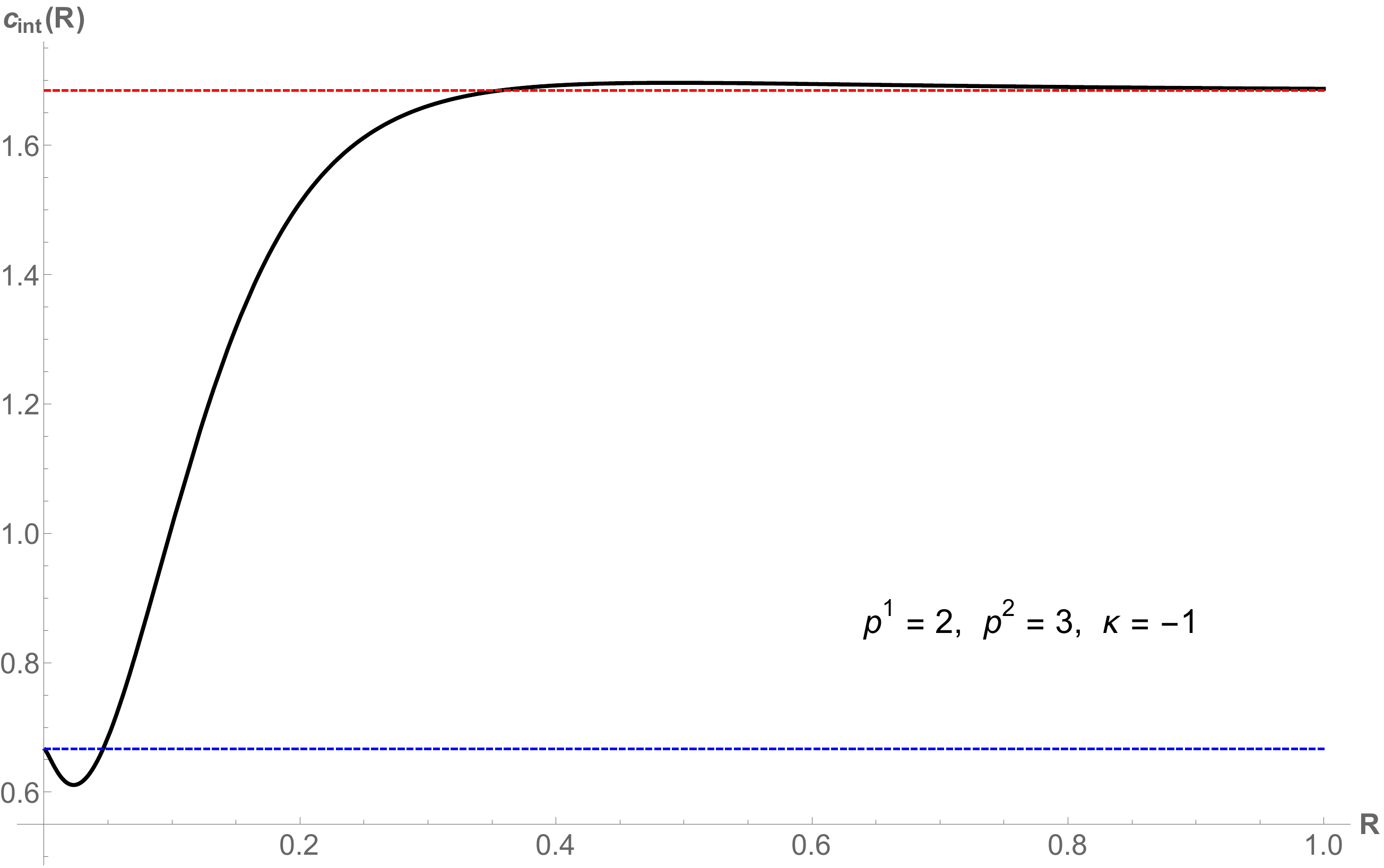}\\
\includegraphics[width=0.46\textwidth]{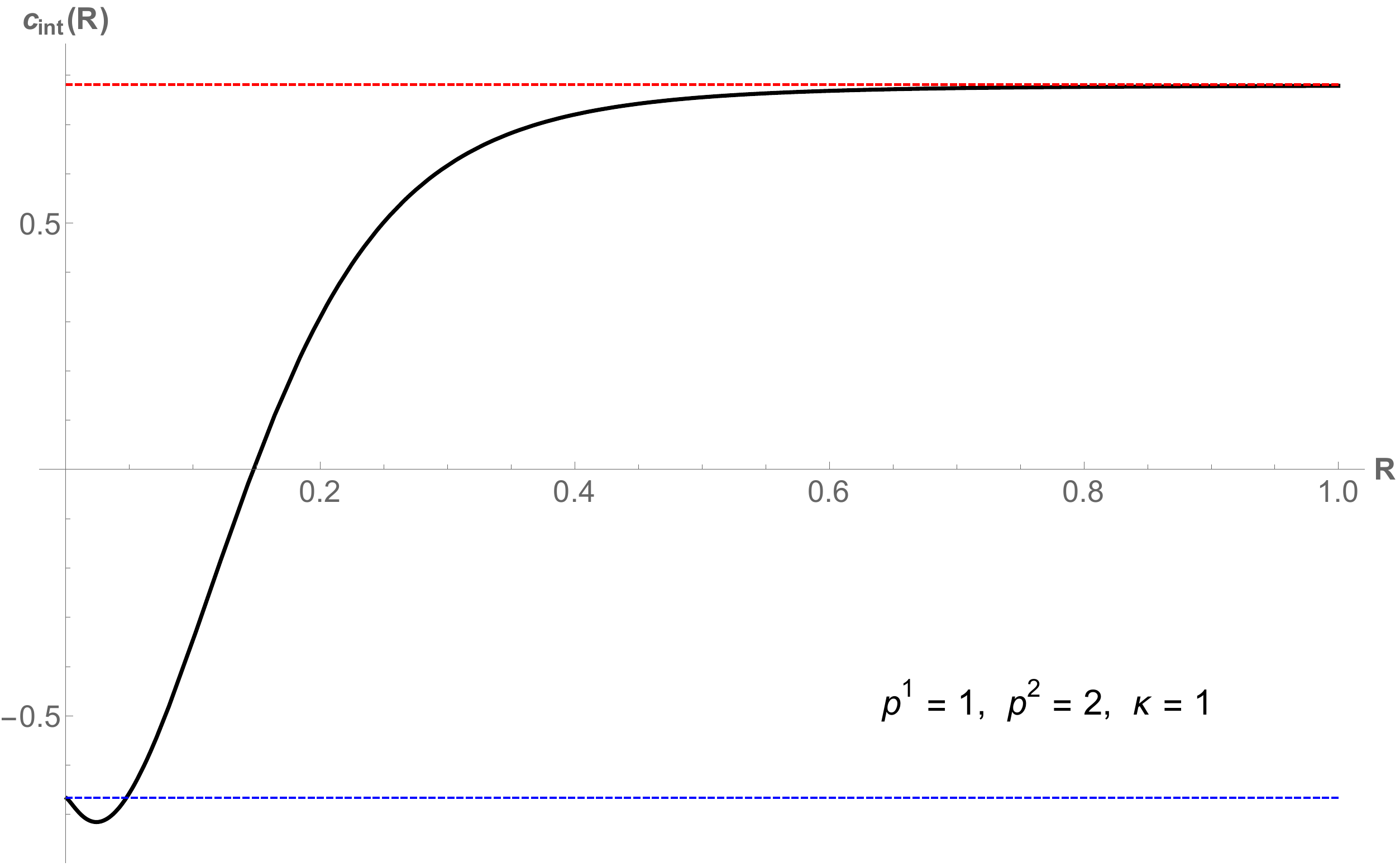}~~
\includegraphics[width=0.46\textwidth]{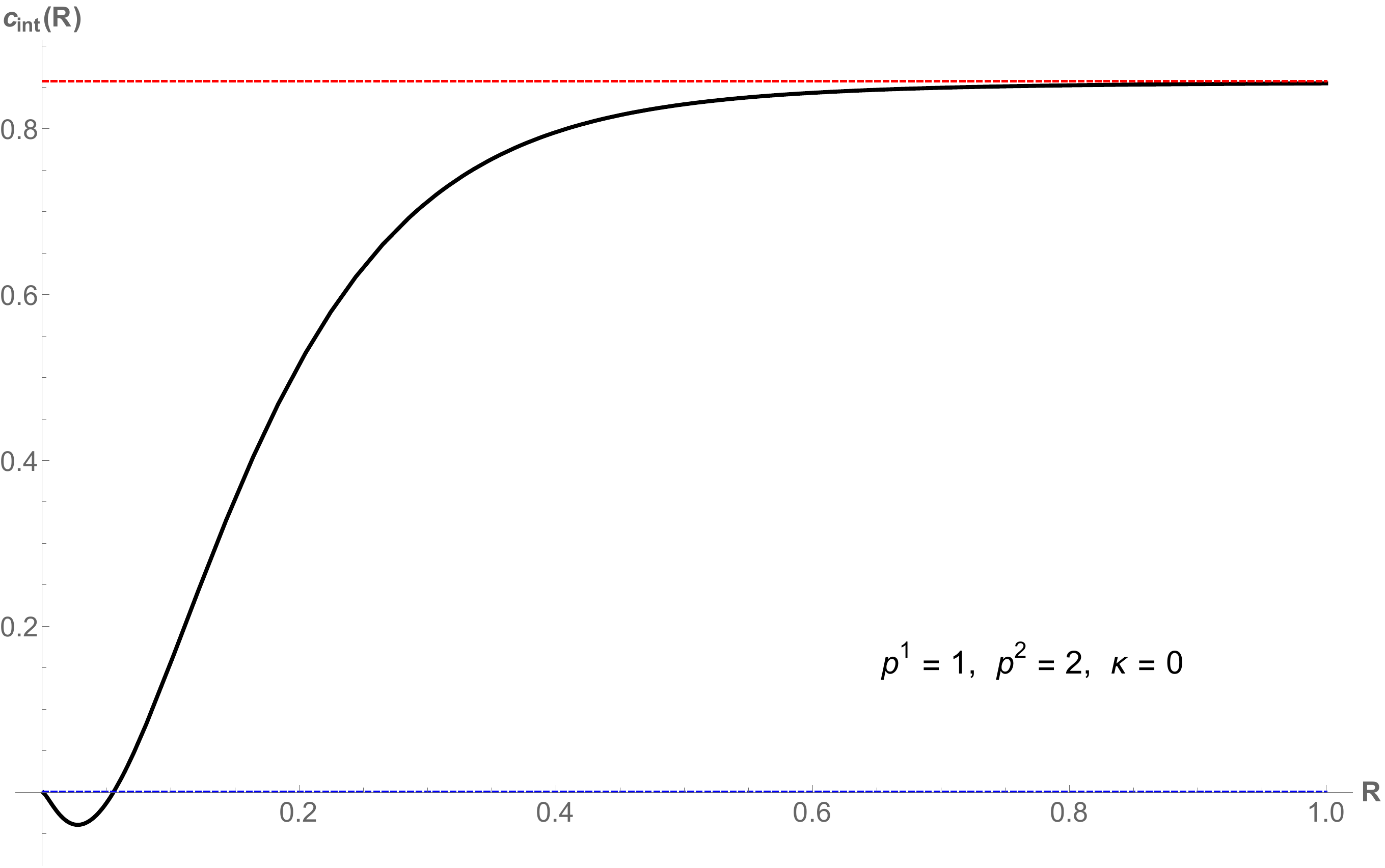}
    \caption{Interpolating holographic $c$-functions (\ref{cc:across:int}) for flows from AdS$_5$ to AdS$_3$ (set $\fft{\text{vol}[\Sigma_{\mathfrak g}]}{4G^{(5)}_N}=1$ for presentation). Red dashed lines represent the asymptotic IR values (\ref{c:int:IR}) and blue dashes curves represent the asymptotic UV values (\ref{c:int:UV}).}\label{cc:across:5to3:int}
\end{figure}

However, the interpolating holographic $c$-function (\ref{cc:across:int}) does \emph{not} behave monotonically. This is not surprising because central charges of different dimensional CFTs count different dimensional degrees of freedom as we have already explained in section \ref{sec:EE}: the 4d central charge of a boundary UV CFT is not necessarily larger than the 2d central charge of a horizon IR CFT as can be seen in some of the panels of Fig.~\ref{cc:across:5to3:int}. 

To present the behavior of an interpolating holographic $c$-function (\ref{cc:across:int}) in flows from AdS$_5$ to AdS$_3$ more explicitly, we consider it in numerical BPS AdS$_5$ black string solutions found by solving (\ref{N=2:sugra:BPS:reduced}). Using the numerical solutions, we evaluated the EE (\ref{EE:area:5d:2}) with $D=4$ for different values of the radius $R$ of the entangling region. Based on these data, we derived an interpolating  holographic $c$-function (\ref{cc:across:int}) for a given magnetically charged BPS AdS$_5$ black string solution. See Fig. \ref{cc:across:5to3:int} for the results. The interpolating holographic $c$-function (\ref{cc:across:int}) connects the IR value
\begin{equation}
	c_\text{int}(R\to\infty)=\fft{\text{vol}[\Sigma_{\mathfrak g}]}{4G_N^{(5)}}\fft{4p^1p^2p^3}{-(p^1)^2-(p^2)^2-(p^3)^2+2(p^1p^2+p^2p^3+p^3p^1)}\label{c:int:IR}
\end{equation}
from (\ref{cc:across:5to3:int:endpoints}) in the large entangling region limit to the UV value
\begin{equation}
	c_\text{int}(R\to0)=-\fft{\kappa}{3}\fft{\text{vol}[\Sigma_{\mathfrak g}]}{2G^{(5)}_N}\label{c:int:UV}
\end{equation}
from (\ref{cc:across:5to3:int:endpoints}) in the small entangling region limit. In (\ref{c:mono:IR}) we have used a near-horizon BPS solution (\ref{BPS:near-horizon}). Also recall that $\ell=\boldsymbol{g}=L_\text{UV}=1$ for numerical BPS black string solutions we have constructed.

\medskip

Here we derive the precise relation between 4d central charges and the coefficient of a universal logarithmic contribution to the EE. The final result will explain the $-\fft{4\kappa\text{vol}[\Sigma_{\mathfrak g}]}{3\pi}$ factor in (\ref{cc:across:5to3:int:endpoints}).

Consider a 4d CFT compactified on a constant-curvature Riemann surface $\Sigma_{\mathfrak g}$. The total geometry is $\RR^2\times\Sigma_{\mathfrak g}$ and a constant-time slice is $\RR\times\Sigma_{\mathfrak g}$. Consider the EE for an interval $B^1$ in $\RR$. That is, the entangling surface $\partial B^1\times\RR$ wraps $\Sigma_{\mathfrak g}$ and consists of two points in $\RR$. The entangling surface has no extrinsic curvature -- it is itself an extremal surface.

As appropriate for a 4d CFT, the EE contains a quadratic area divergence and a subleading logarithmic one, whose coefficient encodes the central charge. We wish to identify the precise combination of the central charges $a$, $c$ that appears as coefficient of the logarithmic term for this particular surface.

We start from equation (4.27) of \cite{Ryu:2006ef}
\begin{equation}
    R \frac{\partial }{\partial R} S_\text{EE}(R;B^1{\times}\Sigma_{\mathfrak g},\epsilon)=\lim_{n\rightarrow 1}\frac{\partial}{\partial n}\left[
    -\frac{c}{16\pi^2}\int_{M_n}d^4x \sqrt{g} W^2+\frac{a}{16\pi^2}\int_{M_n}d^4x\sqrt{g} E_4\right]~.\label{eq:dSA}
\end{equation}
Here $S_\text{EE}(R;B^1,\epsilon)$ is the EE of an entangling surface $\mathcal A=B^1\times\Sigma_{\mathfrak g}$ and $M_n$ is the replica manifold, i.e.,  an $n$-fold cover of the field theory geometry branched at the entangling surface. The radius of a one-dimensional ball $B^1$, $R$, was taken as a length scale associated with the entangling surface, which was denoted by $\ell$ in \cite{Ryu:2006ef}, but can equally be understood as scaling the cut-off (see footnote 13 there). The UV cutoff, $\epsilon$, is introduced as in the main text. The curvature invariants are the squared Weyl tensor and the Euler density (the Ricci scalar $R$ must be distinguished from a length scale $R$)
\begin{align}
    W^2&=W_{\mu\nu\rho\sigma}W^{\mu\nu\rho\sigma}=R_{\mu\nu\rho\sigma}R^{\mu\nu\rho\sigma}-2R_{\mu\nu}R^{\mu\nu}+\frac{1}{3}R^2~,
    \nonumber\\
    E_4&=\tilde R_{\mu\nu\rho\sigma}\tilde R^{\mu\nu\rho\sigma}=R_{\mu\nu\rho\sigma}R^{\mu\nu\rho\sigma}-4R_{\mu\nu}R^{\mu\nu}+R^2~.
\end{align}
The coefficients $a$ and $c$ in (\ref{eq:dSA}) encode the usual geometric contributions to the trace of the energy-momentum tensor of $\mathcal N=4$ SYM, which for the setups considered here also receives contributions from the background gauge fields implementing the twist, see e.g. \cite{Anselmi:1997am,Ohl:2010au}.

The replica manifold $M_n$ for $n$ close to one is the original field theory geometry with a conical deficit at the entangling surface. To evaluate (\ref{eq:dSA}) we need the singular contributions from the conical deficit at the entangling surface. These contributions were evaluated in \cite{Fursaev:1995ef}. The relevant integrated expressions were also collected in (4.32) of \cite{Ryu:2006ef}. They are given by\footnote{In these expressions $\delta_{\partial A}^2$ singularities that lead to the area divergence in the EE have been dropped.}
\begin{align}
    \int_{M_n}\sqrt{g}R^2&=	\int_{M_n\setminus \partial\mathcal A}\sqrt{g}R^2+8\pi(1-n)\int_{\partial\mathcal A}(R_{\partial\mathcal A}+2R_{ii}-R_{ijij})
    \nonumber\\
    \int_{M_n}R^{\mu\nu\rho\sigma}R_{\mu\nu\rho\sigma}&=	\int_{M_n\setminus \partial\mathcal A}R^{\mu\nu\rho\sigma}R_{\mu\nu\rho\sigma}+8\pi(1-n)\int_{\partial\mathcal A} R_{ijij}
    \nonumber\\
    \int_{M_n}R^{\mu\nu}R_{\mu\nu}&=	\int_{M_n\setminus \partial\mathcal A}R^{\mu\nu}R_{\mu\nu}+4\pi(1-n)\int_{\partial\mathcal A} R_{ii}~.
\end{align}
In these expressions $R_{\partial \mathcal A}$ is the intrinsic curvature of the entangling surface $\partial\mathcal A=\partial B^1\times\Sigma_{\mathfrak g}$, while $R_{ii}$ and $R_{ijij}$ are curvature components projected onto the orthogonal directions to $\partial\mathcal A$, e.g.\ $R_{ij}=n^{\mu}_i n^{\nu}_j R_{\mu\nu}$.

For the particular setup we are interested in here, the curvature components normal to $\partial\mathcal A$ are zero. Evaluating the derivative with respect to $n$ leads to 
\begin{equation}
    R \frac{\partial }{\partial R}S_\text{EE}(R;B^1{\times}\Sigma_{\mathfrak g},\epsilon)=\frac{c-3a}{6\pi}\int_{\partial\mathcal A}R_{\partial\mathcal A}~.\label{EE:log:1}
\end{equation}
The entangling surface $\partial\mathcal A$ consists of two copies of the Riemann surface $\Sigma_{\mathfrak g}$, and the curvature integral can be evaluated using Gauss-Bonnet. Integrating for the EE leads to
\begin{equation}
    S_\text{EE}(R;B^1{\times}\Sigma_{\mathfrak g},\epsilon)=\frac{\#}{\epsilon^2} + \frac{c-3a}{3\pi}\int_{\Sigma_{\mathfrak g}}R_{\Sigma_{\mathfrak g}}\log\frac{R}{\epsilon} + \mathcal O(1)~.\label{EE:log:2}
\end{equation}
Using
\begin{equation}
    \int_{\Sigma_{\mathfrak g}}R_{\Sigma_{\mathfrak g}}=2\kappa\,\text{vol}[\Sigma_{\mathfrak g}]
\end{equation}
for the Riemann surface of unit radius where $\kappa=-1,0,1$ for $\Sigma_{\mathfrak g}=H^2,T^2,S^2$ respectively and the fact that $a=c$ in the large-$N$ limit\,(\textit{i.e.} small Newton's constant limit), we can simplify (\ref{EE:log:2}) as
\begin{equation}
    S_\text{EE}(R;B^1{\times}\Sigma_{\mathfrak g},\epsilon)=\frac{\#}{\epsilon^2}-\frac{4\kappa\text{vol}[\Sigma_{\mathfrak g}]}{3\pi}a\,\log\frac{R}{\epsilon} + \mathcal O(1)~.\label{EE:log:3}
\end{equation}
This explains the $-\fft{4\kappa\text{vol}[\Sigma_{\mathfrak g}]}{3\pi}$ factor of difference in (\ref{cc:across:5to3:int:endpoints}) precisely. Note that the above described analysis, however, is not appropriate to keep track of the whole UV expansion of EE associated with an entangling region wrapping $\Sigma_{\mathfrak g}$: for example, the leading $\fft{\ell^2}{R^2}$-order contribution in (\ref{area:UV:4}) is hidden within $\mathcal O(1)$ in (\ref{EE:log:3}). This is why we focused on specific flows from AdS$_5$ to AdS$_3$ in 5d gauged STU model, which allows us to analyze the UV expansion of EE completely.

\subsection{Flows from AdS$_7$ to AdS$_3$: partially compactified 6d $\mathcal N=(2,0)$ theories}\label{sec:EX:7d}
In this subsection we work in the context of 7d maximal gauged supergravity which can, subsequently, be embedded in 11d sugra.  We are interested in a supergravity solution describing a flow that starts in AdS$_7$ towards AdS$_3\times \mathbb{H}_4$, where $\mathbb{H}_4$ is a hyperbolic 4-manifold. Such flows were constructed by Gauntlett, Kim and Waldram in \cite{Gauntlett:2000ng}. The metric takes the expected form:

\be
ds^2 =e^{2f(z)}\left(- dt^2 + dz^2+d r^2\right)+e^{2g(z)}ds^2_{ \mathbb{H}_4}.  \label{eq:metriz7}
\ee
Note that \eqref{metric}  reduces to \eqref{eq:metriz7} for $D=6$, $d=2$ and the asymptotic behavior is still given by \eqref{hor/asymp}. Proceeding parallel to the discussion in subsection \ref{sec:EX:5d} we want to find the effective logarithmic contribution to the EE \eqref{EE:area:5d:2} with $D=6$. In order to study the coefficient of a potential  universal logarithmic contribution to the EE we focus on a particular example of the metric \eqref{eq:metriz7}. The conceptual background for the interpolating solutions between AdS$_7$ in the UV and AdS$_3\times \mathbb{H}_4$ in the IR, can be found in \cite{Gauntlett:2000ng}. The particular BPS flows that we consider here are described in\cite{Gauntlett:2000ng} as an M5 brane wrapping a special Lagrangian four-cycle which we locally described as $\mathbb{H}_4$; an M5 wrapping such cycle gives rise to $(1,1)$ supersymmetry in 2d.  The field theory side corresponds, therefore, to a flow of  the 6d field theory living in the worldvolume of the M5 brane to an effective 2d theory in the worldvolume of the compactified M5 in the  IR. 
In Appendix \ref{App:7to3} we present technical details of the numerical implementation of such flows.

The details of the setup are not terribly important for our analysis and can be found in \cite{Gauntlett:2000ng}. Here we focus on the BPS equations which take the following form: 
\bea
e^{-f} f'&=& -\frac{m}{10}\left(4e^{-2\lambda}+e^{8\lambda}\right)+\frac{\kappa }{5m} e^{2\lambda-2g}-\frac{\kappa^2}{10m^3}e^{-4\lambda-4g}, \label{eq:eff} \\
e^{-f} g'&=& -\frac{m}{10}\left(4e^{-2\lambda}+e^{8\lambda}\right)-\frac{3\kappa }{10m} e^{2\lambda-2g}+\frac{\kappa^2}{15m^3}e^{-4\lambda-4g},\label{eq:egg}\\
e^{-f} \lambda'&=& \frac{m}{5}\left(e^{8\lambda}-e^{-2\lambda}\right)+\frac{\kappa }{10m} e^{2\lambda-2g}+\frac{\kappa^2}{30m^3}e^{-4\lambda-4g}, \label{eq:lambdaeq}
\eea
Besides the metric functions $f(z)$ and $g(z)$, we have the function $\lambda(z)$ which characterizes the single scalar field excited in the 7d maximal gauged supergravity theory. 

The next step is to solve the BPS equations \eqref{eq:eff}, \eqref{eq:egg} and \eqref{eq:lambdaeq}.
The solutions of interest have UV asymptotics ($z\to 0$) of the form $e^{2f}\sim e^{2g}\sim 1/z^2$. This corresponds to an AdS$_7$ region whose slices of constant $z$ take the form $\RR^{1,1}\times \mathbb{H}_4$. This foliation, rather than the traditional $\RR^{1,5}$, allows to interpret the dual of the solution as an M5 brane wrapped on $\mathbb{H}_4$.
It can be checked that the following solution satisfies the BPS equations (\ref{eq:lambdaeq}):
\be\label{eq:IRsolfgl}
e^{10\lambda}=\frac{3}{2}, \qquad e^{2g(z)}=\frac{e^{-6\lambda}}{m^2}, \qquad e^{f(z)}=\frac{e^{2\lambda}}{m}\,\,\,\frac{1}{z}. 
\ee
This solution is naturally interpreted as an IR fixed point. Note that the metric function $g(z)$ is constant and the behavior of $f(z)$ leads to an AdS$_3$ factor, which reinforces its interpretation as the dual of a CFT$_2$ theory. We are interested in studying flows that connect the AdS$_7$ solution in the UV $(z\to 0)$ to the AdS$_3\times \mathbb{H}_4$ region in the IR. Such smooth flows require $\kappa=-1$ above. The parameter $m$  describes the gauge coupling constant, that is, the inverse of the UV AdS radius. We present details of such flows in Appendix \ref{App:7to3}.

\subsubsection{Local Holographic $c$-function}
\begin{figure}[t]
\centering
\includegraphics[width=0.5\textwidth]{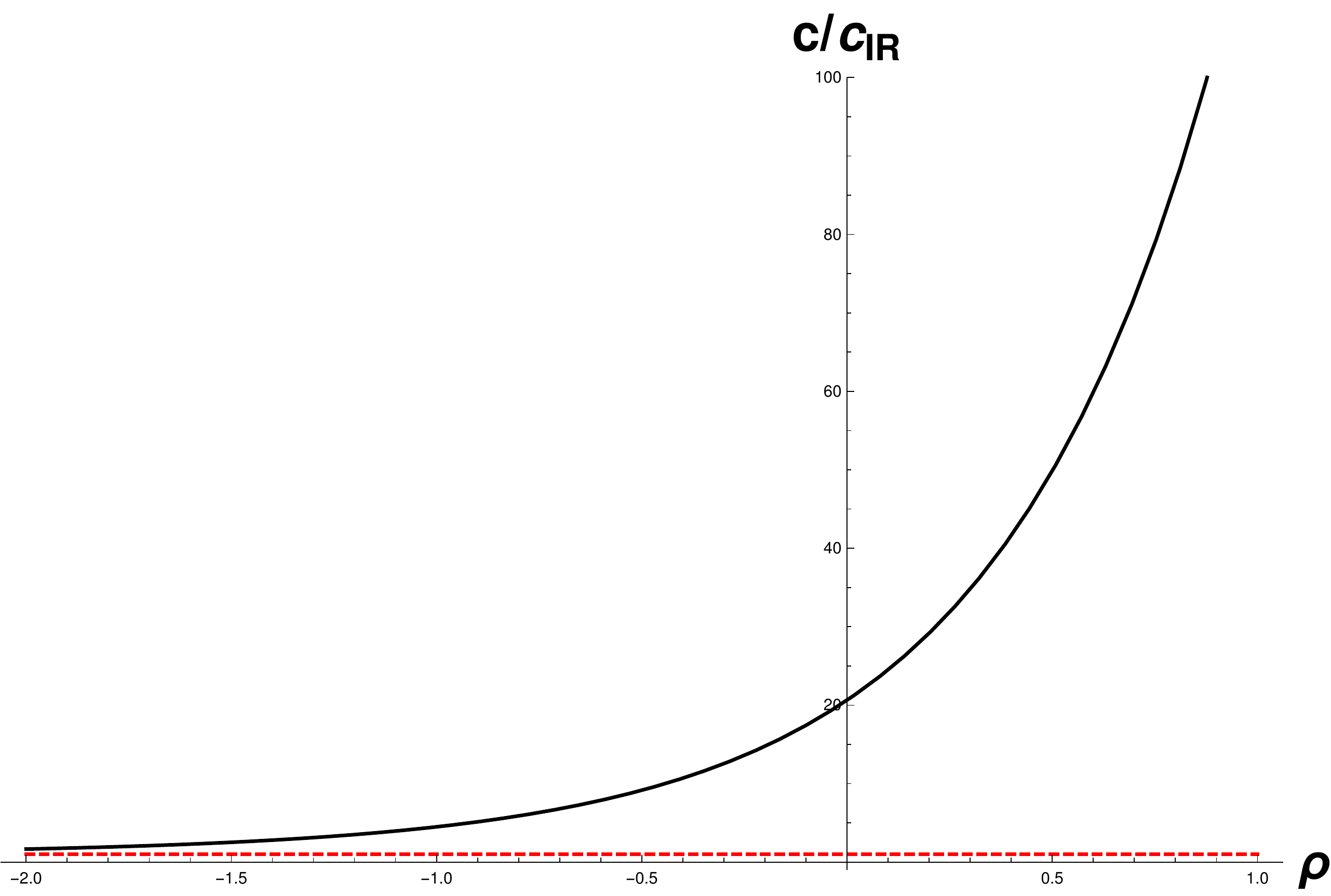}
\caption{LH  $c$-function (\ref{eq:cfunction?}) divided by its corresponding IR asymptotic value $c_\text{IR}$ for flows from AdS$_7$ to AdS$_3$. In the IR the curve approaches to the value $1$, which is indicated by the dashed red line.}\label{cc:across:NEC7d}
\end{figure}

Here we discuss the behavior of a partial $c$-function suggested by our analysis of NEC in section \ref{sec:NEC:across}. In Fig.~\ref{cc:across:NEC7d} we plot the expression (\ref{eq:cfunction?}) with the appropriate normalization. Note that, similar to the AdS$_5$ to AdS$_3$ flow, the function diverges in the UV region. We have advanced a natural explanation according to which as we move to the UV the lower dimensional central charge receives contributions form KK modes descending from the higher dimensional point of view and, therefore, diverges. The plot in Fig.~\ref{cc:across:NEC7d}, is steeper than the plot describing the flow from AdS$_5$ to AdS$_3$ (Fig.~\ref{cc:across:NEC}) and qualitatively supports our thesis for divergence. 

%


\subsubsection{EE $c$-functions}\label{sec:EE:7d:UVIR}

Given the functions $f(z)$ and $g(z)$ we can proceed similarly to section \ref{sec:EX:5d} to evaluate the EE \eqref{EE:area:5d:2} for $D=6$ and expand it in terms of $R$ \eqref{z:to:R} focusing on the small entangling region limit $(R/L_{\text{UV}} \ll1)$ and the large entangling region limit $(R/L_{\text{UV}} \gg 1)$ respectively. Note that in the UV region ($e^{f(z)} \sim e^{g(z)} \sim 1/z$) we have a factor of ${\rm vol}( \mathbb{H}_4)/z^5$ in the area functional which diverges at the same rate as the leading divergence in AdS$_7$. The entangling surface is almost perpendicular to the boundary, meaning that $dz/dr$ is a large number compared to one which leads to a divergence of the form $(R/\epsilon)^4$.


Details about the expansion for small entangling region can be found in Appendix  \ref{App:AdS7Entropy}. Rewriting $z_0$ in terms of $R$ from \eqref{eq:R7dexpansion} and then replacing it in \eqref{eq:SE7dexp}, we obtain
\begin{align}
    \begin{split}
        &S_\text{EE}(R;B^{1}\times \mathbb{H}_4,\epsilon)\\
        & = \fft{L_{\text{UV}}^5\text{vol}[\mathbb{H}_4]}{2G^{(7)}_N}  \left(\frac{8 \kappa^2}{15 m^4} \log \left(\frac{R}{\epsilon}\right)- \frac{2 \kappa}{3 m^2}\frac{\ell^2}{\epsilon^2}+\frac{\ell^4}{4 \epsilon^4}+A\frac{\ell^2}{R^2}+B\frac{\ell^4}{R^4}+ \mathcal{O}\left(\left(\frac{R}{\ell}\right)^0\right)\right),\label{eq:SEEexpansion7d}
    \end{split}
\end{align}
where \begin{align}A =\frac{\kappa}{m^2} \widetilde{A}\left(\frac{6 \Gamma\left(\frac{11}{10}\right)}{\pi^{\frac{1}{2}}\Gamma\left(\frac{8}{5}\right)}\right)^{-2}, \hspace{3mm} B = \widetilde{B}\left(\frac{6 \Gamma\left(\frac{11}{10}\right)}{\pi^{\frac{1}{2}}\Gamma\left(\frac{8}{5}\right)}\right)^{-4},
\end{align}
and $\tilde{A}, \tilde{B}$ can be found in appendix \ref{App:AdS7Entropy}. The $1/\epsilon^4$ and $1/\epsilon^2$ divergences in (\ref{eq:SEEexpansion7d}) are typical for AdS$_7$. Due to the presence of the scale $\ell$, setting the size of manifold locally represented as $\mathbb{H}_4$, we in addition have terms of the form $\ell^4/R^4$ and $\ell^2/R^2$ which are divergent in the limit of small radius, $R$, of the entangling surface. 

We now consier large entangling region  $(R/L_{\text{IR}}\gg 1)$, where we expect an AdS$_3$-like behavior of the holographic EE. The technical steps are quite analogous to those explained in the case of flows from AdS$_5$ to AdS$_3$ in section \ref{sec:EX:5d} and the details regarding the calculation of holographic EE can be found in Appendix \ref{App:AdS7Entropy}. The final expression takes the form: 
\begin{equation}
\begin{split}
	S_\text{EE}(R;B^{1}\times \mathbb{H}_4,\epsilon)=\fft{\ell^4\text{vol}[\mathbb{H}_4]}{2G^{(7)}_N}L_\text{IR}e^{4\tilde G_0}\log\fft{R}{\Lambda}+\mathcal O((\fft{L_\text{IR}}{R})^0).\label{area:IR:37d}
\end{split}
\end{equation}
The constants $L_{\rm IR}$ and  $\tilde{G}_0$ can readily be read off from  equation (\ref{eq:IRsolfgl});  we see that $L_{\text{IR}} = \frac{1}{m} \left(\frac{3}{2}\right)^{1/5} $ and $e^{4\widetilde{G}_0} = \frac{2}{3 m^4}\left(\frac{2}{3}\right)^{1/5}$ hence:
\begin{equation}
\begin{split}
	S_\text{EE}(R;B^{1}\times \mathbb{H}_4,\epsilon)=\fft{\ell^4\text{vol}[\mathbb{H}_4]}{3 m^5G^{(7)}_N}\log\fft{R}{\Lambda}+\mathcal O((\fft{L_\text{IR}}{R})^0).\label{area:IR:47d}
\end{split}
\end{equation}
upon using $G_N^{(7)} = \frac{3 \pi^2}{16 N^3L_{\text{AdS}_7}^5}$ with  $L_{\text{AdS}_7} =\frac{2}{m}$, we have:
\begin{align}
   	S_\text{EE}(R;B^{1}\times \mathbb{H}_4,\epsilon)=\fft{ N^3\ell^4\text{vol}[\mathbb{H}_4]}{18 \pi^2}\log\fft{R}{\Lambda}+\mathcal O((\fft{L_\text{IR}}{R})^0). 
\end{align}
Note that the coefficient of $\log(R/\Lambda)$ is one third of the IR central charge, as expected. Indeed, the central charge  of the 2d CFT  arising as fixed point can be computed using the Brown-Henneaux formula \cite{Brown:1986nw}:
\be
c_{\text{IR}}=\frac{3\, R_{AdS_3}}{2 G^{(3)}_N},
\ee
where the 3d Newton's constant, $G^{(3)}_N$, is related to the eleven-dimensional one. Working in units where the radius of AdS$_7$ is one  and defining $e^f =e^{f_0}/z$ in the IR fixed points, leads to 
\bea
c_{\text{IR}}&=& \frac{8\, N^3}{\pi^2}e^{f_0+4g}\ell^4\text{vol}[\mathbb{H}_4] \nonumber \\
&=& \frac{N^3}{ 6 \pi^2}\ell^4\text{vol}[\mathbb{H}_4].
\eea
Thus, our result for the IR central charge from the EE perfectly agrees with the Brown-Henneaux prescription.

\subsubsection*{Monotonic $c$-function} \label{sec:EE:7d:mono}
\begin{figure}[t]
    \centering
    \includegraphics[width=0.65\textwidth]{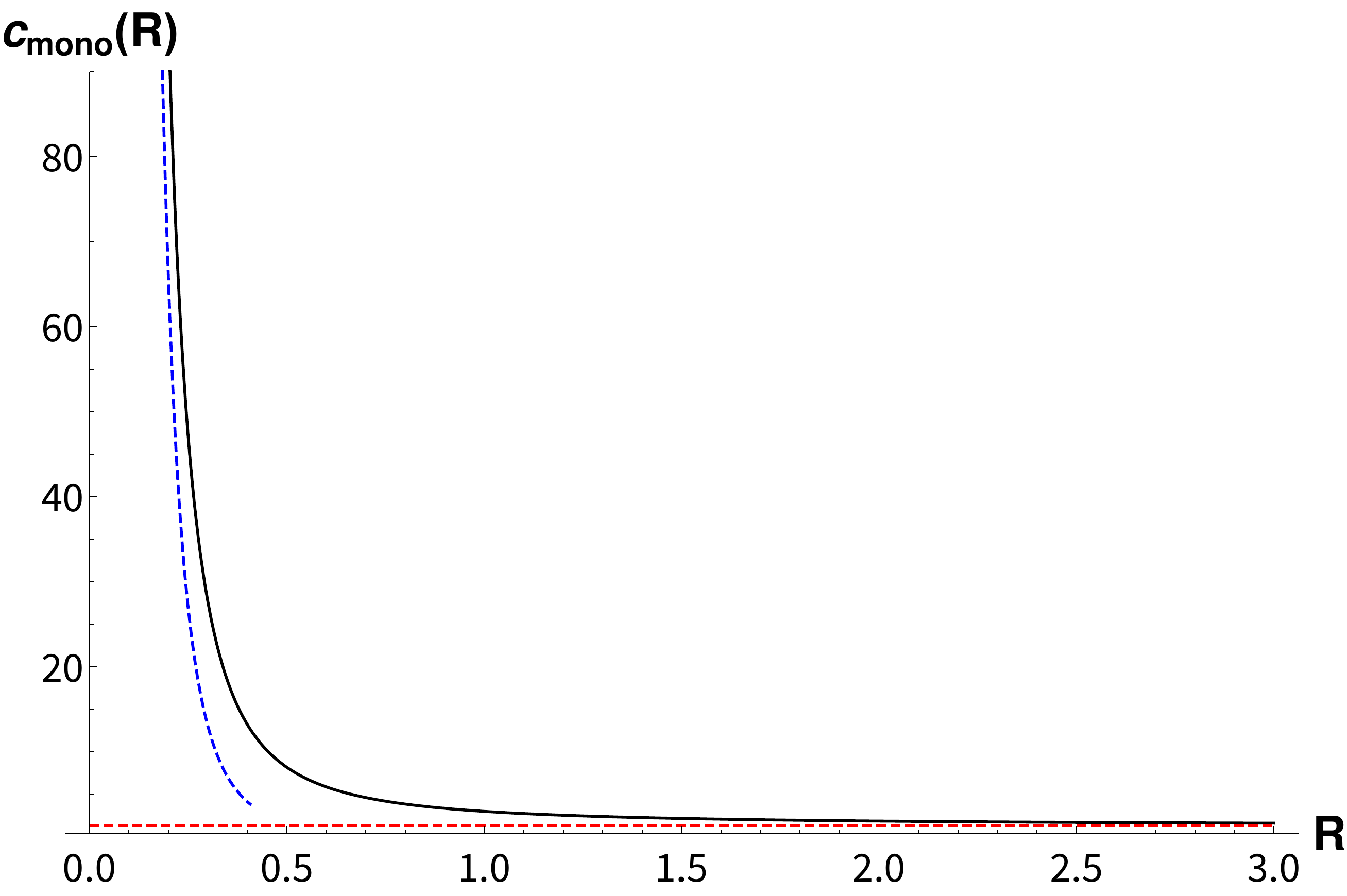}
    \caption{IR motivated $c$-function \eqref{cc:across:mono7d} for flows form AdS$_7$ to AdS$_3$ with $m=1$ ($4 G^{(7)}_N =1$ for presentation). With blue dashed line we have represented the $c$-function obtained form applying \eqref{cc:across:mono7d} to the leading $R^{-4}$ term in \eqref{eq:SEEexpansion7d} whereas with red dashed line we present the IR value of the central charge.}
    \label{fig:cmono7d}
\end{figure}

Analogous to the discussion in section \ref{sec:EX:5d}, we can define a holographic $c$-function across dimension for a flow from AdS$_7$ to AdS$_3$ in terms of the EE for a ``wrapping'' region $B^1 \times \mathbb{H_4}$ as (\ref{eq:SEE-Sq-dR}), namely
\begin{empheq}[box=\fbox]{equation}
	c_\text{mono}(R)\equiv R\,\partial_R S_\text{EE}(R;B^1\times \mathbb{H}_4,\epsilon).\label{cc:across:mono7d}
\end{empheq}
The explicit behavior of (\ref{cc:across:mono7d}) is presented in Fig.~\ref{fig:cmono7d}. We show that in the IR the function approaches the value of the 2d central charge. In the UV, we have an analytic understanding of the divergent terms for small radius, $R$, of the entangling region. The blue dashed curve demonstrates that the $1/R^4$ leading term in the UV expansion of the EE (\ref{eq:SEEexpansion7d}) captures the UV divergent behavior of (\ref{cc:across:mono7d}) precisely, which comes from infinitely many massive KK modes from the 2d field-theoretic point of view. 

\subsubsection*{Interpolating $c$-function}
In this section we aim to track the coefficient of the logarithmic term in the expression for entanglement entropy throughout the flow. Given the generic structures of UV and IR expansions of the EE, namely (\ref{eq:SEEexpansion7d}) and (\ref{area:IR:37d}), we construct an operator that, by acting on $S_\text{EE}(R;B^1 \times \mathbb{H}_4,\epsilon)$, extracts the coefficient of the $\log R$:
\begin{empheq}[box=\fbox]{equation}
c_{\text{int}}(R) \equiv \frac{1}{8}R\partial_R {\left(R\partial_R+2\right)}{\left(R\partial_R+4\right)}S_\text{EE}(R;B^1\times \mathbb{H}_4,\epsilon), \label{eq:Cfunct}
\end{empheq}
where the first operator extracts the constant coefficient of the logarithm and the second \& the third eliminate $1/R^{2}$ \& $1/R^{4}$ divergent contributions in the UV regime respectively.

\begin{figure}[t]
\centering
\includegraphics[width=0.65\textwidth]{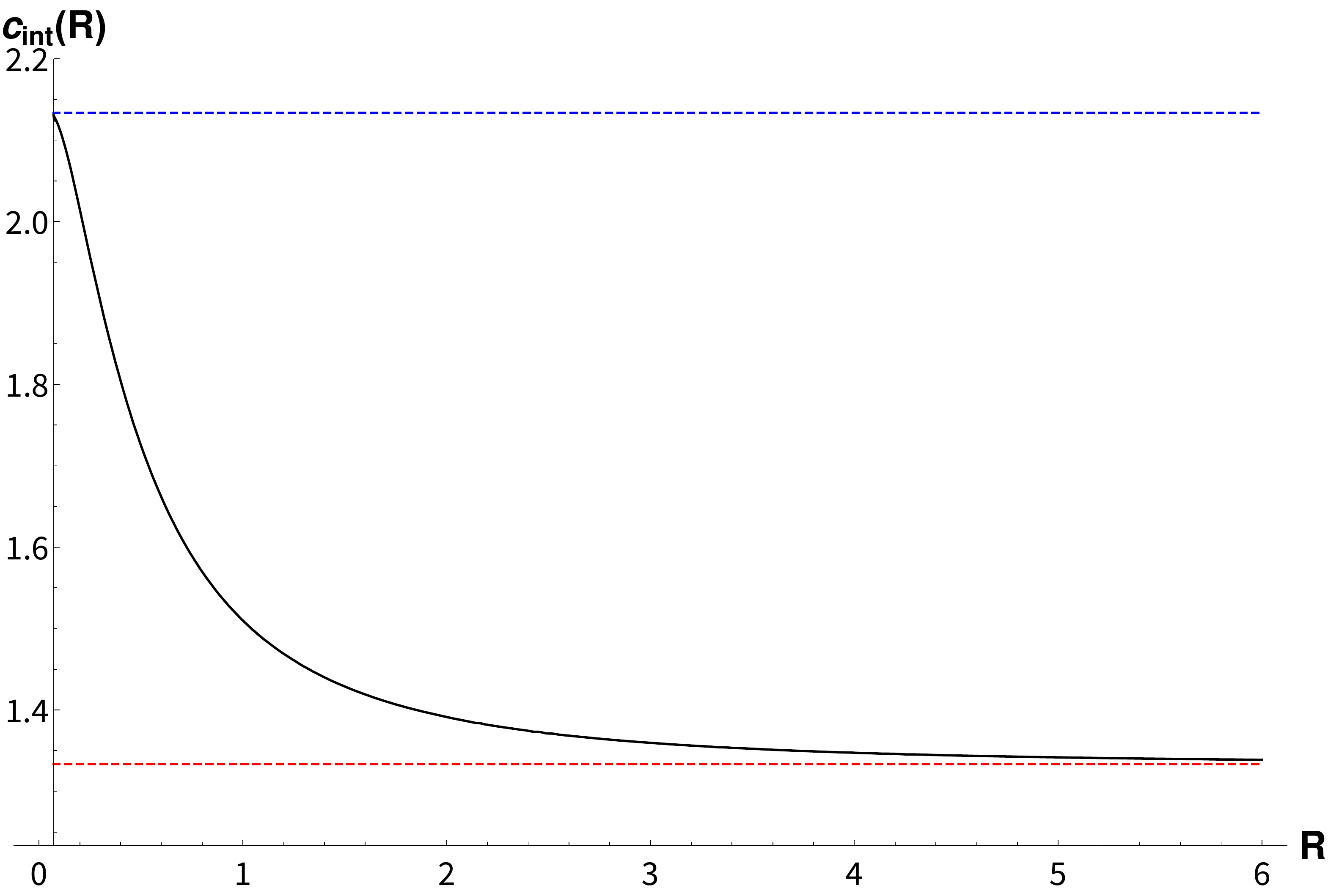}
\caption{In black continuous line we show the interpolating holographic $c$-function (\ref{eq:Cfunct}) for flows from AdS$_7$ to AdS$_3$ (set $4G^{(7)}_N=1$ for presentation). We represent the values of the UV coefficient of the logarithmic term with blue dashed line and the IR value with red dashed line.  We have used $m=1$ and set the ration $\ell/L_{\text{UV}}=1$ for convenience.}\label{cc:across:7to3}
\end{figure}

In Fig.~\ref{cc:across:7to3}, we plot this interpolating function (\ref{eq:Cfunct}) along the flow. 
Notice that, similar to the case of AdS$_5$ to AdS$_3$ flows presented in section \ref{sec:EX:5d}, the function $c_{\text{int}}(R)$ \eqref{eq:Cfunct} probes the coefficient of the logarithm in both the UV and the IR regions and smoothly interpolates between the two values. The coefficient of the logarithmic term in the UV will be proportional to a linear combination of the 6d central charges \cite{Safdi:2012sn, Miao:2015iba}.

\subsection{Flows from AdS$_7$ to AdS$_5$: partially compactified 6d $\mathcal N=(2,0)$ theories}\label{sec:EX:7-5d}
In this section we consider compactifications of the 6d $\mathcal N=(2,0)$ theory on a torus, $T^2$. The geometries of the holographic duals interpolate between AdS$_7$ solutions in the UV and AdS$_5$ solutions in the IR. Compared to the AdS$_3$ IR asymptotics encountered in the previous examples this brings about new aspects. For example, the definition of the central charge which is natural from the perspective of the lower-dimensional theory differs from (\ref{c:CT}), which was taylored to flows to 2d CFTs in the IR. The generalization to 4d CFTs in the IR comes with technical issues regarding the regularization of the EE, which appear already in flows within the same dimension \cite{Casini:2017vbe} and carry additional subtleties for flows between dimensions, as we will discuss.

The holographic duals can be described in 7d gauged supergravity, with a metric of the form 
\be
    ds^2 =e^{2f(z)}\left(- dt^2 + dz^2+d r^2+r^2 d\Omega_2^2\right)+e^{2g(z)}ds^2_{ T^2}~.\label{metric:7to5}
\ee
This corresponds to (\ref{metric}) with $D=6$ and $d=4$. 
The solutions for torus compactifications were constructed in \cite{Bah:2012dg}, extending earlier work in \cite{Maldacena:2000mw}, and we will follow the presentation in \cite{Uhlemann:2021itz}.\footnote{In \cite{Bah:2012dg,Uhlemann:2021itz} the metric was written as $ds^2 =e^{2f(z)}ds^2_{\mathbb{R}^{1,3}}+e^{2h(z)} dz^2 +e^{2g(z)}ds_{T^2}^2$. The function $h$ defines a gauge, or holographic radial coordinate. We will use $h=f$ with radial coordinate $z$ and $h=0$ with radial coordinate $\rho$, related to $z$ by (\ref{z:to:rho}). Also note that we introduce an explicit length scale $\ell$ for a torus in (\ref{metric:7to5}).}
The backgrounds also include two scalars $\lambda_i$ and two Abelian gauge fields $A_\mu^{(i)}$ whose field strengths along $T^2$ are denoted by $F^{(i)}$.
The BPS equations are \cite{Bah:2012dg} 
\bea
f'+\lambda_1'+\lambda_2' +e^{h-4\lambda_1-4\lambda_2}&=&0\,,\nonumber \\
g' -4\lambda_1'-4\lambda_2' + 2e^{h+2\lambda_1}+2e^{h+2\lambda_2} -3e^{h-4\lambda_1-4\lambda_2}&=&0\,,\nonumber \\
3\lambda_1'+ 2\lambda_2' -2 e^{h+2 \lambda_1} 
+ 2e^{ h - 4 \lambda_1-4\lambda_2 }- e^{h-2g -2\lambda_1}F^{(1)} &=&0\,,\nonumber \\
2\lambda_1'+ 3\lambda_2'-2e^{ h+ 2 \lambda_2} + 2e^{h-4 \lambda_1 - 4\lambda_2} -e^{h-2g -2\lambda_2} F^{(2)} &=&0\,.
\end{eqnarray}
The AdS$_5\times T^2$ IR fixed point solution is given by 
\begin{equation}\label{eq:BPS-AdS7-5}
	\begin{alignedat}{2}
		e^{f(z)}&=\frac{3^{2/5}}{2^{6/5}}\,\,\frac{1}{z},&\qquad e^{2g(z)}&=\frac{3^{3/10}}{2^{2/5}}\frac{|n|}{8},\\
		\lambda_1&=\frac{1}{10}\ln\left(\frac{33}{4}-\frac{19\sqrt{13}\, n}{4|n|}\right),&\qquad \lambda_2&= \frac{1}{10}\ln\left(\frac{33}{4}+\frac{19\sqrt{13}\, n}{4|n|}\right),\\
		F^{(1)}&=\frac{n}{8},&\qquad F^{(2)}&=-\frac{n}{8},\\
	\end{alignedat}
\end{equation}
where  $n\in \mathbb{Z}$. We have a one-parameter family of IR fixed point solutions characterized by the integer $n$. However, in the BPS equations the dependence on $n$ can simply be absorbed into a shift of $g$, by setting $g=\tilde g+\frac{1}{2}\ln(n/8)$.  In Appendix \ref{App:7to5} we construct numerical flows which connect an AdS$_7$ region in the UV with the above IR AdS$_5\times T^2$ fixed points.

\subsubsection{Local holographic $c$-function}
\begin{figure}[t]
\centering
\includegraphics[width=0.5\textwidth]{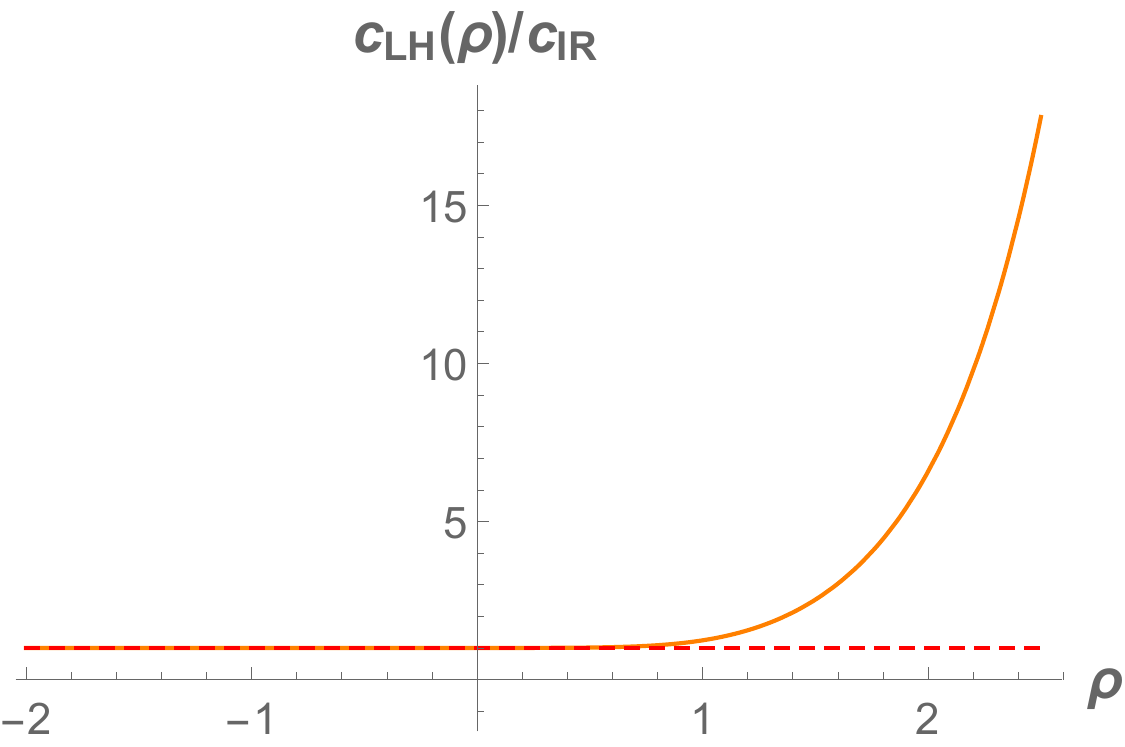}
\caption{Local Holographic  $c$-function (\ref{eq:c_LH--7-5}) divided by its corresponding IR asymptotic value, $c_\text{IR}$, for flows from AdS$_7$ to AdS$_5$. In the IR the curve approaches one, shown as dashed red line. }\label{c_LH::NEC7-5d}
\end{figure}

In this subsection we present the LH $c$-function (\ref{eq:cfunction?}) which reduces, in flows from AdS$_7$ to AdS$_5$ (\ref{metric:7to5}), to
\be
\label{eq:c_LH--7-5}
c_{\textrm{LH}}(z)=\frac{1}{\Big[\big(e^{-f(z)-\frac{2}{3}g(z)}\big)'\Big]^3}.
\ee
In Fig. \ref{c_LH::NEC7-5d} we have plotted this function with respect to the holographic coordinate $\rho$ defined in terms of $z$ through (\ref{z:to:rho}). The LH $c$-function behaves analogously to those defined in previous flows: in the IR it approaches a constant which is set by the central charge of the IR fixed point CFT corresponding to the solution (\ref{eq:BPS-AdS7-5}), $c_{\rm IR}$, and in the UV it diverges.

\subsubsection{EE $c$-functions}\label{sec:EE:7-5:UVIR}
We now turn to entanglement entropies and the functional (\ref{EE:area}). For $D=6,d=4$ it takes the form 
\begin{equation}
    S_\text{EE}(R;B^{3}{\times}T^2,\epsilon)
    =\fft{\text{vol}[S^2]\text{vol}[T^2]}{4G^{(7)}_N}\min_{r(z=\epsilon)=R}\bigg[\int_\epsilon^{z_0}dz\,r^{2}\, e^{3f(z)+2g(z)}\sqrt{1+r'(z)^2}\bigg]\,,\label{eqn:c75}
\end{equation}
%
As discussed above, the dependence on the parameter $n$ specifying the fluxes and labeling the one-parameter family of IR fixed points can be absorbed into a shift of $g$ by a constant. Such shifts only change the EE by an overall factor.

The EE is divergent for $\epsilon\to 0$, and we note that already for flows within the same dimension in 4d, \cite{Casini:2017vbe} considered EE's regularized by subtracting the EE associated with the UV fixed point. We will follow a similar approach here.
To discuss the structure of the divergences, we focus on a simple special case: the solution obtained by taking AdS$_7$ in Poincar\'e coordinates with identifications imposed on two of the field theory directions to obtain an $\RR^{1,4}\times T^2$ slicing. This corresponds to $f_{UV}(z)=g_{UV}(z)=-\log(z/L)$. 
This locally AdS$_7$ solution does not end in an AdS$_5$ fixed point in the IR; we only use it to discuss the UV structure of the EE in the twisted compactifications described above.
For the EE in the locally AdS$_7$ solution we obtain
\begin{equation}\label{SUV-AdS7}
	S_\text{UV}(R;B^{3}{\times}T^2,\epsilon)
	=\fft{L^5\text{vol}[S^2]}{4G^{(7)}_N}\fft{\text{vol}[T^2]}{R^2}
	\left(\frac{R^4}{\epsilon^4} -\frac{3R^2}{16 \epsilon^2} -\frac{1}{128}\log \left(\frac{R}{\epsilon}\right)+0.0136377 +{\cal O}(\left(\frac{\epsilon}{R}\right)^2) \right).
\end{equation}
The divergent terms were derived in closed form. The $\mathcal O(1)$ constant was obtained numerically by solving the extremality equation for the minimal surfaces and subsequently fitting the resulting areas. All given figures in the numerical constant are significant (for the discussions below we used a fit up to $\mathcal O(10^{-32})$).
Let us point out some salient features. For a maximally spherical entangling region in AdS$_7$, the leading divergence would be $\mathcal O(R^4/\epsilon^4)$, where $R$ is the radius of the entangling region (see e.g.\ eq.~(\ref{Eq:EE-General})).  The leading divergence in the result (\ref{SUV-AdS7}) for surfaces wrapping $T^2$, on the other hand, is of the form $R^2 \ell^2 /\epsilon^4$, where $\ell$ is the size of $T^2$.
Similar modifications apply for the subleading divergences and finite terms -- they are all proportional to the volume of $T^2$, with the remaining powers of $R$ determined by dimensional analysis.

\begin{figure}[t]
\centering
\includegraphics[width=0.5\textwidth]{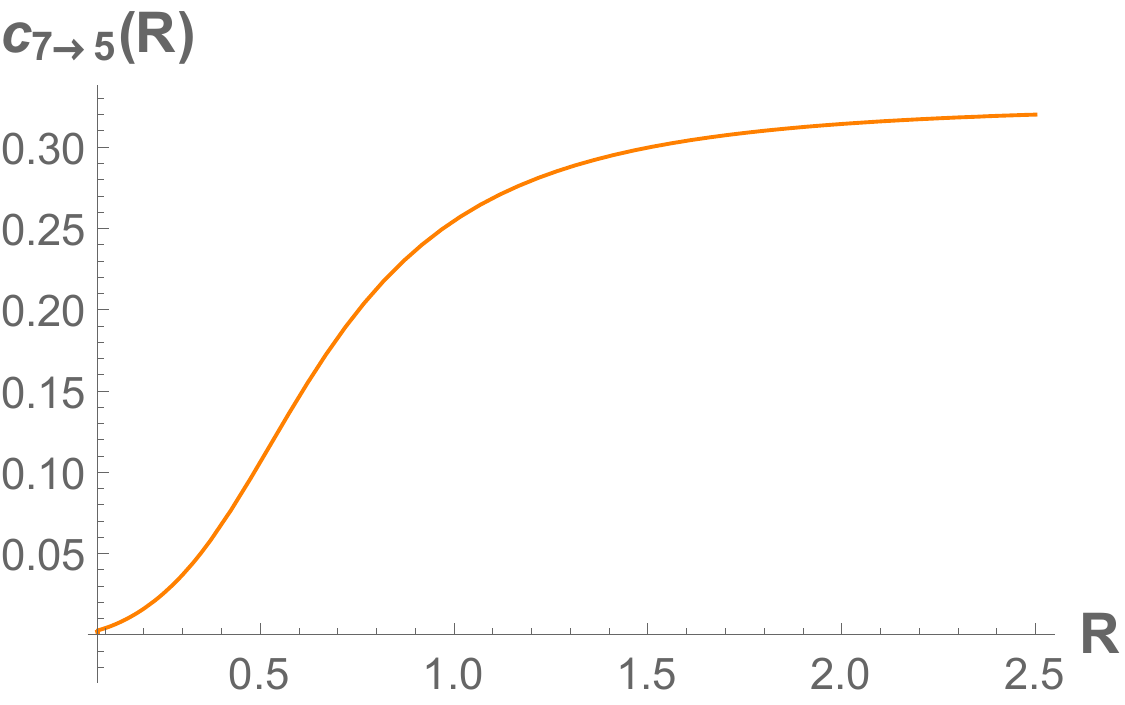}
\caption{ The $c$-function (\ref{eqn:c7-5}) for flows from AdS$_7$ to AdS$_5$. In the IR (large $R$) the curve approaches a constant value.}\label{c7-5d}
\end{figure}

As regularized EE we then consider the difference between the EE in the twisted compactifications following from the BPS conditions (\ref{eq:BPS-AdS7-5}) with non-zero $n$ on the one hand, and the UV result (\ref{SUV-AdS7}) on the other,
\begin{equation}
 \Delta S_{EE}(R;;B^{3}{\times}T^2)=  S_\text{EE}(R;B^{3}{\times}T^2,\epsilon)-S_\text{UV}(R;B^{3}{\times}T^2,\epsilon)\,. \label{Eq:7-5-reg}
\end{equation}
From (\ref{Eq:7-5-reg}) we construct the following candidate $c$-function:
\begin{equation}
	c_{7\to 5}(R)\equiv R\, \partial_R \left(R\,\partial_R -2\right)\Delta S_\text{EE}(R;B^3\times T^2).\label{eqn:c7-5}
\end{equation}
The differential operator on the right hand side is the natural generalization of that in (\ref{c:CT}) to capture the coefficient of the logarithmic term of 4d IR fixed points, along the lines of \cite{Liu:2012eea}.
This function is plotted in Fig. \ref{c7-5d}. The fact that this candidate $c$-function vanishes for small $R$ is consistent with the claim that the UV behavior is captured by (\ref{SUV-AdS7}). We note that, in the sense that the subtracted EE $\Delta S_{\rm EE}$ vanishes in the UV by construction, this candidate $c$-function is different from the interpolating $c$-functions discussed in the previous examples.

We close with general remarks on analogs of the monotonic $c$-function $c_\text{mono}$. 
Attempts to construct such a function would naturally start from the combinations (\ref{c:CTT}) motivated by strong subadditivity with $D=4$. The idea would be to regard the 6d theory on $T^2$ as a 4d theory comprising infinite towers of KK modes, in analogy to the perspective we have previously taken for flows to 2d CFTs in the IR. The flow can be consistently interpreted from the 4d perspective, with more KK modes becoming relevant as the energy scale is increased. The hope would be that this leads to a (generalized) central charge which is finite in the IR, monotonic along the flow and divergent in the UV. 
For flows to 4d, however, this IR-motivated perspective becomes subtle. 
A crucial part in the arguments of \cite{Casini:2017vbe} exploiting the implications of strong subadditivity for flows within 4d was a proper treatment of the UV divergences -- they were canceled by subtracting the EE of the UV fixed point theory.
Implementing a similar regularization by subtracting a UV EE for flows across dimensions creates tension with a purely lower-dimensional point of view, since the flow does not have a lower-dimensional UV fixed point. One could subtract the EE associated with the AdS$_5\times T^2$ IR fixed point instead. This would capture the UV strucutre in the EE of the 4d IR theory, but would fail to capture the 6d divergence structure in (\ref{SUV-AdS7}). The definition we used in (\ref{Eq:7-5-reg}), on the other hand, subtracts the full divergence structure of the 6d theory, but the subtracted EE (\ref{SUV-AdS7}) is not scale invariant in the same sense as spherical EE's in CFTs usually are -- 
for a spherical surface the EE is invariant under simultaneous rescalings of the size $R$ and the cut-off $\epsilon$, while for (\ref{SUV-AdS7}) we have $S_{\rm EE,wrap}\rightarrow \alpha^{-2}S_{\rm EE,wrap}$ for $(R,\epsilon)\rightarrow\alpha(R,\epsilon)$. We will leave a more detailed exploration of these issues for the future.

\section{Summary and Conclusions}\label{Sec:SumCon}

In this manuscript we explored the notion of $c$-function in RG flows across dimensions. One might have hoped for functions which change monotonically along the flow and coincide with the natural central charges at the fixed points. For flows across dimensions, however, we found that we had to settle for functions which satisfy one of these two criteria but not both.

Our first exploration focused on the implications of the NEC on holographic RG flows across dimensions satisfying the Einstein's equations. We were able to identify a monotonic combination of metric functions, thus establishing a holographic $c$-theorem across dimensions. This combination agrees with the IR central charge and diverges in the region of the solution corresponding to the UV. In Figs. \ref{cc:across:NEC}, \ref{cc:across:NEC7d} and \ref{c_LH::NEC7-5d}, we presented plots of such functions for holographic RG flows from AdS$_5$ to AdS$_3$, from AdS$_7$ to AdS$_3$ and from AdS$_7$ to AdS$_5$, respectively. The plotted functions are monotonic as expected from the implications of the NEC studied in subsection \ref{sec:NEC:across}. We interpret the divergent behavior in the UV regions as natural from the lower-dimensional perspective. Roughly speaking, in the IR the theory is lower dimensional because  almost all KK modes are massive. As we depart from the IR, the infinite towers of  massive KK modes begin to contribute and render this candidate $c$-function divergent.  One aspect that we learn from our explorations, and that can be qualitatively seen by comparing Figs. \ref{cc:across:NEC}, \ref{cc:across:NEC7d} and \ref{c_LH::NEC7-5d}, is the role of the dimensionality of the compatification manifold which controls the rate of divergence of this candidate $c$-function. 

Our second approach to exploring potential $c$-functions is based on entanglement. It is motivated by the understanding of universal irreversable functions obtained from the EE, e.g.\ in \cite{Casini:2017roe,Casini:2017vbe}. We focused on extending these studies of EE's associated with spherical regions to flows across dimensions. 

We qualitatively discussed entangling surfaces which generalize spherical surfaces in flat space to compactifications on possibly curved manifolds, in the sense that they enclose all points with geodesic distance less than a given upper bound to a reference point. We argued that these surfaces detect the topology of the compact space through corner contributions. At the technical level, the extremality condition for these surfaces leads to PDEs, whose explicit solution we hope to address elsewhere.
For the explicit computations we focused on entangling regions which take a simple product form, with one factor being the entire compact part of the field theory geometry and the other a ball on a constant-time slice of the non-compact part of the field theory geometry. As illustrated in Fig.~\ref{fig:wrapping region} for a toy model, even in the $R\to 0$ limit these  entropies capture some physics down to the compactification scale, since the regions include the entire compact manifold.
The entropic candidates for $c$-functions that we explored are:
\begin{enumerate}
    \item {\bf IR-motivated monotonic $c$-function:} In two  of the types of flows that we explored the lower-dimensional theory was a 2d CFT where the EE follows the expression in (\ref{Eq:EE-General}). Motivated by the analysis leading to equation (\ref{c:CTT}), we find it natural to consider 
    \begin{equation}
        c_{\rm mono}(R) =  R\,\partial_R\,  S_\text{EE}(R).
    \end{equation}
    We plotted this function in Fig.~\ref{cc:across:5to3:mono} for the AdS$_5$ to AdS$_3$ flow and in Fig.~\ref{fig:cmono7d} for the AdS$_7$ to AdS$_3$ flow. We treated the IR and UV asymptotics analytically.  In the IR ($R\to\infty$) this function precisely agrees with the central charge of the dual two-dimensional theory. In the UV ($R\to 0$) this function blows up as $1/R^2$ for flows from AdS$_5$ to AdS$_3$ and as a combination of $1/R^4$ and $1/R^2$ for flows from AdS$_7$ to AdS$_3$. We analytically computed the coefficients and showed agreement with the numerical plots in Fig.~\ref{cc:across:5to3:mono} and Fig.~\ref{fig:cmono7d} respectively. 
    
    \item {\bf Interpolating $c$-function:} For the entangling region wrapping the compact space, the dependence of the entropy on the radius acquires cut-off independent terms proportional inverse powers of $R$ (see, for example, equations  
    (\ref{area:UV:4}), (\ref{eq:SEEexpansion7d}) and (\ref{SUV-AdS7})). We can construct an interpolating candidate $c$-function by tracking the coefficient of the logarithmic term. The combinations of derivatives that isolate such terms are:
    \begin{equation}
    \begin{split}
        c^{4d\rightarrow 2d}_{\rm int}(R) & \equiv \fft{1}{2} R\partial_R{\left(R\partial_R+2\right)}S_\text{EE}^{4d}(R), \\
        c^{6d\rightarrow 2d}_{\rm int}(R) &\equiv \frac{1}{8}R\partial_R {\left(R\partial_R+2\right)}{\left(R\partial_R+4\right)} S_\text{EE}^{6d}(R).
    \end{split}\label{c:4d6d}
    \end{equation}
    We note that in 4d this interpolating $c$-function captures the combination of $a$ and $c$ type central charges given in (\ref{EE:log:2}) in the UV. The hopes for monotonicity are tenuous: we presented a number of cases where this interpolating  $c$-function not only fails to be monotonic but its value at the IR fixed point is also greater than its value at the UV fixed point. \\
    The interpolating $c$-functions (\ref{c:4d6d}) are constructed in terms of differential operators,  reminiscent of the work of \cite{Liu:2012eea}. A difference here is that (\ref{c:4d6d}) are constructed for RG flows from 4d/6d CFTs to 2d CFTs and therefore the differential operators are supposed to eliminate UV-divergent terms of the form $\sim\fft{1}{R^2},\fft{1}{R^4}$ for the resulting $c$-function to capture the $\log R$ coefficients at the fixed points, while only positive powers of $R$ appear in the analysis of \cite{Liu:2012eea} for interpolating $c$-functions in flows within the same dimension. 
\end{enumerate}

There are a number of interesting problems following from our explorations that we find worth highlighting. In this manuscript we have studied both generic flows and supersymmetric ones but did not explore the precise role of supersymmetry in the behavior of partial $c$-functions carefully. It would be interesting to understand if there are fundamental differences between generic flows and supersymmetric ones elaborating on the supersymmetric attractor approach taken in \cite{Bhattacharyya:2014oha,Bhattacharyya:2014gsa,Amariti:2016mnz}. For example, the treatment via the NEC applies to generic flows but it is likely that supersymmetric flows satisfy a more stringent condition that we have not yet exploited. It is also possible to envision that the stability properties of supersymmetric backgrounds translate ultimately into properties of the EE through the metric functions. 

The two main examples in this manuscript explored flows from AdS$_5$ to AdS$_3$ and AdS$_7$ to AdS$_3$, although we also briefly discussed flows form AdS$_7$ to AdS$_5$. It would be interesting to consider flows between AdS spacetimes of different parity. For example, flows between odd-dimensional and even-dimensional AdS will present challenges for our interpolating $c$-functions as one would need to track universality between a logarithmic and  a constant terms. 

More ambitiously would be to attempt a direct field-theoretic approach to flows across dimensions. Most of the approaches in field theory have Lorentz-invariance embedded as one of the axioms but we hope to have provided evidence that relaxing Lorentz-invariance might lead to interesting possibilities.

\section*{Acknowledgments}
We are thankful to Antonio Amariti, Ibou Bah, Francesco Benini, Nikolay Bobev, Evan Deddo, Lorenzo Di Pietro, Sebastian Grieninger, Max Jerdee, Shiraz Minwalla and particularly to  Carlos N\'u\~nez. JH is supported in part by an Odysseus grant G0F9516N from the FWO, by the KU Leuven C1 grant ZKD1118 C16/16/005, and by the Research Programme of the Research Foundation grant G.0926.17N from the FWO. JTL, LPZ and CFU are supported in part by the U.S. Department of Energy
under grant DE-SC0007859. AGL is supported by an appointment to the JRG Program at the APCTP through the Science and Technology Promotion Fund and Lottery Fund of the Korean Government and from the National Research Foundation of Korea(NRF) grant funded by the Korea government(MSIT) (No. 2021R1F1A1048531)
\appendix


\section{Hologrpahic RG flow solutions}
In this Appendix we provide concrete plots of RG flow solutions used in the main part, for flows from AdS$_5$ to AdS$_3$, from AdS$_7$ to AdS$_3$ and from AdS$_7$ to AdS$_5$.

\subsection{5d $\mathcal N=2$ gauged STU model}\label{App:STU}
In this Appendix we provide the details of the supergravity theory that provides holographic flows discussed in section  \ref{sec:EX:5d}. The bosonic action of 5d $\mathcal N=2$ gauged supergravity known as the STU model is given as
\begin{equation}
\begin{split}
	S&=\fft{1}{16\pi G_{(5)}}\int d^5x\sqrt{|g|}\bigg[R+4\boldsymbol{g}^2\sum_{I=1}^3\fft{1}{X^I}-\fft12\sum_{x=1}^2\partial_\mu\phi^x\partial^\mu\phi^x-\fft14\sum_{I=1}^3(X^I)^{-2}F^I_{\mu\nu}F^I{}^{\mu\nu}\\
	&\kern10em+\fft{1}{4}\varepsilon^{\mu\nu\rho\sigma\lambda}F^1_{\mu\nu}F^2_{\rho\sigma}A^3_\lambda\bigg],
\end{split}\label{N=2:sugra:action:5d}
\end{equation}
where $x \in\{1,2\}$  and $I,J,K\in\{1,2,3\}$. Our convention for the Levi-Civita symbol is 
\begin{equation}
\varepsilon^{\mu\nu\rho\sigma\lambda}=\begin{cases}
	-|g|^{-1/2} & (\text{even permutation})\\
	+|g|^{-1/2} & (\text{odd permutation})
\end{cases}.
\end{equation}
The physical scalars $\phi^x$  parametrize the sections
\begin{equation}
    X^I=e^{\sum_{x=1}^2c^I{}_x\phi^x},
\end{equation}
with the constraint $\sum_{I=1}^3c^I{}_x=0$. Typical values of $c^I{}_x$ are given as
\begin{equation}
    X^1=e^{-\fft{1}{\sqrt6}\phi^1-\fft{1}{\sqrt2}\phi^2},\qquad X^2=e^{-\fft{1}{\sqrt6}\phi^1+\fft{1}{\sqrt2}\phi^2},\qquad X^3=e^{\fft{2}{\sqrt6}\phi^1}.
\end{equation}
From the bosonic action (\ref{N=2:sugra:action:5d}), the Einstein equations are given as
\begin{equation}
\begin{split}
	R_{\mu\nu}-\fft12g_{\mu\nu}(R+4\boldsymbol{g}^2\sum_{I=1}^3(X^I)^{-1})&=\fft12\sum_{x=1}^2\partial_\mu\phi^x\partial_\nu\phi^x-\fft14g_{\mu\nu}\sum_{x=1}^2\partial_\rho\phi^x\partial^\rho\phi^x\\
	&\quad+\fft12\sum_{I=1}^3(X^I)^{-2}F^I_{\mu\rho}F^I{}_\nu{}^\rho-\fft18g_{\mu\nu}\sum_{I=1}^3(X^I)^{-2}F^I_{\rho\sigma}F^I{}^{\rho\sigma}.
\end{split}\label{N=2:sugra:Einstein}
\end{equation}
The scalar equations of motion are given as
\begin{equation}
\begin{split}
	0&=\nabla_\mu\nabla^\mu\phi^x-\fft14\sum_{I=1}^3\partial_{\phi^x}(X^I)^{-2}F^I_{\mu\nu}F^I{}^{\mu\nu}+4\boldsymbol{g}^2\sum_{I=1}^3\partial_{\phi^x}(X^I)^{-1}.\label{N=2:sugra:scalar}
\end{split}
\end{equation}
The Bianchi identity and the vector equations of motion are given as
\begin{equation}
    0=\partial_{[\rho} F^I_{\lambda\sigma]},\label{N=2:sugra:Bianchi}
\end{equation}
and ($\epsilon_{IJK}=1$ for even permutations)
\begin{equation}
\begin{split}
	0&=\nabla_\mu((X^I)^{-2}F^I{}^{\mu\nu})+\fft14\sqrt{|g|}|\epsilon_{IJK}|\varepsilon^{\mu\lambda\rho\sigma\nu}F^J_{\mu\lambda}F^K_{\rho\sigma},
\end{split}\label{N=2:sugra:vector}%
\end{equation}
respectively. Finally, the BPS equations are given as
\begin{subequations}
\begin{align}
	0&=\left[\partial_\mu+\fft14\omega^{ab}_\mu\gamma_{ab}+\fft{i}{24}(\gamma_\mu{}^{\nu\rho}-4\delta_\mu^\nu\gamma^\rho)\sum_{I=1}^3(X^I)^{-1}F^I_{\nu\rho}+\fft{\boldsymbol{g}}{6}\sum_{I=1}^3X^I\gamma_\mu-\fft{i\boldsymbol{g}}{2}\sum_{I=1}^3A^I_\mu\right]\epsilon,\\
	0&=\left[-\fft{i}{4}\partial_\mu\phi^x\gamma^\mu+\fft18\sum_{I=1}^3(\partial_{\phi^x}(X^I)^{-1})F^I_{\mu\nu}\gamma^{\mu\nu}+\fft{i\boldsymbol{g}}{2}\sum_{I=1}^3\partial_{\phi^x}X^I\right]\epsilon.
\end{align}\label{N=2:sugra:BPS}%
\end{subequations}
The Ansatz (\ref{ansatz:BS}) satisfies the vector equations of motion (\ref{N=2:sugra:vector}) and the Bianchi identity (\ref{N=2:sugra:Bianchi}) automatically. The Einstein equations (\ref{N=2:sugra:Einstein}) and the scalar equations of motion (\ref{N=2:sugra:scalar}) are reduced for the Ansatz (\ref{ansatz:BS}) as
\begin{subequations}
\begin{align}
	0&=2f'{}^2+8f'g'-4g'{}^2-2f''-4g''-(\phi^1{}')^2-(\phi^2{}')^2,\\
	0&=-\fft{\kappa}{\ell^2} e^{-2g}+e^{-2f}(2f'{}^2+5f'g'+2g'{}^2+2f''+g'')-4\boldsymbol{g}^2\sum_I\fft{1}{X^I},\\
	0&=\fft{\kappa}{\ell^2}e^{-2g}+e^{-2f}(f'{}^2+f'g'-2g'{}^2+f''-g'')-\fft{e^{-4g}}{2\ell^4}\sum_I\fft{(p^I)^2}{(X^I)^2},
\end{align}\label{BS:Einstein}%
\end{subequations}
and
\begin{equation}
    0=e^{-2f}(\phi^x{}''+(f+2g)'\phi^x{}')-4\boldsymbol{g}^2\sum_I\fft{c^I{}_x}{X^I}+\fft{e^{-4g}}{\ell^4}\sum_I\fft{c^I{}_x(p^I)^2}{(X^I)^2}\label{BS:scalar}
\end{equation}
respectively. Although we have focused on BPS solutions in the main text we have preliminarily explored potential differences between supersymmetric flows and flows that only satisfy the equations of motion. At this early exploratory stage we did not find qualitative differences worth reporting.

\subsubsection*{Numerical black string solutions}
\begin{figure}[tp]
\centering
    \includegraphics[width=0.46\textwidth]{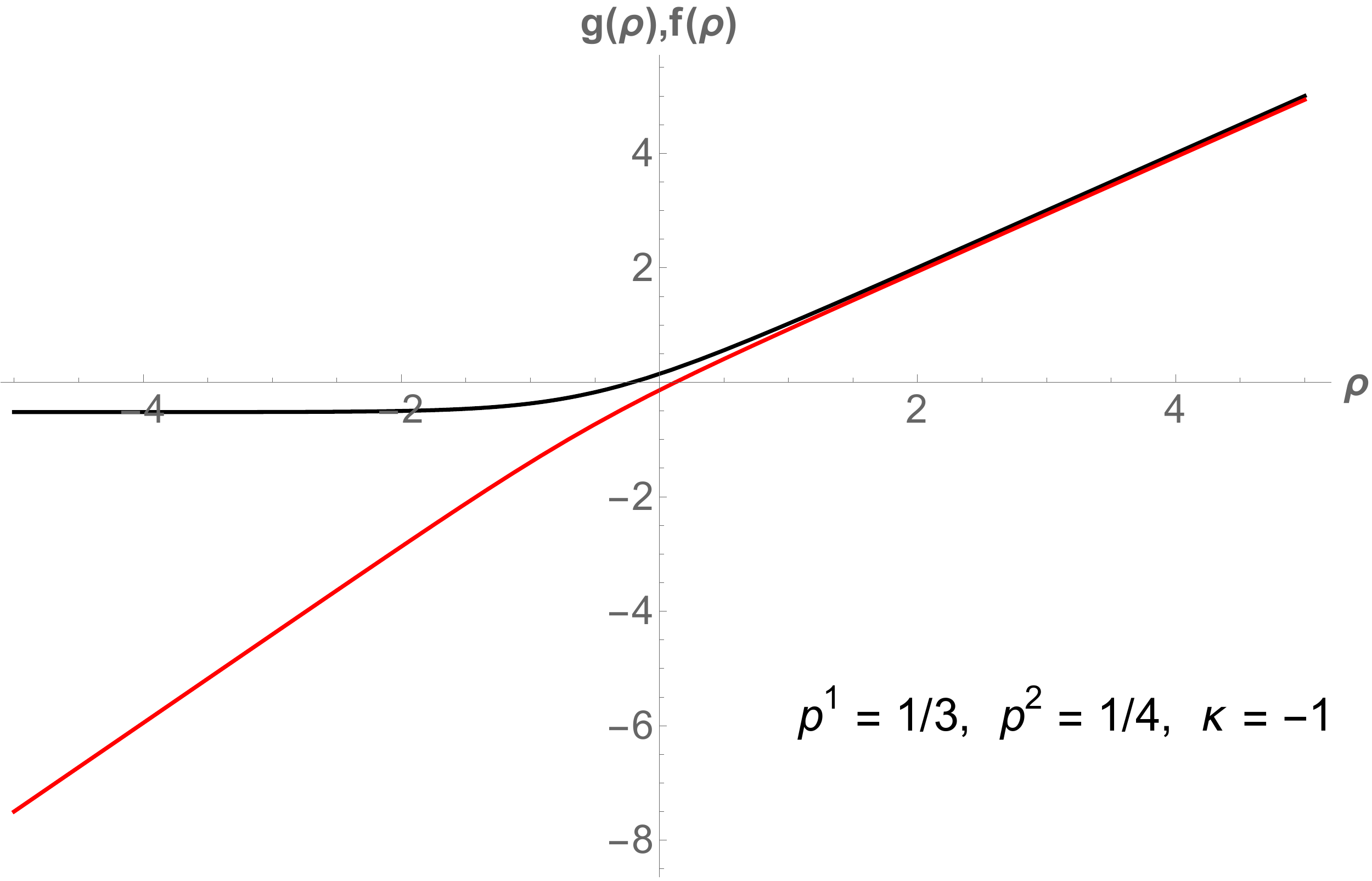}~~
    \includegraphics[width=0.46\textwidth]{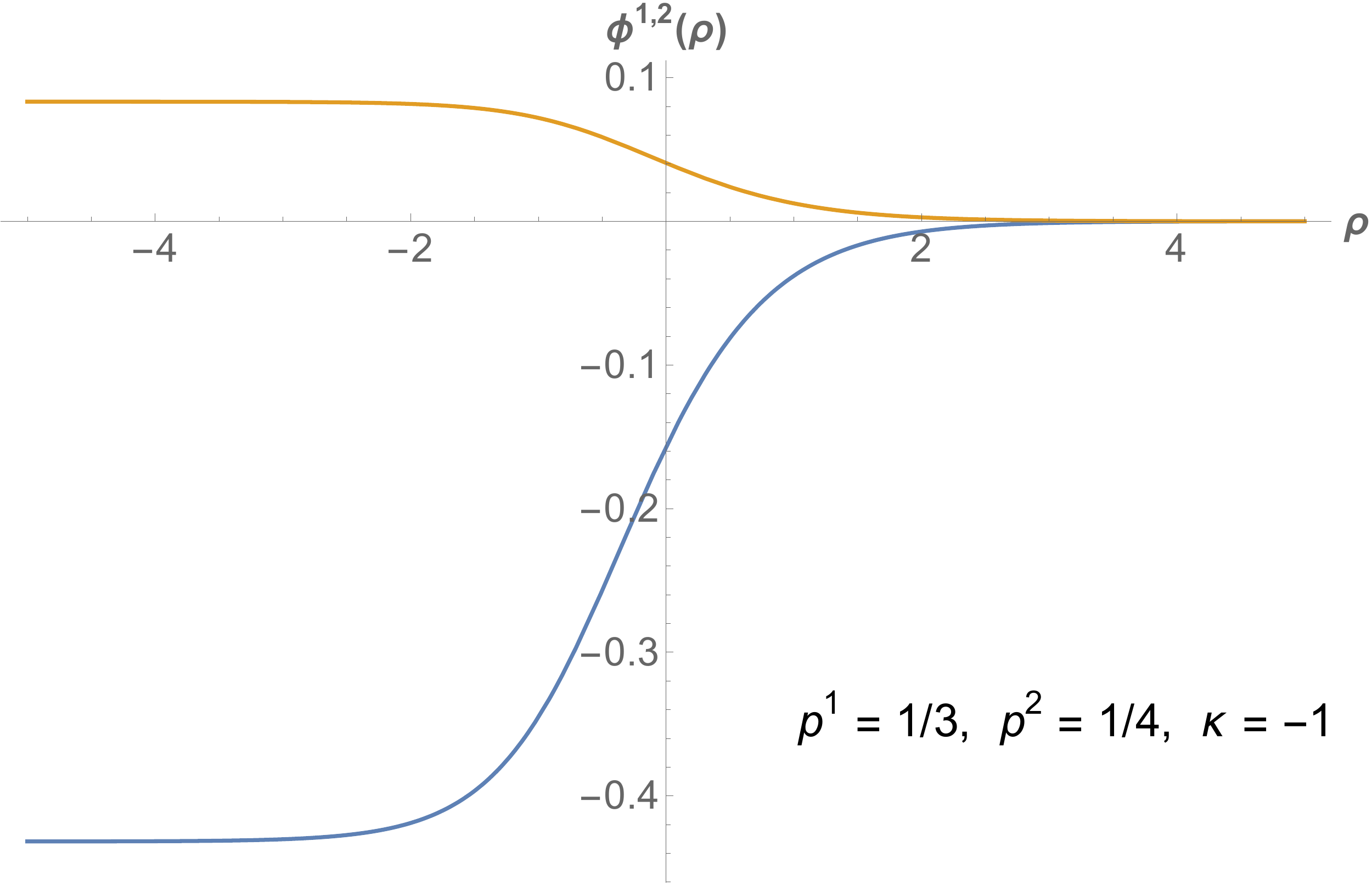}\\
    \includegraphics[width=0.46\textwidth]{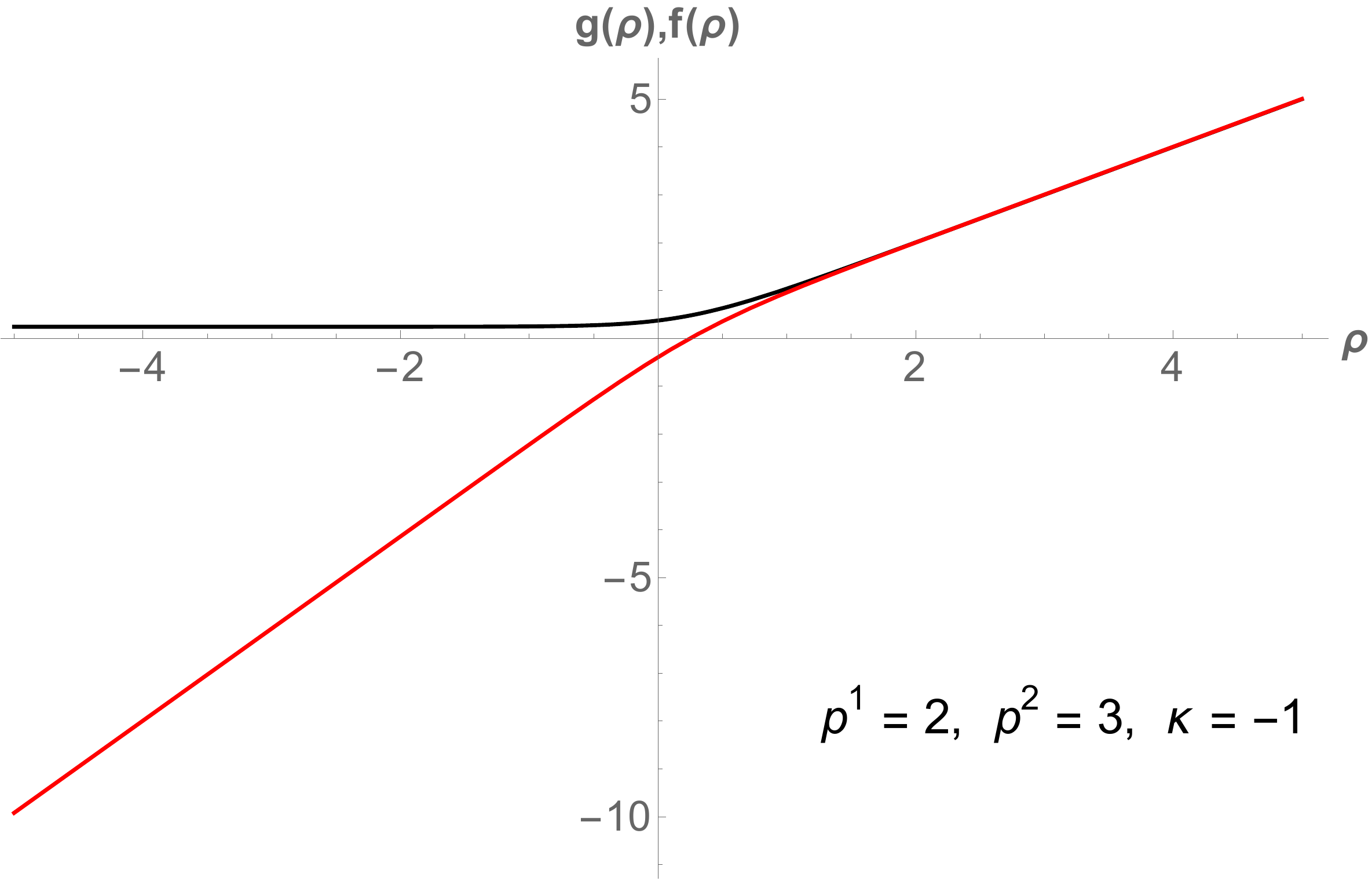}~~
    \includegraphics[width=0.46\textwidth]{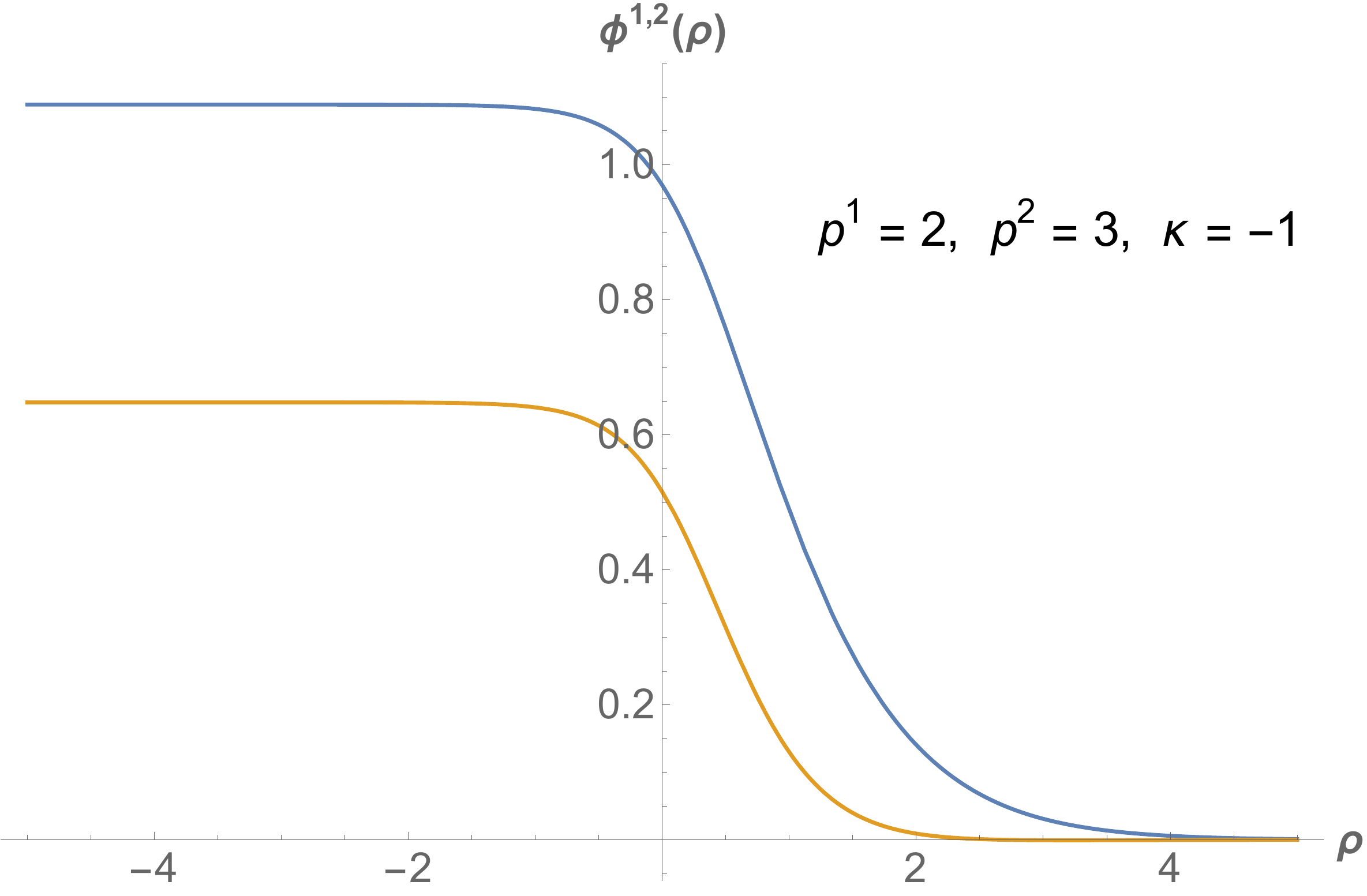}\\
    \includegraphics[width=0.46\textwidth]{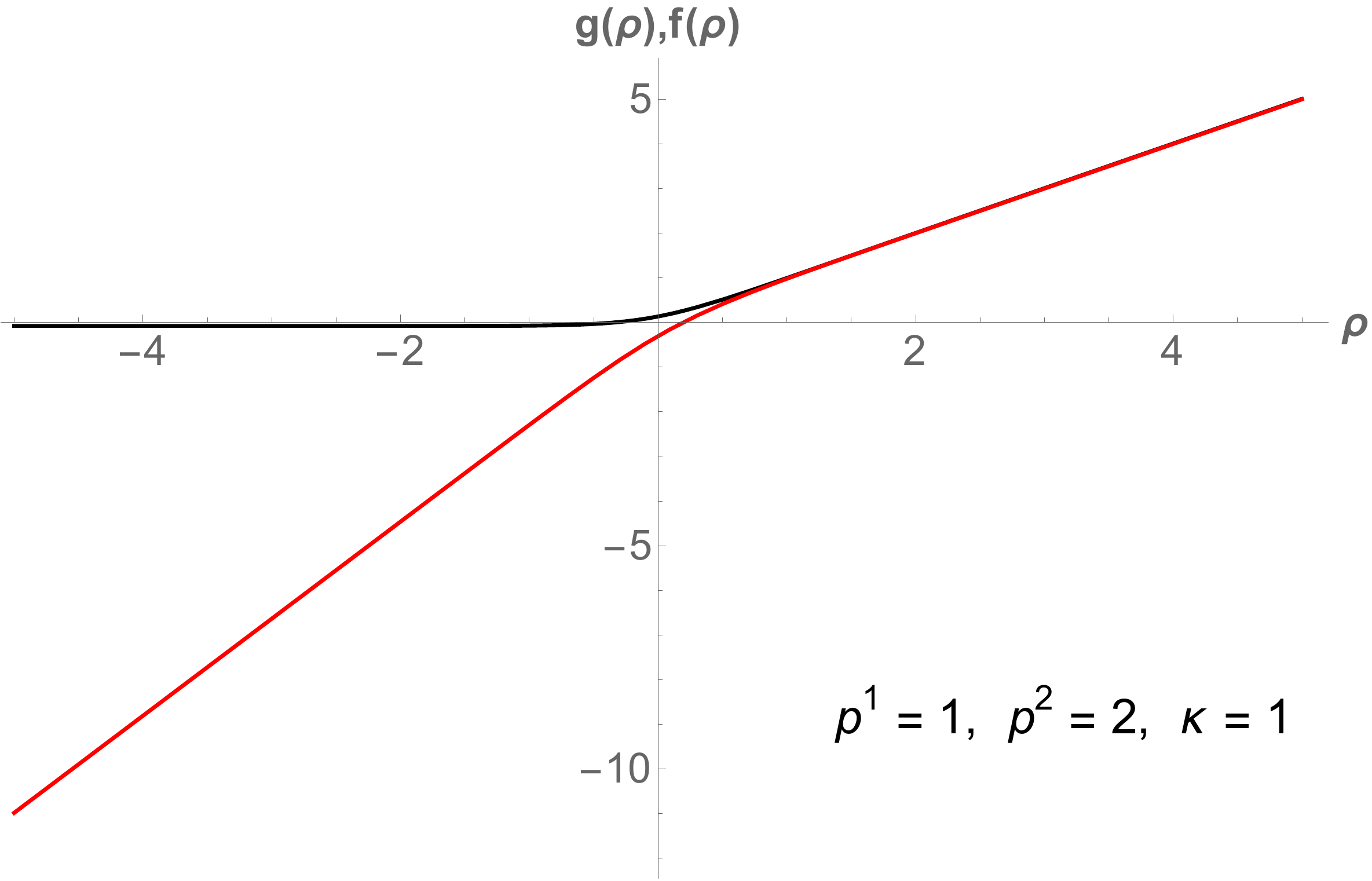}~~
    \includegraphics[width=0.46\textwidth]{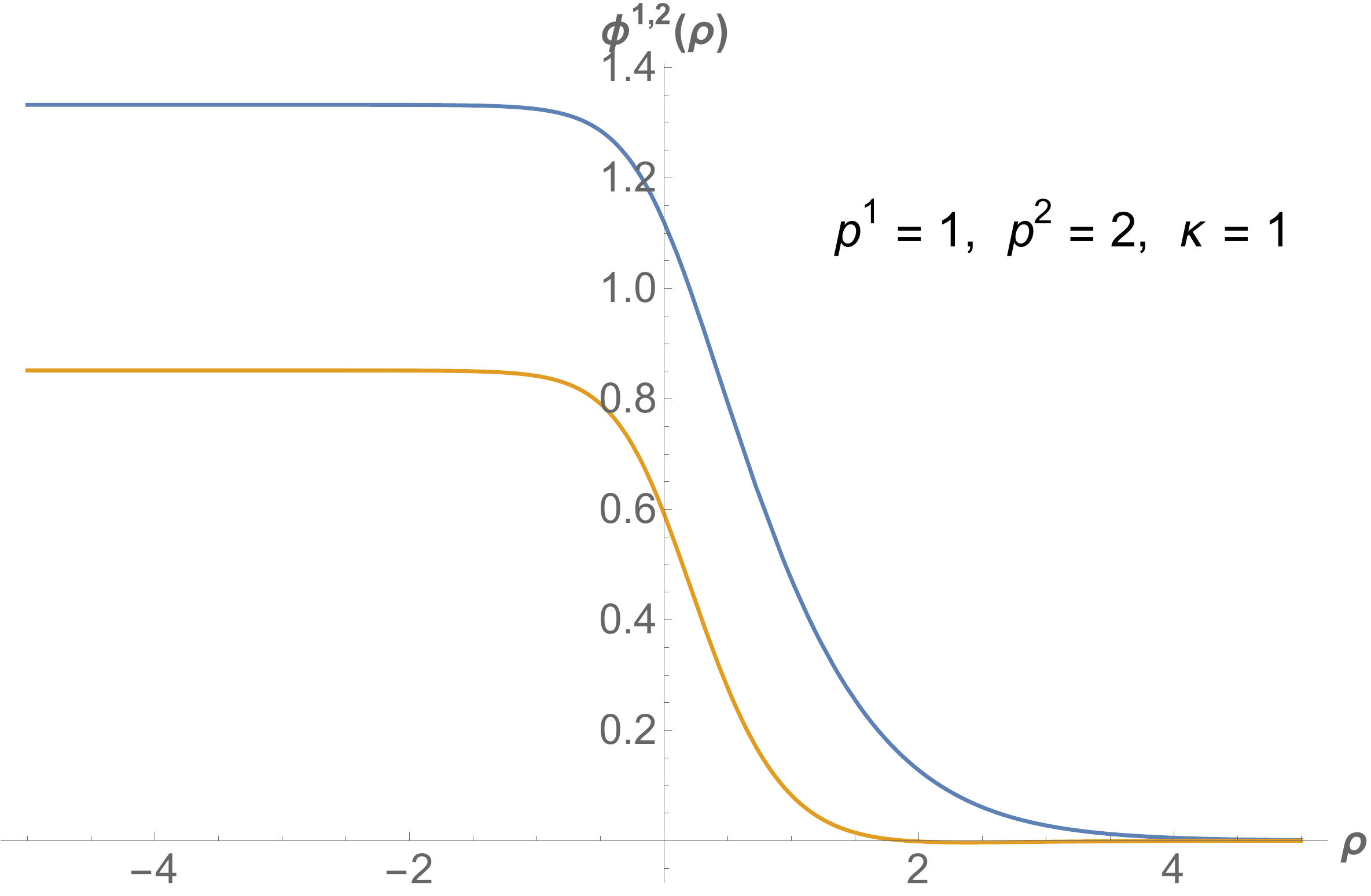}\\
    \includegraphics[width=0.46\textwidth]{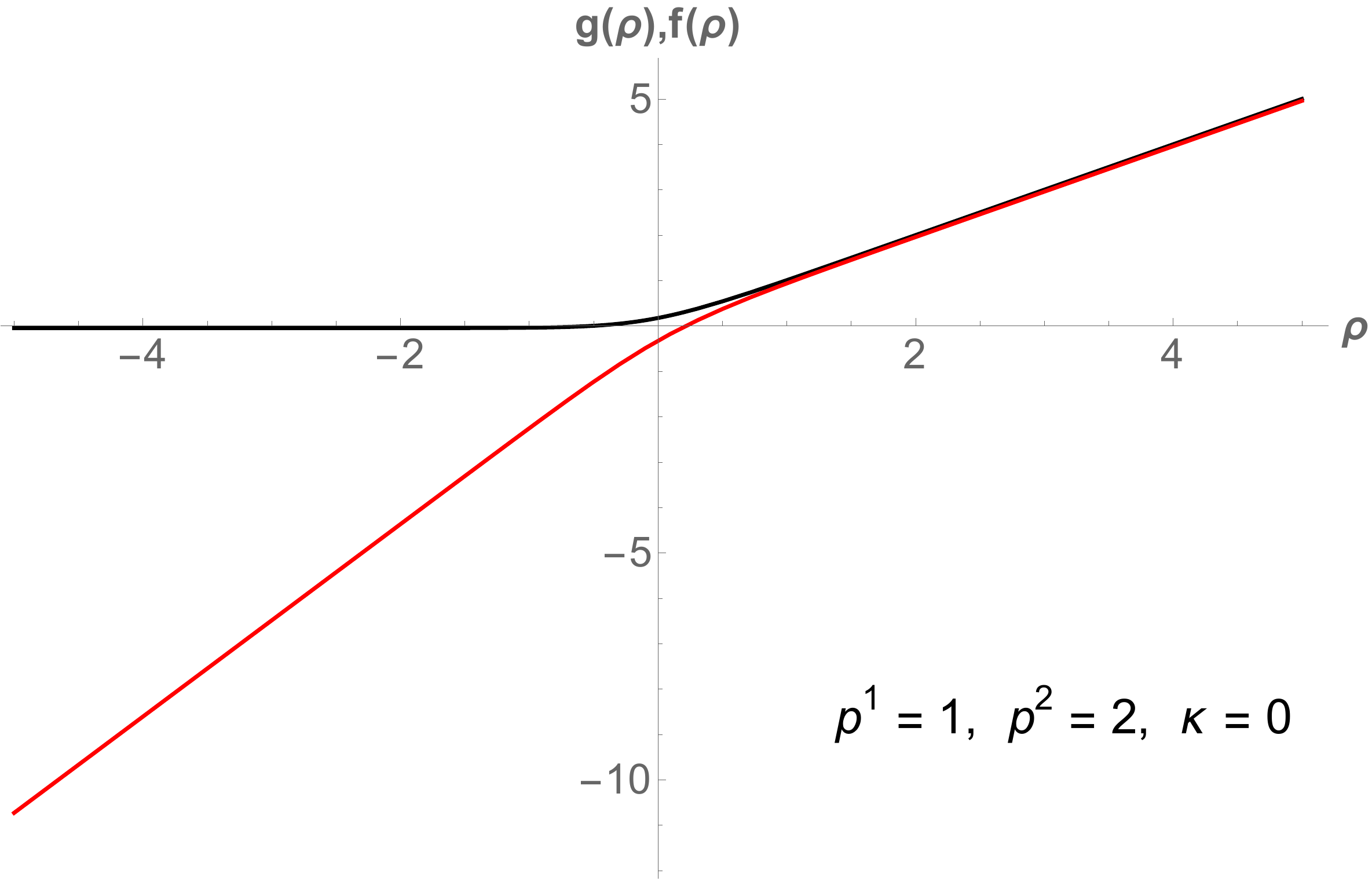}~~
    \includegraphics[width=0.46\textwidth]{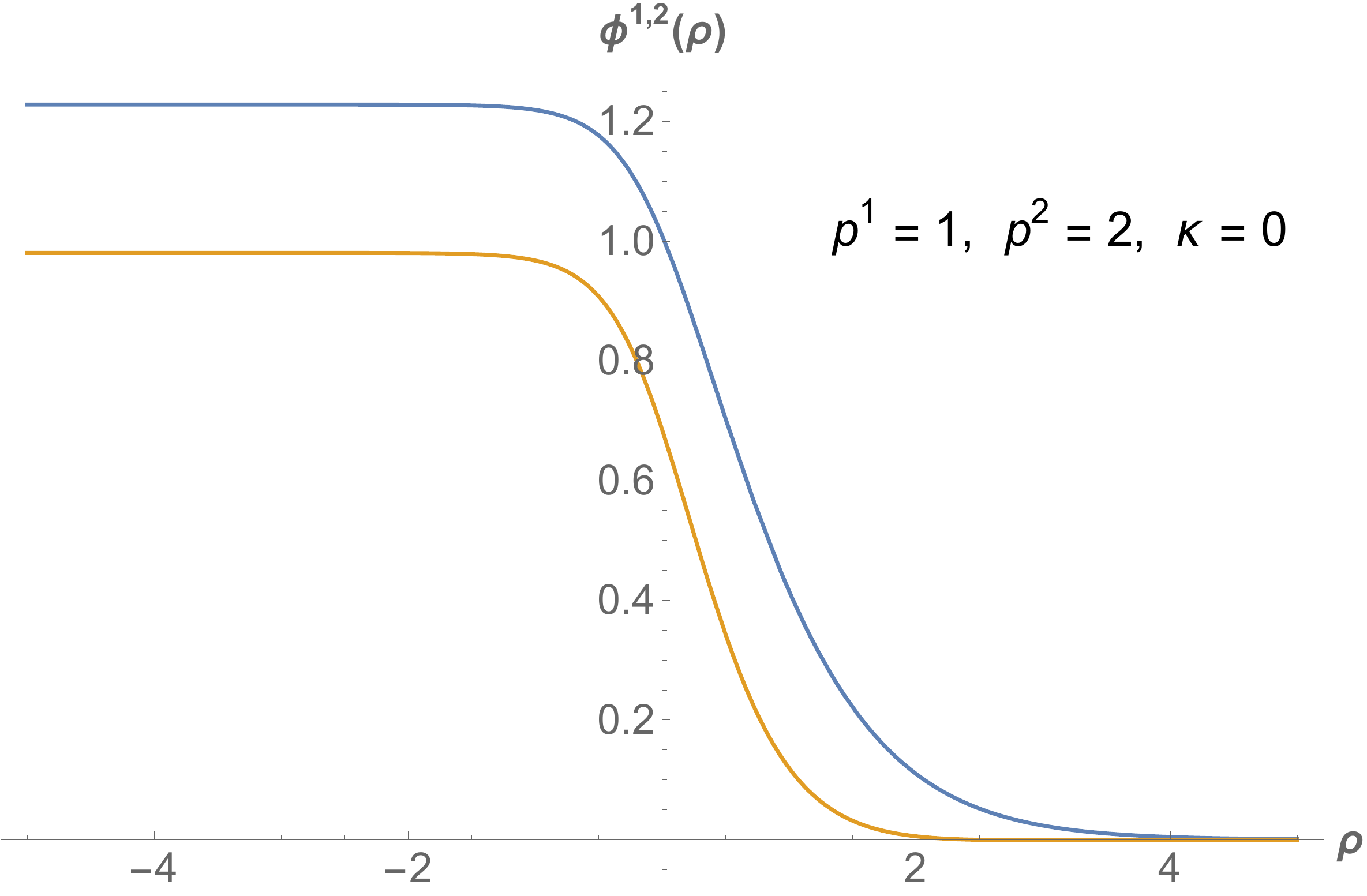}\\
\caption{Interpolating numerical solutions of the BPS equations (\ref{N=2:sugra:BPS:reduced}). In the left panels, black and red curves represent $g(\rho)$ and $f(\rho)$ flows, respectively. In the right panels, blue and orange curves represent $\phi^1(\rho)$ and $\phi^2(\rho)$ flows respectively.}\label{gphi:across:5to3}
\end{figure}

The BPS equations (\ref{N=2:sugra:BPS:reduced}) are obtained by reducing  (\ref{N=2:sugra:BPS}) for a magnetic black string ansatz (\ref{ansatz:BS}) in terms of a new holographic radial coordinate $\rho$ as (\ref{z:to:rho}). Numerical flow solutions are shown in Fig.~\ref{gphi:across:5to3}. The numerical solutions were constructed by using \texttt{NDSolve} in \textit{Mathematica} at \texttt{WorkingPrecision} $=50$. 
%

\subsection{Maximal gauged supergravity in $D=7$ compactified on $\mathbb{H}^4/\Gamma$}\label{App:7to3}
\begin{figure}[t]
	\centering
	\includegraphics[width=0.46\textwidth]{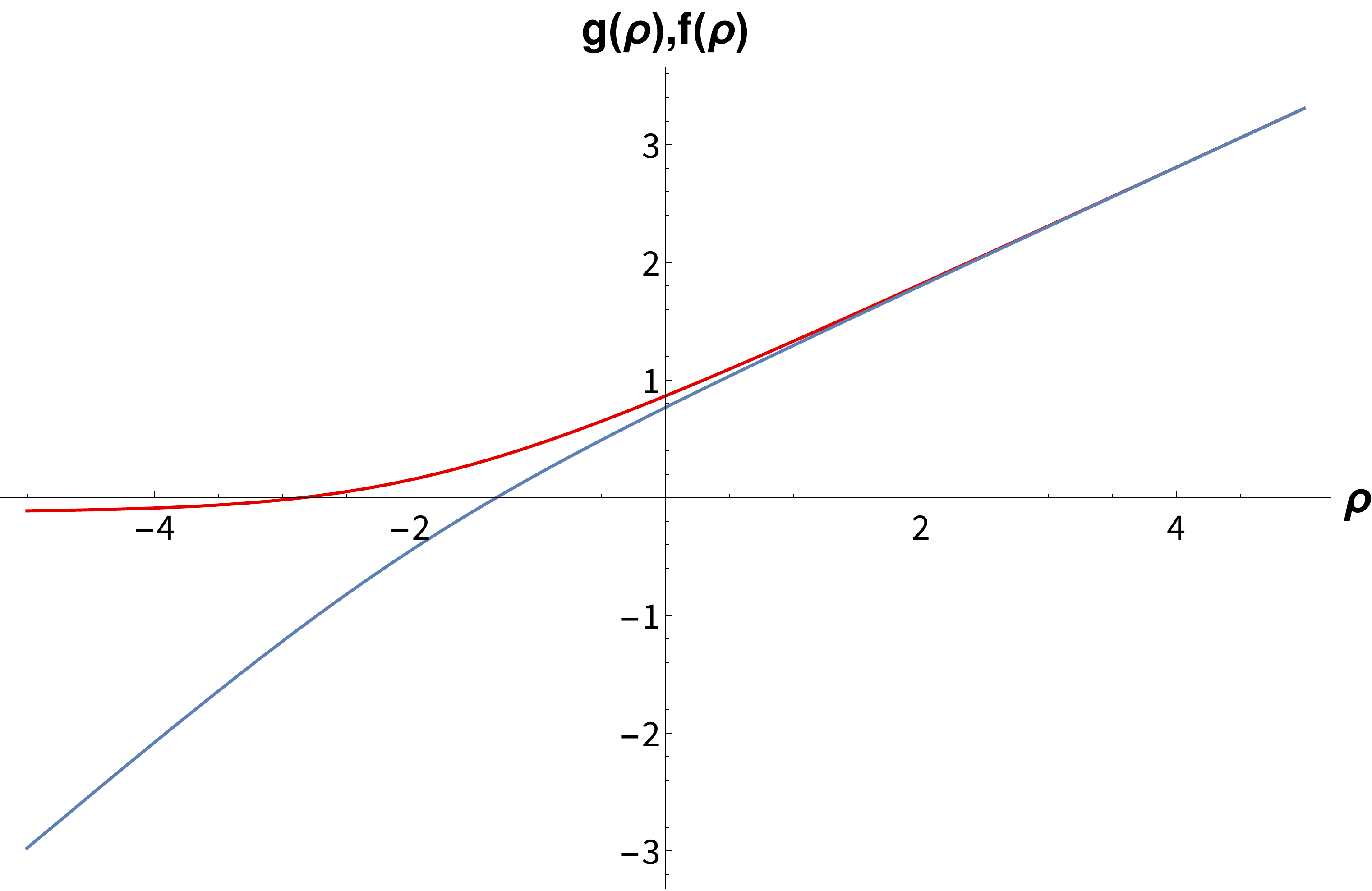}~~
    \includegraphics[width=0.468\textwidth]{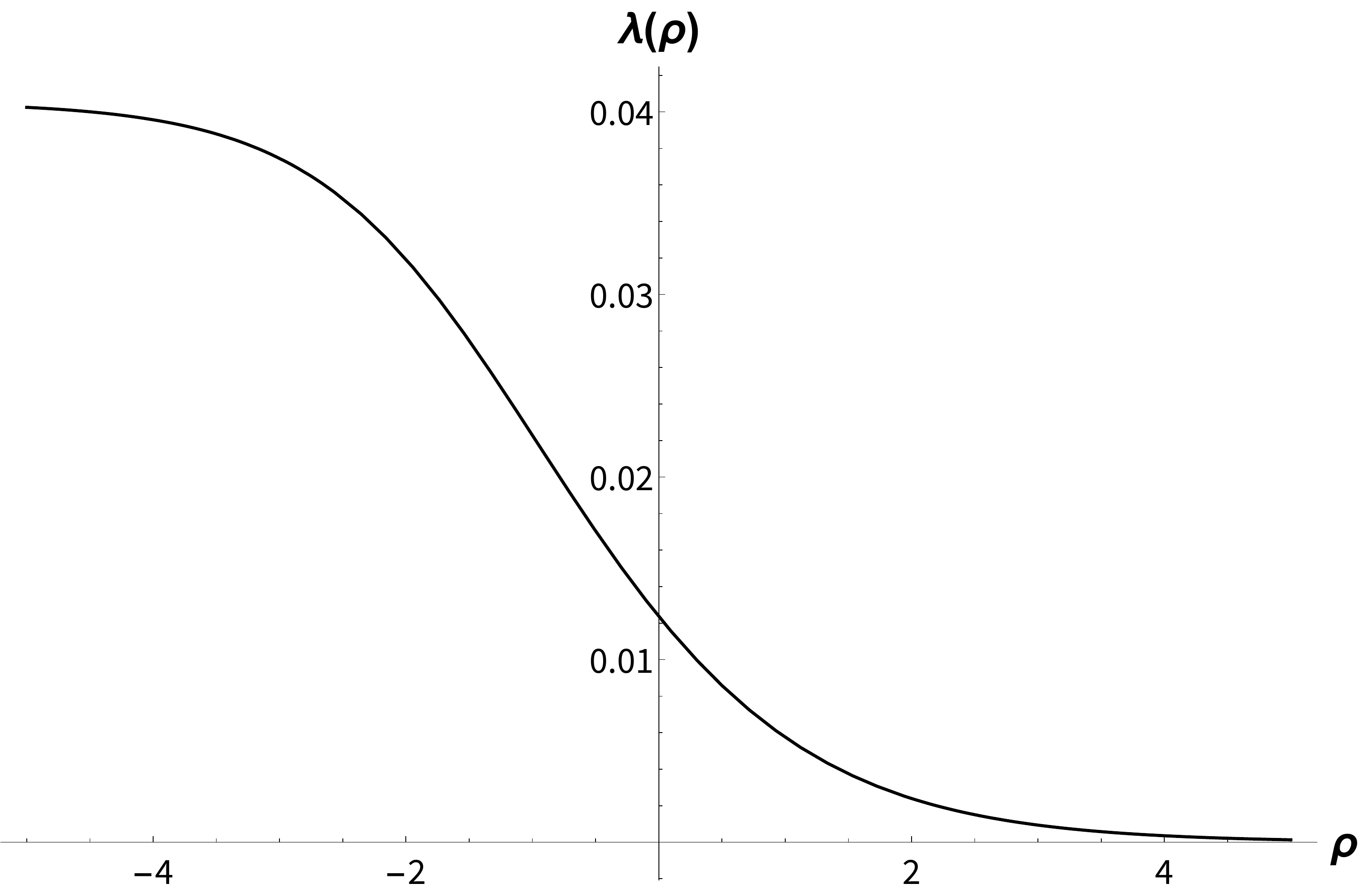}\\
	\caption{On the left panel we show a joint plot of the functions $f(\rho)$(blue) and $g(\rho)$(red) for a range of the coordinate $-5\leq \rho \leq 5$ and $m=1, \kappa= -1$. The UV behavior of $f(\rho)$ and $g(\rho)$ is $g(\rho)\simeq f(\rho) \simeq \frac{1}{2} \rho$ as $\rho \rightarrow \infty$. On the right panel we show the function $\lambda(\rho)$(black). }\label{Fig:fglambda:for7d}
\end{figure}

Let us collect some details about the 7d maximal gauged supergravity described in \cite{Gauntlett:2000ng}. For convenience we rewrite the BPS equations \eqref{eq:eff}, \eqref{eq:egg} and \eqref{eq:lambdaeq}: 
\bea
e^{-f} f'&=& -\frac{m}{10}\left(4e^{-2\lambda}+e^{8\lambda}\right)+\frac{\kappa }{5m} e^{2\lambda-2g}-\frac{\kappa^2}{10m^3}e^{-4\lambda-4g},  \nonumber \\
e^{-f} g'&=& -\frac{m}{10}\left(4e^{-2\lambda}+e^{8\lambda}\right)-\frac{3\kappa }{10m} e^{2\lambda-2g}+\frac{\kappa^2}{15m^3}e^{-4\lambda-4g},\nonumber \\
e^{-f} \lambda'&=& \frac{m}{5}\left(e^{8\lambda}-e^{-2\lambda}\right)+\frac{\kappa }{10m} e^{2\lambda-2g}+\frac{\kappa^2}{30m^3}e^{-4\lambda-4g}. \label{eq:App:lambdaeq}
\eea
In what follows we present the numerical solutions to \eqref{eq:App:lambdaeq} that we use in main body of the paper to evaluate the EE.
We first rewrite \eqref{eq:App:lambdaeq} using a new coordinate $\rho$ related to the usual holographic $z$-coordinate as: $d\rho =- e^{- f(z)} dz$, which yields:

\bea
 f'(\rho) -\frac{m}{10}\left(4e^{-2\lambda(\rho)}+e^{8\lambda(\rho)}\right)+\frac{\kappa }{5m} e^{2\lambda(\rho)-2g(\rho)}-\frac{\kappa^2}{10m^3}e^{-4\lambda(\rho)-4g(\rho)}&=& 0,  \\
 g'(\rho) -\frac{m}{10}\left(4e^{-2\lambda(\rho)}+e^{8\lambda(\rho)}\right)-\frac{3\kappa }{10m} e^{2\lambda(\rho)-2g}+\frac{\kappa^2}{15m^3}e^{-4\lambda(\rho)-4g(\rho)}&=&0,\\
\lambda'(\rho)+ \frac{m}{5}\left(e^{8\lambda(\rho)}-e^{-2\lambda(\rho)}\right)+\frac{\kappa }{10m} e^{2\lambda(\rho)-2g(\rho)}+\frac{\kappa^2}{30m^3}e^{-4\lambda(\rho)-4g(\rho)}&=& 0. \label{eq:App:lambdaeqRho}
\eea
Note that the UV region is located at $\rho \rightarrow \infty$ whereas the infrared corresponds to $\rho \rightarrow - \infty$. 
  In Fig.~\ref{Fig:fglambda:for7d} we present plots for the metric functions $(f,g)$ that we used on the main text. The numerical solutions were constructed with \texttt{NDSolve} in \textit{Mathematica} setting \texttt{WorkingPrecision} $=60$. It turns out that one can fix the function $g(\rho)$ up to some constant shift, however to make the numerical solution compatible with the Fefferman-Graham expansion we have used in the UV (see \ref{eq:expansionfg7d}) we need to guarantee that no constant shift in $g(\rho)$ appears in the UV. 
  The strategy we follow is to impose boundary conditions in the IR region and then introduce appropriate small perturbations to the expected IR behavior such that the desired UV behavior is reproduced. 
    
\subsection{Maximal gauged supergravity in $D=7$ compactified on $T^2$ }\label{App:7to5}

\begin{figure}[t]
	\centering
	\includegraphics[width=0.46\textwidth]{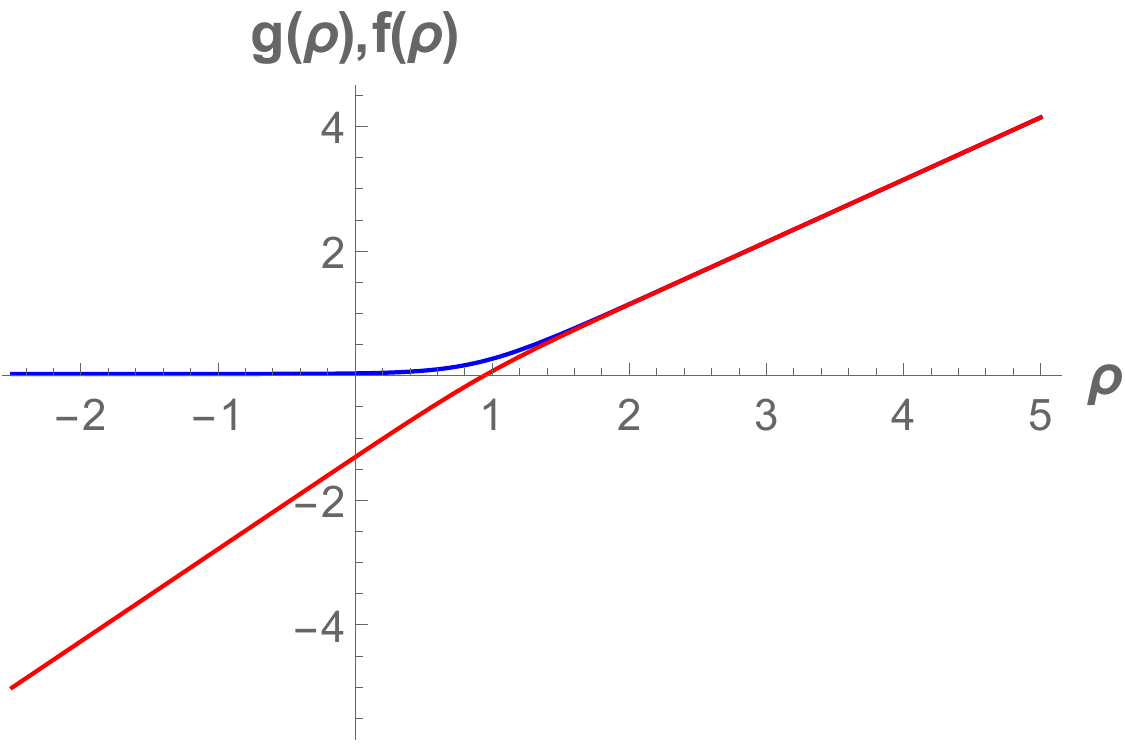}~~
    \includegraphics[width=0.468\textwidth]{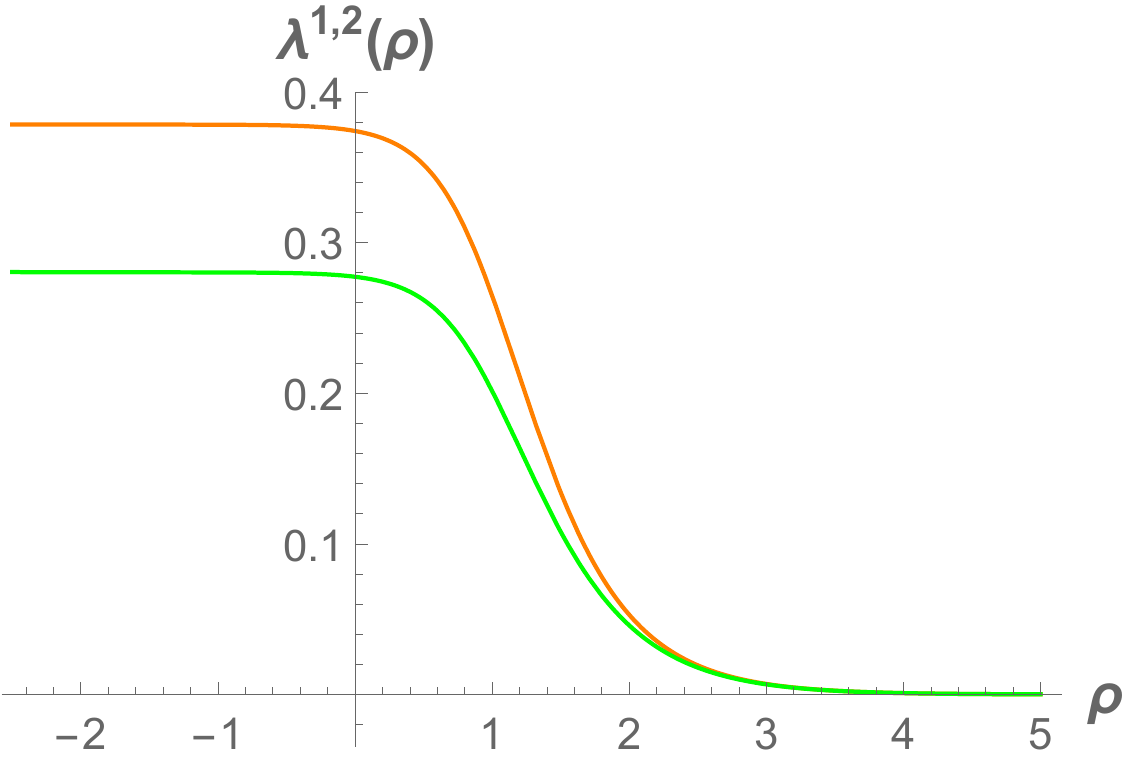}\\
	\caption{On the left panel we show a joint plot of the functions $f(r)$(red) and $g(r)$(blue) for a range of the coordinate $-2.5\leq \rho \leq 5$. The UV behavior of $f(\rho)$ and $g(r)$ is $g(r)\simeq f(r) \simeq \frac{1}{2} r$ as $r\rightarrow \infty$. On the right panel we show the functions $\lambda^{1,2}(\rho)$. }\label{Fig:fglambda1-2:for7to5d}
\end{figure}

In this section we present details about the interpolating AdS$_7\to$ AdS$_5$ discussed in section \ref{sec:EX:7-5d} and discussed in \cite{Maldacena:2000mw, Bah:2012dg,Uhlemann:2021itz}. We actually work in the gauge where $h=0$, the equations of motion we numerically solved are: 
\bea
f'+\lambda_1'+\lambda_2' +e^{-4\lambda_1-4\lambda_2}&=&0,\nonumber \\
g' -4\lambda_1'-4\lambda_2' + 2\,\, e^{2\lambda_1}+2\,\, e^{2\lambda_2} -3 \,\,e^{-4\lambda_1-4\lambda_2}&=&0,\nonumber \\
3\lambda_1'+ 2\lambda_2' -2 e^{2 \lambda_1} 
+ 2 \,\,e^{ - 4 \lambda_1-4\lambda_2 }- e^{-2g -2\lambda_1}F_1 &=&0,\nonumber \\
2\lambda_1'+ 3\lambda_2'-2 \,\,e^{2 \lambda_2} + 2\,\,e^{-4 \lambda_1 - 4\lambda_2} -e^{-2g -2\lambda_2} F_2 &=&0, 
\end{eqnarray}
The radial direction is related to the canonical $z$-radial direction as $e^{f(z)}dz=-d\rho$ and prime in the system above refers to derivative with respect to $\rho$.  The numerical solution is shown in Fig.~\ref{Fig:fglambda1-2:for7to5d}.

\section{Entanglement Entropy expansions}\label{App:EE}
In this Appendix we present technical details regarding the computation of the EE in the two flows that we presented in \ref{sec:EX:5d} and \ref{sec:EX:7d}. Our emphasis is to demonstrate how we achieve analytical control for the EE in the regime of small, as well as large, entangling regions. 

\subsection{Flows from AdS$_5$ to AdS$_3$}\label{App:EE:5d}
\textbf{Small entangling region:}
To consider the small entangling region limit, we first expand metric functions and scalar fields in the Ansatz (\ref{ansatz:BS}) around $z=0$ as\,\footnote{\label{fn:expansion} According to the Fefferman-Graham expansion, the UV expansion of metric functions and scalar fields, (\ref{fgphi:UV}), should also have logarithmic terms in general; for example, one must include $\sim(z/L_\text{UV})^4\log(z/L_\text{UV})$ terms into (\ref{f:UV}) and (\ref{g:UV}) (see \cite{deHaro:2000vlm} for example). We can ignore these logarithmic terms, however, since they do not affect the first coefficients of the UV expansion, (\ref{fgphi:UV:perturbative}), used to derive the UV expansion of the EE, (\ref{area:UV:3}).} ($f,g,\phi^x$ are all dimensionless in our convention)
\begin{subequations}
\begin{align}
	f(z)&=-\log(z/L_\text{UV})+\sum_{n=1}^\infty F_n(z/L_\text{UV})^n,\label{f:UV}\\
	g(z)&=-\log(z/L_\text{UV})+\sum_{n=1}^\infty G_n(z/L_\text{UV})^n,\label{g:UV}\\
	\phi^x(z)&=\sum_{n=1}^\infty\phi^x_n(z/L_\text{UV})^n,
\end{align}\label{fgphi:UV}%
\end{subequations}
such that the ansatz (\ref{ansatz:BS}) corresponds to an AdS$_5$ solution of radius $L_\text{UV}$ with vanishing scalars in the asymptotic region $z\to0$. The $\mathcal F$ function (\ref{eq:F}) is then expanded as
\begin{equation}
\begin{split}
	\mathcal F(z_0s)&=s^3\bigg(1+(1-s)(F_1+2G_1)\fft{z_0}{L_\text{UV}}\\
	&\kern3em+\fft12((s-1)^2(F_1+2G_1)^2+2(1-s^2)(F_2+2G_2))\fft{z_0^2}{L_\text{UV}^2}+\mathcal O(\fft{z_0^3}{L_\text{UV}^3})\bigg),
\end{split}\label{F:UV}
\end{equation}
in terms of which the EE (\ref{EE:area:5d:2}) reads
\begin{equation}
    S_\text{EE}(R;B^1{\times}\Sigma_{\mathfrak g},\epsilon)=\fft{\ell^2\text{vol}[\Sigma_{\mathfrak g}]}{2G^{(5)}_N}\,e^{f_0+2g_0}z_0\int_{\epsilon/z_0}^1ds\,\fft{1}{\mathcal F(z_0s)\sqrt{1-\mathcal F(z_0s)^2}}.\label{area:UV:1}
\end{equation}

To evaluate the EE (\ref{area:UV:1}) explicitly, we need to know the  perturbative coefficients in the UV expansions (\ref{fgphi:UV}). They are determined by solving the Einstein equations (\ref{BS:Einstein}) and the scalar equations of motion (\ref{BS:scalar}) perturbatively with the UV expansions (\ref{fgphi:UV}) as 
\begin{subequations}
\begin{align}
	z^0\text{-order: }&L_\text{UV}=\boldsymbol{g}^{-1}\\
	z^1\text{-order: }&F_1=G_1=0,\quad\phi^x_1=0\\
	z^2\text{-order: }&F_2=\fft{\kappa}{18\boldsymbol{g}^2\ell^2},\quad G_2=-\fft{7\kappa}{36\boldsymbol{g}^2\ell^2},\\
	z^3\text{-order: }&F_3=0,\quad G_3=0.
\end{align}\label{fgphi:UV:perturbative}%
\end{subequations}
Note that the radius of a Riemann surface $\ell$ is not constrained by the equations of motion: hence we can fix the ratio $\ell/L_\text{UV}$ finite and use $R/\ell$ (or $z_0/\ell$) as an alternative expansion parameter instead of $R/L_\text{UV}$ (or $z_0/L_\text{UV}$). 

Substituting (\ref{F:UV}) and (\ref{fgphi:UV:perturbative}) into the EE (\ref{area:UV:1}) and using the following integrals,
\begin{equation}
\begin{split}
	\int_{\epsilon/z_0}^1ds\fft{1}{s^3\sqrt{1-s^6}}&=\fft{z_0^2}{2\epsilon^2}-\fft{\pi^\fft12\Gamma(\fft23)}{2\Gamma(\fft16)}+\mathcal O(\fft{\epsilon^4}{z_0^4}),\\
	\int_{\epsilon/z_0}^1ds\fft{1+s^2+2s^4}{s(1+s^2+s^4)\sqrt{1-s^6}}&=-\log\fft{\epsilon}{z_0}+\fft{1+\log2}{3}-\fft{\pi^\fft12\Gamma(\fft23)}{3\Gamma(\fft16)}+\mathcal O(\fft{\epsilon^4}{z_0^4}),
\end{split}\label{integrals:UV:1}
\end{equation}
we obtained the UV expansion
\begin{equation}
\begin{split}
    &S_\text{EE}(R;B^1{\times}\Sigma_{\mathfrak g},\epsilon)\\
    &=\fft{L_\text{UV}^3\text{vol}[\Sigma_{\mathfrak g}]}{2G^{(5)}_N}\,\left(\fft{\ell^2}{2\epsilon^2}-\fft{\pi^\fft12\Gamma(\fft23)}{2\Gamma(\fft16)}\fft{\ell^2}{z_0^2}-\fft{\kappa}{3}\bigg(-\log\fft{\epsilon}{z_0}+\fft{1+\log2}{3}-\fft{\pi^\fft12\Gamma(\fft23)}{3\Gamma(\fft16)}\bigg)+\mathcal O(\fft{z_0^2}{\ell^2})\right).\label{area:UV:3}
\end{split}
\end{equation}
Here the $\epsilon/z_0\to0$ limit is taken independently from the small entangling region limit $z_0/L_\text{UV}\ll1$. To rewrite (\ref{area:UV:3}) as a function of the radius $R$ of an entangling region $B^1$ on the boundary, we expand the relation between $R$ and $z_0$ given in (\ref{z:to:R}) using the UV expansion (\ref{fgphi:UV}) and the integrals (\ref{integrals:UV:2}) as
\begin{equation}
	R=z_0\left(\fft{\pi^\fft12\Gamma(\fft23)}{\Gamma(\fft16)}-\fft{\kappa}{9}\left(1-\fft{\pi^\fft12\Gamma(\fft23)}{\Gamma(\fft16)}\right)\fft{z_0^2}{\ell^2}+\mathcal O(\fft{z_0^4}{\ell^4})\right).\label{z:to:R:UV}
\end{equation}
Substituting (\ref{z:to:R:UV}) into (\ref{area:UV:3}) finally gives the UV expansion of EE with respect to $R$ in the main text, (\ref{area:UV:4}).

To derive the relation between $R$ and $z_0$ in UV, (\ref{z:to:R:UV}), we have substituted the UV expansion (\ref{fgphi:UV}) into (\ref{z:to:R}) and then used the following integrals
\begin{equation}
\begin{split}
	\int_0^1ds\fft{s^3}{\sqrt{1-s^6}}&=\fft{\pi^\fft12\Gamma(\fft23)}{\Gamma(\fft16)},\\
	\int_0^1ds\fft{s^3}{(1+s^2+s^4)\sqrt{1-s^6}}&=\fft13\left(1-\fft{\pi^\fft12\Gamma(\fft23)}{\Gamma(\fft16)}\right).
\end{split}\label{integrals:UV:2}
\end{equation}
%

\textbf{Large entangling region:}
To consider the large entangling region limit, first we expand metric functions and scalar fields in the Ansatz (\ref{ansatz:BS}) around $z=\infty$ as
\begin{equation}
\begin{split}
	f(z)&=\log(L_\text{IR}/z)+\sum_{n=1}^\infty\tilde F_n(L_\text{IR}/z)^n,\\
	g(z)&=\sum_{n=0}^\infty\tilde G_n(L_\text{IR}/z)^n,\\
	\phi^x(z)&=\sum_{n=0}^\infty\tilde\phi^x_n(L_\text{IR}/z)^n,
\end{split}\label{fgphi:IR}
\end{equation}
so that the ansatz (\ref{ansatz:BS}) corresponds to an AdS$_3\times\Sigma_{\mathfrak g}$ solution with the AdS$_3$ radius $L_\text{IR}$ at the horizon $z\to\infty$. We also introduce the radius of convergence of the IR expansion (\ref{fgphi:IR}), namely $\Lambda$, such that the IR expansion (\ref{fgphi:IR}) is valid under $z>\Lambda$. We then split the integral for an EE (\ref{EE:area:5d:2}) as
\begin{equation}
    S_\text{EE}(R;B^1{\times}\Sigma_{\mathfrak g},\epsilon)=\fft{\ell^2\text{vol}[\Sigma_{\mathfrak g}]}{2G^{(5)}_N}\,e^{f_0+2g_0}\bigg(\int_{\Lambda}^{z_0}+\int_{\epsilon}^{\Lambda}\bigg)dz\,\fft{1}{\mathcal F(z)\sqrt{1-\mathcal F(z)^2}}.\label{area:IR:1}
\end{equation}
We will estimate the two integrals in (\ref{area:IR:1}) separately. The 1st integral in (\ref{area:IR:1}) can be expanded by substituting the IR expansion (\ref{fgphi:IR}) as
\begin{equation}
\begin{split}
	S_\text{EE}^{(1)}(R;B^1{\times}\Sigma_{\mathfrak g},\epsilon)&=\fft{\ell^2\text{vol}[\Sigma_{\mathfrak g}]}{2G^{(5)}_N}\sum_{n=0}^\infty{-\fft12\choose n}(-1)^n\left(\fft{L_\text{IR}e^{2\tilde G_0}}{z_0}(1+\mathcal O(\fft{L_\text{IR}}{z_0}))\right)^{2n}\\
	&\quad\times\int_{\Lambda}^{z_0}dz\,\left(\fft{L_\text{IR}e^{2\tilde G_0}}{z}(1+\mathcal O(\fft{L_\text{IR}}{z}))\right)^{1-2n}\\
	&=\fft{\ell^2\text{vol}[\Sigma_{\mathfrak g}]}{4G^{(5)}_N}L_\text{IR}e^{2\tilde G_0}\log\fft{z_0}{\Lambda}+\mathcal O((\fft{L_\text{IR}}{z_0})^0),
\end{split}\label{area:IR:1:1}
\end{equation}
where the logarithmic term in (\ref{area:IR:1:1}) comes from the $n=0$ contribution only. One can also estimate the 2nd integral in (\ref{area:IR:1}) easily as
\begin{equation}
\begin{split}
	S_\text{EE}^{(2)}(R;B^1{\times}\Sigma_{\mathfrak g},\epsilon)&=\mathcal O((\fft{L_\text{IR}}{z_0})^0).
\end{split}\label{area:IR:1:2}
\end{equation}
Substituting (\ref{area:IR:1:1}) and (\ref{area:IR:1:2}) into the EE (\ref{area:IR:1}) gives the IR expansion
\begin{equation}
    S_\text{EE}(R;B^1{\times}\Sigma_{\mathfrak g},\epsilon)=\fft{\ell^2\text{vol}[\Sigma_{\mathfrak g}]}{2G^{(5)}_N}L_\text{IR}e^{2\tilde G_0}\log\fft{z_0}{\Lambda}+\mathcal O((\fft{L_\text{IR}}{z_0})^0).\label{area:IR:2}
\end{equation}
To rewrite (\ref{area:IR:2}) as a function of the radius $R$ of an entangling region $B^1$ on the boundary, we expand the relation between $R$ and $z_0$ given in (\ref{z:to:R}) using the IR expansion (\ref{fgphi:IR}) as
\begin{equation}
	R=z_0+\mathcal O((\fft{L_\text{IR}}{z_0})^0).\label{z:to:R:IR}
\end{equation}
Substituting (\ref{z:to:R:IR}) into (\ref{area:IR:2}) finally gives the IR expansion of EE with respect to $R$ in the main text, (\ref{area:IR:3}).

\subsection{Flow from AdS$_7$ to AdS$_3$}\label{App:AdS7Entropy}

\textbf{Small entangling region:}
Let us denote as $L_{\text{UV}}$ the radius of AdS$_7$ and consider an entangling region whose tip probes a point $z_0$ satisfying $z_0/L_{\text{UV}} \ll 1$ and $R/L_{\text{UV}} \ll 1$, this will be a small entangling region. 
Let us then proceed to expand the functions $f(z), g(z)$ and $\lambda(z)$ around the UV region $z = 0$:
\bea 
f(z)  &= & - \log(z/L_{\text{UV}}) + \sum_{n=1}^{\infty} F_n (z/L_{\text{UV}})^n  \label{eq:expansionfg7d}\nonumber \\
g(z) &= & - \log(z/L_{\text{UV}}) + \sum_{n=1}^{\infty} G_n (z/L_{\text{UV}})^n \nonumber \\ 
\lambda(z) &= & \sum_{n=1}^{\infty}\Lambda_n(z/L_{\text{UV}})^n .\nonumber \\ 
\eea
Defining the $z =z_0 s$ we can write the expansion of $\mathcal{F}(z_0 s)$  as:
\begin{align}
    \mathcal{F}(z_0 s) & =s^5\bigg(1+
(1-s^2)(F_2+4G_2)\fft{z_0^2}{L_\text{UV}^2}\\
&\frac{1}{2} \left(s^2-1\right) \left( \left(s^2-1\right)(F_2+4 G_2)^2 -\left(s^2+1\right)( F_4+4 G_4) \right)+\mathcal O(\fft{z_0^5}{L_\text{UV}^5})\bigg), \label{eq:FF7d}
\end{align}
Let us then write the EE as follows:
\begin{align}
\begin{split}
     S_\text{EE}(R;B^{1},\epsilon)& =\fft{\ell^4\text{vol}[\mathbb{H}_4]}{4G^{(7)}_N} 2 e^{f_0 +4 g_0}z_0\int_{\epsilon/z_0}^{1} \frac{ds}{\mathcal{F}(z_0 s)\sqrt{1 - \mathcal{F}(z_0 s)^2}} \label{eq:SE7d}\\
     \end{split}
\end{align}

The coefficients $F_n, \hspace{2mm} G_n, \hspace{2mm} \Lambda_n$ in the expansion \eqref{eq:expansionfg7d} can be found  by perturbatively solving the BPS equations \eqref{eq:eff}, \eqref{eq:egg} and \eqref{eq:lambdaeq}, to obtain:
\begin{align}
    \begin{split}
        z^0 & - \text{order}: L_{\text{UV}} = \frac{2}{m} \label{eq:pert7d} \\
        z^1 & -  \text{order}: F_1 = G_1  = \Lambda_1 =0 \\
        z^2 & - \text{order}: F_2 =  \frac{2 \kappa}{15 m^2}, \hspace{3mm} G_2 = - \frac{11 \kappa}{30 m^2}, \hspace{3mm} \Lambda_2 = - \frac{\kappa}{10 m^2} \\
        z^3 & - \text{order}: F_3 = G_3=\Lambda_3=0 \\
        z^4 & -\text{order}: F_4 = - \frac{\kappa^2}{225 m^2}, \hspace{3mm} G_4 =- \frac{79 \kappa^2}{900 m^2}.
    \end{split}
\end{align}
 We should point out that the value of $\Lambda_4$ is not successfully captured by our perturbative approach, however this is to be expected since, as pointed out in footnote \ref{fn:expansion}, we should have included extra logarithmic terms into the Fefferman-Graham expansion, but not such terms nor the function $\lambda(\rho)$ affect our calculations so we can safely ignore such extra logarithmic terms.
 
We have used the following integrals in UV expansion around AdS$_7$:
\begin{align}
    \int_{\epsilon/z_0}^1ds \frac{1}{s^5 \sqrt{1- s^{10}}} & =  \frac{z_0^4}{4 \epsilon^4} -\frac{\pi^\fft12 \Gamma \left(\frac{3}{5}\right)}{4 \Gamma \left(\frac{1}{10}\right)} + \mathcal{O}\left(\left(\frac{\epsilon}{z_0}\right)^4\right)\label{eq:int7d} \\
  \int_{\epsilon/z_0}^1ds \frac{ \left(1+s^2+s^4+s^6+ 2 s^8\right)}{s^3\left(1+s^2+s^4+s^6+  s^8\right) \sqrt{1- s^{10}}}    & = \frac{z_0^2}{2 \epsilon^2} +\frac{1}{240} \pi^\fft12 \left(-\frac{8 \Gamma \left(\frac{8}{5}\right)}{\Gamma \left(\frac{11}{10}\right)}-\frac{27 \Gamma \left(\frac{9}{5}\right)}{\Gamma \left(\frac{13}{10}\right)}\right) +\mathcal{O}\left(\left(\frac{\epsilon}{z_0}\right)^4\right) \nonumber\\
 \int_{\epsilon/z_0}^1ds \frac{ \left(3+3s^2+3s^4-4s^6+16 s^8\right)}{s\left(1+s^2+s^4+s^6+  s^8\right) \sqrt{1- s^{10}}}   & = - 3 \log\left(\frac{\epsilon}{z_0}\right) +\frac{13}{5} +\frac{3}{5}\log 2 \nonumber\\
 & + \frac{1}{30} \pi^\fft12 \left(\frac{45 \Gamma \left(\frac{9}{5}\right)}{\Gamma \left(\frac{13}{10}\right)}-\frac{7 \Gamma \left(\frac{8}{5}\right)}{\Gamma \left(\frac{11}{10}\right)}\right) + \mathcal{O}\left(\left(\frac{\epsilon}{z_0}\right)^4\right)
 \nonumber\\
  \int_{0}^1ds \frac{s^5}{\sqrt{1- s^{10}}}& = \frac{\pi^\fft12 \Gamma\left(\frac{8}{5}\right)}{6 \Gamma\left(\frac{11}{10}\right)}.
\end{align}
Replacing the expansion coefficients \eqref{eq:pert7d} into \eqref{EE:area:5d:2} for $D=6$ and using the integrals listed in \ref{eq:int7d} yields:

\begin{align}
\begin{split}
    S_\text{EE}(R;B^{1}\times \mathbb{H}_4,\epsilon)& = \fft{L_{\text{UV}}^5\text{vol}[\mathbb{H}_4]}{2G^{(7)}_N}  \left(-\frac{8 \kappa^2}{15 m^4} \log \left(\frac{\epsilon}{z_0}\right)- \frac{2 \kappa}{3 m^2}\frac{\ell^2}{\epsilon^2}+\frac{\ell^4}{4 \epsilon^4}+\frac{\kappa}{m^2} \widetilde{A}\frac{\ell^2}{ z_0^2}+\widetilde{B} \frac{\ell^4}{z_0^4}+ \mathcal{O}\left(\left(\frac{z_0}{\ell}\right)^0\right)\right) \label{eq:SE7dexp} \\
   \widetilde{B}  & = -\frac{\pi^{\frac{1}{2}} \Gamma \left(\frac{3}{5}\right)}{4 \Gamma \left(\frac{1}{10}\right)}, \hspace{3mm} \widetilde{A} =\frac{1}{180} \pi^{\frac{1}{2}} \left(\frac{8 \Gamma \left(\frac{8}{5}\right)}{\Gamma \left(\frac{11}{10}\right)}+\frac{27 \Gamma \left(\frac{9}{5}\right)}{\Gamma \left(\frac{13}{10}\right)}\right).
    \end{split}
\end{align}
To obtain a relation between $R$ and $z_0$ we expand the following expression:
\begin{align}
    R & = z_0 \int_{0}^1 ds \frac{\mathcal{F}(z_0 s)}{\sqrt{1- \mathcal{F}(z_0 s)^2}} \label{eq:R7dexpansion} 
    = z_0 \left(\frac{\pi^{\frac{1}{2}} \Gamma \left(\frac{8}{5}\right)}{6 \Gamma \left(\frac{11}{10}\right)} + \mathcal{O}\left(\frac{z_0}{\ell}\right)^2\right).
\end{align}

\textbf{Large entangling region:} We expand the functions $f(z)$, $g(z)$ and $\lambda(z)$ around $z \rightarrow \infty$:
\begin{align}
    \begin{split}
        f(z) & = \log\left(L_{\text{IR}}/z\right) + \sum_{n=1}^{\infty} \widetilde{F}_n \left(L_{\text{IR}}/z\right)^n \label{eq:IRfgl}\\
        g(z) & = \sum_{n=0}^{\infty} \widetilde{G}_n\left(L_{\text{IR}}/z\right)^n\\
        \lambda(z) & = \sum_{n=0}^{\infty} \widetilde{\Lambda}_n\left(L_{\text{IR}}/z\right)^n.
    \end{split}
\end{align}
The ansatz \eqref{eq:IRfgl} will correspond to AdS$_3 \times  \mathbb{H}_4$ with AdS$_3$ radius $L_{\text{IR}}$ at $z \rightarrow \infty$. Once again introducing a scale $\Lambda$ characterizing the radius of convergence of the expansion \eqref{eq:IRfgl}, the EE \eqref{EE:area:5d:2} can be written as:

\begin{equation}
S_\text{EE}(R;B^1,\epsilon)=\fft{\ell^4\text{vol}[\mathbb{H}_4]}{2G^{(7)}_N}\,e^{f_0+4g_0}\bigg(\int_{\Lambda}^{z_0}+\int_{\epsilon}^{\Lambda}\bigg)dz\,\fft{1}{\mathcal F(z)\sqrt{1-\mathcal F(z)^2}}.\label{area:IR:17d}
\end{equation}
The analysis to estimate the two integrals in (\ref{area:IR:17d}) goes through completely analogous to the AdS$_5$ to AdS$_3$ case, therefore we summarize the main results here:

The 1st integral in (\ref{area:IR:17d}) yields:
\begin{equation}
\begin{split}
	S_\text{EE}^{(1)}(R;B^1,\epsilon)&=\fft{\ell^4\text{vol}[\mathbb{H}_4]}{4G^{(7)}_N}L_\text{IR}e^{4\tilde G_0}\log\fft{z_0}{\Lambda}+\mathcal O((\fft{L_\text{IR}}{z_0})^0),
\end{split}\label{area:IR:1:17d}
\end{equation}
where the logarithmic term in (\ref{area:IR:1:17d}) comes from the $n=0$ contribution only. Similarly, the 2nd integral in (\ref{area:IR:17d}) can be estimated as:
\begin{equation}
\begin{split}
	S_\text{EE}^{(2)}(R;B^1,\epsilon)&=\mathcal O((\fft{L_\text{IR}}{z_0})^0).
\end{split}\label{area:IR:1:27d}
\end{equation}
 Substituting the estimation (\ref{area:IR:1:17d}) and (\ref{area:IR:1:27d}) back into the EE (\ref{area:IR:17d}), we obtain
\begin{equation}
S_\text{EE}(R;B^1,\epsilon)=\fft{\ell^4\text{vol}[\mathbb{H}_4]}{2G^{(7)}_N}L_\text{IR}e^{4\tilde G_0}\log\fft{z_0}{\Lambda}+\mathcal O((\fft{L_\text{IR}}{z_0})^0).\label{area:IR:27d}
\end{equation}

The EE \eqref{area:IR:1:27d} can be expressed as a function of the radius $R$ of an entangling region $B^1$ on the boundary, where :
\begin{equation}
\begin{split}
	R&=z_0+\mathcal O((\fft{L_\text{IR}}{z_0})^0).
\end{split}\label{z:to:R:IR7d}
\end{equation}
%





\section{Monotonicity of $c_{\mathrm{mono}}$ from the null energy condition}

For the holographic EE, we start with the expression for the area of a minimal surface of a region wrapping $M_{D-d}$ given in (\ref{EE:area:5d}), which we write here as
\begin{equation}
    S_\text{EE}(z_0,\epsilon)=\fft{2\text{vol}[M_{D-2}]}{4G^{(D+1)}_N}\int_\epsilon^{z_0}dz\,e^{\tilde f(z)}\sqrt{1+r'(z)^2}.
\label{eq:SEErz}
\end{equation}
The minimal surface is parametrized by the profile $r(z)$ where $z_0$ is the cap-off point in the bulk.  In order to obtain the EE as a function of the radius $R$ of the entangling surface, we also need the relation
\begin{equation}
    R(z_0)=-\int_0^{z_0}r'(z)dz.
\end{equation}
The negative sign arises since $r'(z)<0$ in this parametrization.

At the cap-off point, the slope $r'(z)$ reaches $-\infty$, which makes these expressions somewhat awkward to deal with.  However, we can follow \cite{Myers:2012ed} and instead parametrize the minimal surface by $z(r)$ with boundary conditions
\begin{equation}
z(0)=z_0,\qquad z'(0)=0,\qquad z(R)=0.
\end{equation}
The area of the minimal surface, (\ref{eq:SEErz}), then takes the form
\begin{equation}
    S_\text{EE}(R,\epsilon)=\fft{2\text{vol}[M_{D-2}]}{4G^{(D+1)}_N}\int_0^{R_c}dr\,e^{\tilde f(z(r))}\sqrt{1+z'(r)^2}.
\label{eq:SEEzr}
\end{equation}
with boundary condition $z'(0)=0$.  Here $R_c$ is the cutoff value of the radius $R$ in the sense that $z(R_c)=\epsilon$.

The profile of the minimal surface is of course independent of our choice of parametrization.  In either case, the equation of motion governing the surface admits a first integral
\begin{equation}
    r'(z)=-\fft{\mathcal F}{\sqrt{1-\mathcal F^2}}\qquad\mbox{or}\qquad z'(r)=-\fft{\sqrt{1-\mathcal F^2}}{\mathcal F}.
\label{eq:fint}
\end{equation}
where $\mathcal F$ was defined in (\ref{eq:F}), and takes the simple form
\begin{equation}
    \mathcal F=e^{\tilde f(z_0)-\tilde f(z)}
\end{equation}
when expressed in terms of the effective $\tilde f$ given in (\ref{eq:tildef}).  Note that $\mathcal F$ goes from zero at the UV boundary ($z=0$) to one at the cap-off point $z_0$.

By using the $z(r)$ parametrization, we may more directly compute the holographic $c$-function $c_{\mathrm{mono}}(R)$ defined in (\ref{cc:across:mono})
\begin{equation}
    c_{\mathrm{mono}}(R)=R\partial_RS_{EE}(R,\epsilon).
\end{equation}
Following \cite{Myers:2012ed}, the $R$ derivative of (\ref{eq:SEEzr}) acts both on the upper limit of the integral and through the dependence of $z(r)$ on $R$.  Varying the integrand with respect to $z(r)$ gives the equation of motion in the bulk, but leaves a surface term.  So we find
\begin{align}
    c_{\mathrm{mono}}(R)&=\fft{2\text{vol}[M_{D-2}]}{4G^{(D+1)}_N}R\left[\fft{dR_c}{dR}e^{\tilde f}\sqrt{1+z'^2}+e^{\tilde f}\fft{z'}{\sqrt{1+z'^2}}\fft{dz}{dR}\right]_{r=R_c}\nn\\
    &=\fft{2\text{vol}[M_{D-2}]}{4G^{(D+1)}_N}e^{\tilde f}R\left[\fft{dR_c}{dR}\sqrt{1+z'^2}+\fft{z'}{\sqrt{1+z'^2}}\fft{dz}{dR}\right]_{r=R_c}.
\label{eq:cmonoR}
\end{align}
While the surface is parametrized by $z(r)$, this profile implicitly depends on the radius $R$.  Thus we can consider the UV boundary condition to have the form $z_R(r=R_c)=\epsilon$.  Varying this with respect to $R$ then yields the relation
\begin{equation}
z'\fft{dR_c}{dR}+\fft{dz}{dR}=0,
\end{equation}
which allows us to rewrite (\ref{eq:cmonoR}) as
\begin{equation}
    c_{\mathrm{mono}}(R)=-\fft{2\text{vol}[M_{D-2}]}{4G^{(D+1)}_N}\left.\fft{e^{\tilde f}}{\sqrt{1+z'^2}}\fft{R}{z'}\fft{dz}{dR}\right|_{r=R_c}.
\end{equation}
Note that this expression is evaluated at the cutoff point $z=\epsilon$.  Assuming AdS$_{D+1}$ asymptotics, $e^f\sim e^g\sim L_{UV}/z$, we can show that $z'=-dz/dR$ at the boundary.  Hence
\begin{equation}
   c_{\mathrm{mono}}(R)=\fft{2\text{vol}[M_{D-2}]}{4G^{(D+1)}_N}\left.\fft{e^{\tilde f}}{\sqrt{1+z'^2}}R\right|_{r=R_c}.
\end{equation}
Finally, using the first integral, (\ref{eq:fint}), we find the compact expression
\begin{equation}
    c_{\mathrm{mono}}(R)=\fft{2\text{vol}[M_{D-2}]}{4G^{(D+1)}_N}e^{\tilde f(z_0)}R,
\label{eq:cmonofin}
\end{equation}
which generalizes the result of \cite{Myers:2012ed} to the case where the surface wraps an internal space $M_{D-d}$.  While this looks relatively simple, it is still a non-local expression in that it depends on both $R$ and the cap-off point $z_0$.  The latter is given by inverting the integral expression (\ref{z:to:R}), namely
\begin{equation}
    R=\int_0^{z_0}dz\fft{\mathcal F(z)}{\sqrt{1-\mathcal F(z)^2}}.
\label{eq:Rz0}
\end{equation}

As in \cite{Myers:2012ed}, it is possible to demonstrate monoticity of $c_{\mathrm{mono}}$ as a function of the cap-off point $z_0$.  To do so, we start with
\begin{equation}
    \fft{dc_{\mathrm{mono}}}{dz_0}=\fft{2\text{vol}[M_{D-2}]}{4G^{(D+1)}_N}e^{\tilde f(z_0)}\left(\fft{dR}{dz_0}+R\tilde f'(z_0)\right).
\label{eq:dcmdz0}
\end{equation}
In order to compute $dR/dz_0$ it is useful to remove the square-root singularity in the integrand of (\ref{eq:Rz0}) by integrating by parts
\begin{equation}
    R=\lim_{\varepsilon\to0}\left(\fft1{\mathcal F'(\varepsilon)}-\int_\varepsilon^{z_0}\sqrt{1-\mathcal F^2}\fft{\mathcal F''}{(\mathcal F')^2}dz\right).
\label{eq:Rz0n}
\end{equation}
Note that this expression has to be regulated because the individual terms diverge at the UV boundary, but combine to give a finite result.  Using
\begin{equation}
    \fft{\partial\mathcal F}{\partial z_0}=\mathcal F\tilde f'(z_0),
\end{equation}
we can show that
\begin{equation}
    \fft{dR}{dz_0}=\lim_{\varepsilon\to0}\left(-\fft1{\mathcal F'(\varepsilon)}+\int_\varepsilon^{z_0}\fft1{\sqrt{1-\mathcal F^2}}\fft{\mathcal F''}{(\mathcal F')^2}dz\right)\tilde f'(z_0).
\label{eq:dRdz0}
\end{equation}
Inserting (\ref{eq:dRdz0}) and (\ref{eq:Rz0n}) into (\ref{eq:dcmdz0}) then gives
\begin{equation}
    \fft{dc_{\mathrm{mono}}}{dz_0}=-\fft{2\text{vol}[M_{D-2}]}{4G^{(D+1)}_N}e^{2\tilde f(z_0)}(e^{-\tilde f'(z_0)})'\int_0^{z_0}\fft{\mathcal F^2}{\sqrt{1-\mathcal F^2}}\fft{\mathcal F''}{(\mathcal F')^2}dz.
\end{equation}
Note that the factor of $\mathcal F^2$ in the numerator makes the integrand finite at $z=0$ so the regulator $\varepsilon$ can be removed.

We now recall that NEC1, (\ref{eq:NEC1}), directly implies the inequality (\ref{eq:tf''}), which is equivalent to $\mathcal F''\ge0$.  Hence the integrand and the resulting integral is non-negative.  In addition, since $(e^{-\tilde f})'$ starts positive in the UV and is concave up, it remains positive throughout the flow.  This is now sufficient to demonstrate the monoticity of $c_{\mathrm{mono}}$, namely
\begin{equation}
	\fft{dc_{\mathrm{mono}}}{dz_0}\le0.
\end{equation}
As a two-dimensional central charge function, this flows from infinity in the UV to $c_{\mathrm{IR}}$ in the IR.

\bibliographystyle{JHEP}
\bibliography{RG-Across-D}

\end{document}